%% file: HP-arxiv.tex
\documentclass[12p]{imsart}
\usepackage{amsmath,amssymb,amsthm}
\usepackage[titletoc,title]{appendix}
\usepackage[english]{babel}
\newtheorem{mytheorem}{Theorem}
\usepackage[section]{placeins} 
\usepackage{trimspaces,enumerate,subfig, xcolor, float}
\usepackage{graphicx}
\graphicspath{{img1/}}
\usepackage{url}

\RequirePackage{natbib}
\usepackage{hyperref}
\hypersetup{colorlinks=true,citecolor=blue,urlcolor=blue}

\startlocaldefs

\def\parencite{\citep}
\def\textcite{\citet}

\def\P{\mathbb{P}}
\def\p{\mathbf{p}}
\def\E{\mathbf{E}}
\def\I{\mathbb{I}}

\def\Z{\mathbf{Z}}

\def\z{\mathbf{z}}
\def\S{\mathbf{S}}
\def\s{\mathbf{s}}

\def\d{\mathbf{d}}

\def\zhj{\(z_{hj}\)}
\def\shj{\(s_{hj}\) }
\def\h{\(h\) }
\def\j{\(j\) }
\def\yh{\(\gamma_h\) }
\def\pj{\(\rho_j\) }
\def\yv{\boldsymbol{\gamma}}
\def\pv{\boldsymbol{\rho}}

\AtBeginDocument{}

\endlocaldefs

\begin{document}
\begin{frontmatter}
  \title{A hierarchical Bayesian model for predicting ecological interactions using scaled evolutionary relationships}
\runtitle{Bayesian model for predicting ecological interactions}









\author{Mohamad Elmasri$^{1,}\thanks{Corresponding author {mohamad.elmasri@mail.mcgill.ca}}$,
  Maxwell J. Farrell$^{2}$, \\ T. Jonathan Davies $^{3}$
  David A. Stephens$^1$\\
  $^1$Department of Mathematics and Statistics, \\
  $^2$Department of Biology \\
  McGill University \\
  $^{3}$Botany, Forest \& Conservation Sciences, \\ University of British Columbia
}


\begin{abstract}\hspace{1em} Identifying undocumented or potential future interactions among species is a challenge facing modern ecologists. Recent link prediction methods rely on trait data, however large species interaction databases are typically sparse and covariates are limited to only a fraction of species. On the other hand, evolutionary relationships, encoded as phylogenetic trees, can act as proxies for underlying traits and historical patterns of parasite sharing among hosts. We show that using a network-based conditional model, phylogenetic information provides strong predictive power in a recently published global database of host-parasite interactions. By scaling the phylogeny using an evolutionary model, our method allows for biological interpretation often missing from latent variable models. To further improve on the phylogeny-only model, we combine a hierarchical Bayesian latent score framework for bipartite graphs that accounts for the number of interactions per species with the host dependence informed by phylogeny. Combining the two information sources yields significant improvement in predictive accuracy over each of the submodels alone. As many interaction networks are constructed from presence-only data, we extend the model by integrating a correction mechanism for missing interactions, which proves valuable in reducing uncertainty in unobserved interactions.
\end{abstract}


\end{frontmatter}

\section{Introduction}
\label{firstpage}
As we enter into a data revolution in the study of biodiversity \parencite{LaSalle2016}, global databases of species interactions are becoming readily available \parencite{wardeh2015database,Stephens2017,POELEN2014148}. However, most ecological networks that represent the interactions among organisms are only partially observed, and fully characterizing all interactions via systematic sampling involves substantial effort that is not feasible in most situations \parencite{Jordano2015}. Approaches to predict highly probable, yet previously undocumented links in ecological networks will help to expand our understanding of biodiversity, and can aid in the proactive surveillance of pathogens that infect multiple host species \parencite{Farrell2013}. 

Many potential approaches exist for link prediction in networks, a large group of them can be classified under covariates or feature models, where covariates of a pair of nodes are used to determine the likelihood of their interaction. The latent space model, introduced by \cite{hoff2002latent}, came to augment the former approach by representing each node (\(i\)) as a point \(s_{i}\) in a latent low dimensional space. The likelihood of the edge \((i,j)\) is driven by the individual covariates of each node, and a form of distance \(d(s_i,s_j)\) of the corresponding pairs in the latent space. Such an approach proved valuable in link prediction for social networks for many reasons, including i) the abundance of covariate data in social networks, and ii) most applications favour predictive power over interpretability.

A number of recent approaches for link prediction in ecological networks rely on trait data and node-specific features, such as body size or similarity of trophic interactions \parencite{Williams2000,Petchey2008,Gravel2013,Bartomeus2013,Stock2017,dallas2017predicting,Bastazini2017,Olival2017}. While these approaches work well for small scale datasets, they scale poorly to large-scale ecological datasets in which traits determining species interactions are often unknown or are available only for a limited subset of species \parencite{MoralesCastilla2015}. When trait information is limited, evolutionary relationships among species may be used as a proxy to study species interactions \parencite{Webb2002}. Phylogenetic trees are a representation of the evolutionary relationships among species, which provide means to quantify ecological similarity \parencite{Wiens2010} and co-evolutionary history \parencite{Davies2008}. Just as many species traits co-vary with phylogeny, species interactions are also phylogenetically structured \parencite{Gomez2010}. Incorporating phylogeny into ecological link prediction has the added benefit that it is universally applicable across all systems, and offers added biological interpretability over current latent variable models.

Different approaches have been proposed to incorporate phylogeny-based similarity in link prediction \citep{Ovaskainen2016,Ovaskainen2017,Chiu2011,Bastazini2017,Pearse2013}. Despite the emerging interest in this topic, currently proposed models treat the phylogeny as fixed or linearly scaled, and do not offer approaches to capture the underlying evolutionary processes that determine species differences.

Evolutionary biologists have developed methods of transforming phylogenies to represent alternative modes of evolution \parencite{pagel1999inferring, harmon2010early}. Rescaling the tree using these approaches alters the dependence structure among hosts, yielding improved predictions that can also be interpreted in the context of a model of trait evolution. This allows for added flexibility in the incorporation of phylogenetic information, as the dissimilarity of potential traits underlying ecological interactions may evolve under different processes than that expected by the inferred phylogeny.

In this work, we show that single-parameter (non-linear) tree scaling based on evolutionary models improves predictive performance and allows for predictions that would otherwise be overlooked by contemporary link prediction models. Shifting away from linearity results in theoretical and computational issues. Theoretically, the conditional nature of phylogenies forces interaction probabilities to be specified conditionally on other interactions, hence, the joint distribution (if it exists) might be inaccessible. As a consequence, efficient and scalable sampling methods are required, as proposed in this work. To our knowledge, this work is the first to attempt incorporating phylogenetic evolutionary scaling in link prediction, by incorporating non-linear phylogenetic scaling.

We develop a phylogeny-based framework for predicting undocumented links using a recent global database of host-parasite interactions \parencite{Stephens2017}. In host-parasite networks, parasite community similarity is often constrained by evolutionary distances among hosts \parencite{Gilbert2007, Davies2008, Streicker2010, Braga2014, Huang2015}. We focus on wild mammal hosts that are most closely related to domesticated ungulates and carnivores, as these species are known to harbour diseases of concern for humans and livestock \parencite{Cleaveland2001}, and include many species that are threatened with extinction due to infectious diseases \parencite{Pedersen2007}. We incorporate phylogenetic information as a weighted network, where weights quantify pairwise host similarities. This approach allows for easy expansion to different forms of dependency, if phylogenetic information is unavailable, or if other dependency structures are preferred. However, we show that phylogenetic information alone can generate accurate point estimate predictions. We improve our initial point estimate by incorporating a single-parameter tree scaling model which results in posterior distributions for the probabilities of each host-parasite interaction.

We then show that this phylogeny-only model can be extended by using node-specific affinity (sociability) parameters, mimicking that of covariate-based network models such as \cite{hoff2002latent, hoff2005bilinear, chung2006complex, bickel2009nonparametric}.

To facilitate the construction of the full joint distribution, we first augment the model using a hierarchical latent variable framework. The latent variable acts as an underlying scoring system, with higher scores attributed to more probable links. Second, we apply a method similar to the iterated conditional modes approach in auto-dependent models of \cite{besag1974spatial} to deal with the conditional dependency imposed by phylogeny, and include a method to account for uncertainty in unobserved interactions. Our approach allows for robust predictions for large species interaction networks with limited covariate data, and can be extended to any bipartite network with a dependency structure for one of the interacting classes.

\section{Data}\label{sec:data}
We illustrate our framework on the Global Mammal Parasite Database version 2.0 (GMPD), described in \cite{Stephens2017}. The GMPD contains over 24,000 documented associations between hosts and their parasites collected from published reports and scientific studies. The assumed interactions are based on empirical observations of associations between host-parasite pairs using a variety of evidence types (visual identification, serological tests, or detection of genetic material from a parasite species in one or more host individuals). Associations are reported along with their publication reference. The GMPD gathers data on wild mammals and their parasites (including both micro and macroparasites), which are separated into three primary databases based on host taxonomy: Primates, Carnivora, and ungulates (terrestrial hooved mammals in the orders Artiodactyla and Perissodactyla). We restricted our analyses to the ungulate and Carnivora subsets because of prior experience with these data \citep{JANE:JANE12342}, and tractability of the size of the resulting network.

The GMPD was used to construct a bipartite binary matrix, where rows represent hosts and columns parasites and documented associations (at least one piece of evidence that a parasite infects a given host species) are indicated by \(1\). We construct host pairwise similarities as the inverse of phylogenetic distances calculated from the mammal phylogeny of \textcite{Fritz2009}, which involved collapsing host subspecies to species. We excluded parasites that were not reported to species level. This resulted in a GMPD subset with 4178 pairs of interactions among 236 hosts and 1308 parasites. Out of these 1308 parasites, 695 were found to associate with a single host (\(\approx 54\%\) of parasites, and \(\approx 17\%\) of total interactions) 

One of the models proposed in Section \ref{sec:Network} (the phylogeny-only model) can only be specified for multi-host parasites. Thus, for the purpose of model comparison, we remove single-host parasites, reducing the GMPD to 3483 interactions among 229 hosts and 613 parasites. In subsequent analyses we refer to the database without single-host parasites, unless otherwise specified.

\begin{figure}[ht!]
  \centering
  \subfloat[][host phylogeny]{\includegraphics[width=0.5\textwidth]{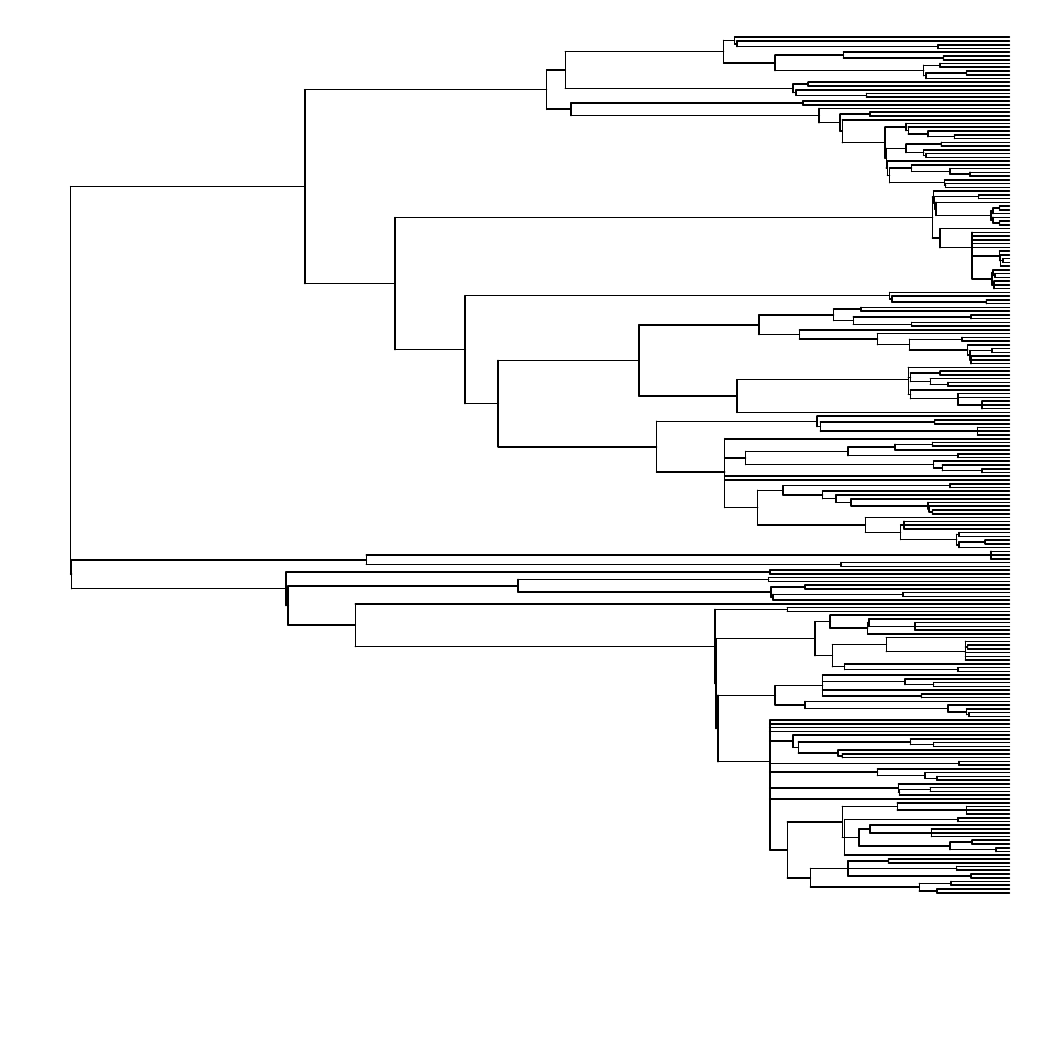}}
  \subfloat[][GMPD]{\includegraphics[width=0.5\textwidth]{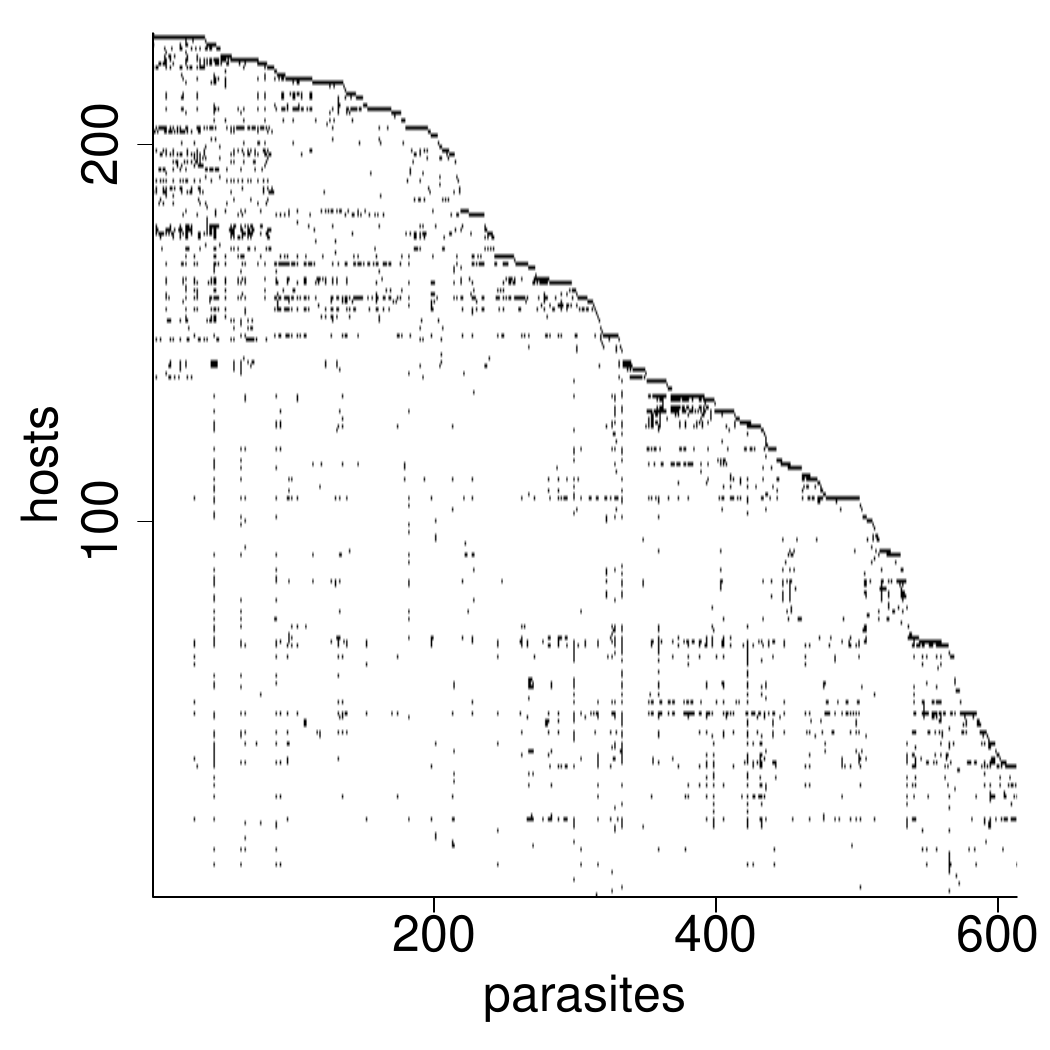}}
  \caption{ a) The host phylogeny and b) the left ordered interaction matrix $\mathbf Z$  of the GMPD, without single-host parasites.}
  \label{fig:GMP-Z}
\end{figure}

Figure \ref{fig:GMP-Z} shows the left-ordered interaction matrix \(\Z\) of GMPD, and corresponding host phylogeny. The matrix \(\Z\) is sparse, and the degree distributions of both hosts and parasites exhibit a power-law structure (Online Supplement Figure \ref{fig:GMP-Degree-G}). 

\section{Bayesian hierarchical model for prediction of ecological interactions}
\subsection{Network-based latent score model}\label{sec:Network}

Conditional modelling is common in many biological network models, where the class of auto-models of \textcite{besag1974spatial} and latent space models of \cite{hoff2002latent} are widely applied. One example is the use of a network-based auto-probit model in \textcite{jiang2011network}, where a protein-protein association network is used as a prior to predict protein functional roles conditional on the roles of neighbouring proteins. Such network-based models rely on a pre-existing binary or weighted network with a clearly defined neighbourhood structure. Probabilities are then derived by averaging over neighbouring nodes.

Evolutionary distances among species, represented by phylogenies, translate to a fully connected weighted network. Since pairwise distances among species are measured relative to their most recent common ancestor, the same distance may be assigned to multiple host pairs. A neighbourhood structure can be constructed with weights on the fully connected network, or a threshold method can be applied, but with two main drawbacks: i) the complexity of inferring the threshold parameter, and ii) the interpretation of the threshold with respect to evolutionary distance.

In the case of host-parasite interactions, parasites are often found to interact with closely related hosts, but in some cases may make large jumps in phylogeny and interact with distantly related hosts \citep{Parrish2008,park2018characterizing}. To account for such behaviour and to overcome the drawbacks of the threshold method, we let the probability of a host-parasite interaction be driven by the sum of evolutionary distances to the documented hosts of the parasite.

Let \(\Z\) be an \(H \times J\) host-parasite interaction matrix, where the binary variable \(z_{hj}\) denotes whether an interaction between host \h and parasite \j has been observed. Quantifying divergences starting from the root of the tree, let \(T_{hi}\) be a unit-free pairwise phylogenetic distances among hosts \(h\) and \(i\), and their common ancestor \(k\), such that \(T_{hi} = T_{hk} + T_{ik} = (t_{h} - t_{k}) + (t_{i} - t_{k})\). Phylogenetic distances are commonly measured in millions of years, so to arrive at the unit-free distance we divide all distances by the total depth of the tree. 

A valid and basic conditional probability distribution of host \(h\) interacting with parasite \(j\) can be defined in terms of the pairwise phylogenetic distances from host \(h\) to all other hosts interacting with parasite \(j\), as

\begin{equation}\label{eq:phylogeny-only-basic}
   \P(z_{hj}=1 \mid \z_{(-h)j}) = 1- \exp(-\delta_{hj}), \quad \delta_{hj} = \sum_{\stackrel{i=1}{i\neq h}}^{H}\frac{z_{hj}}{T_{hi}}, 
\end{equation}
where \(\z_{(-h)j} \) is the set of interactions of the \(j\)-th parasite among the \(H\) hosts (\(\z_{.j}= (z_{1j}, \dots, z_{Hj})\)), excluding that of the \(h\)-th host.

The conditional distribution \eqref{eq:phylogeny-only-basic} allocates higher probabilities when closely related hosts interact with a given parasite, or when many distantly related hosts also interact. The more distantly related the hosts are, the smaller the value of \(1/T_{hi}\). Of course, the probability distribution in \eqref{eq:phylogeny-only-basic} is conditional on a probabilistic model for \(T\).

The exponential choice in \eqref{eq:phylogeny-only-basic} is motivated by the power-law structure shown in Figure \ref{fig:GMP-Z} and Online Supplement Figure \ref{fig:GMP-Degree-G}, thus we expect interaction probabilities to decay exponentially with respect to the parameters. Other probability structures are viable, though with no tractability guarantees. We later show that, under such construction, a tractable probabilistic framework is possible for a class of latent variables with tail probabilities as in \eqref{eq:phylogeny-only-basic}. The next section introduces a family of single-parameter models that have meaningful biological interpretation.

\subsubsection{Evolutionary models and phylogeny transformations}
A focus of macroevolutionary research has been to develop models of trait evolution. A well-known model and default in many ecological applications is Brownian motion, however, transformations of the phylogenetic tree can be made to reflect alternatives in the tempo and mode of evolution. Common evolutionary models that can be defined by a transformational single parameter transformation include the early-burst (EB), delta, kappa, lambda, and the Ornstein-Uhlenbeck transformation \citep{pagel1999inferring,harmon2010early}, each scales phylogenetic distances according to a model of evolution. We term this the phylogeny-only model and define it as

\begin{equation}\label{eq:phylogeny-only}
  \P(z_{hj}=1 \mid \z_{(-h)j}) = 1- \exp(-\delta_{hj}), \quad \delta_{hj} = \sum_{\stackrel{i=1}{i\neq h}}^{H}\frac{z_{hj}}{\phi(T_{hi},\eta)}, 
\end{equation}
where \(\phi(T_{hi},\eta)\) is the transformed distance under a given evolutionary model controlled by a single parameter \(\eta\).

\begin{figure}[ht!]
  \centering
  \includegraphics[width=0.35\textwidth]{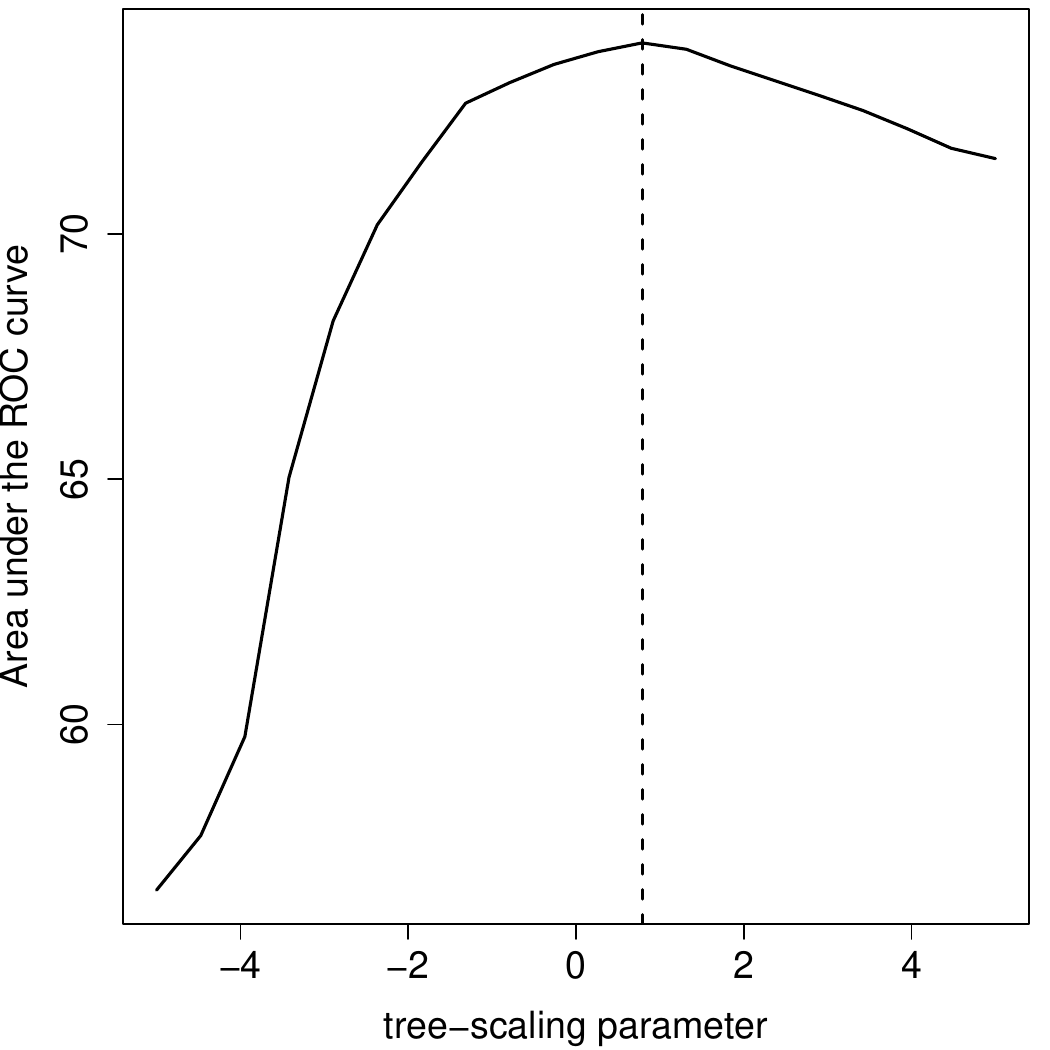}
  \caption{Area under the ROC curve evaluated over a fine grid under the phylogeny-only model \eqref{eq:phylogeny-only} with early-burst tree transformational method, with GMPD (including single-host parasites).}
  \label{fig:tree-scaling-eb}
\end{figure}

With further investigation, we find that the EB model stands-out by displaying a non-trivial convex relationship with predictive power, as shown in Figure \ref{fig:tree-scaling-eb}. In this figure we are evaluating the potential predictive accuracy of a simple phylogeny-only model (Eq. \eqref{eq:phylogeny-only}) with phylogeny scaled according to the early-burst method, for the database in Section \ref{sec:data} (for more details refer to Online Supplement Figure \ref{fig:AUC-transformational-models}).

\begin{figure}[ht!]
  \centering
\subfloat[][\(EB(T,0.02)\)]{\includegraphics[width=0.37\textwidth]{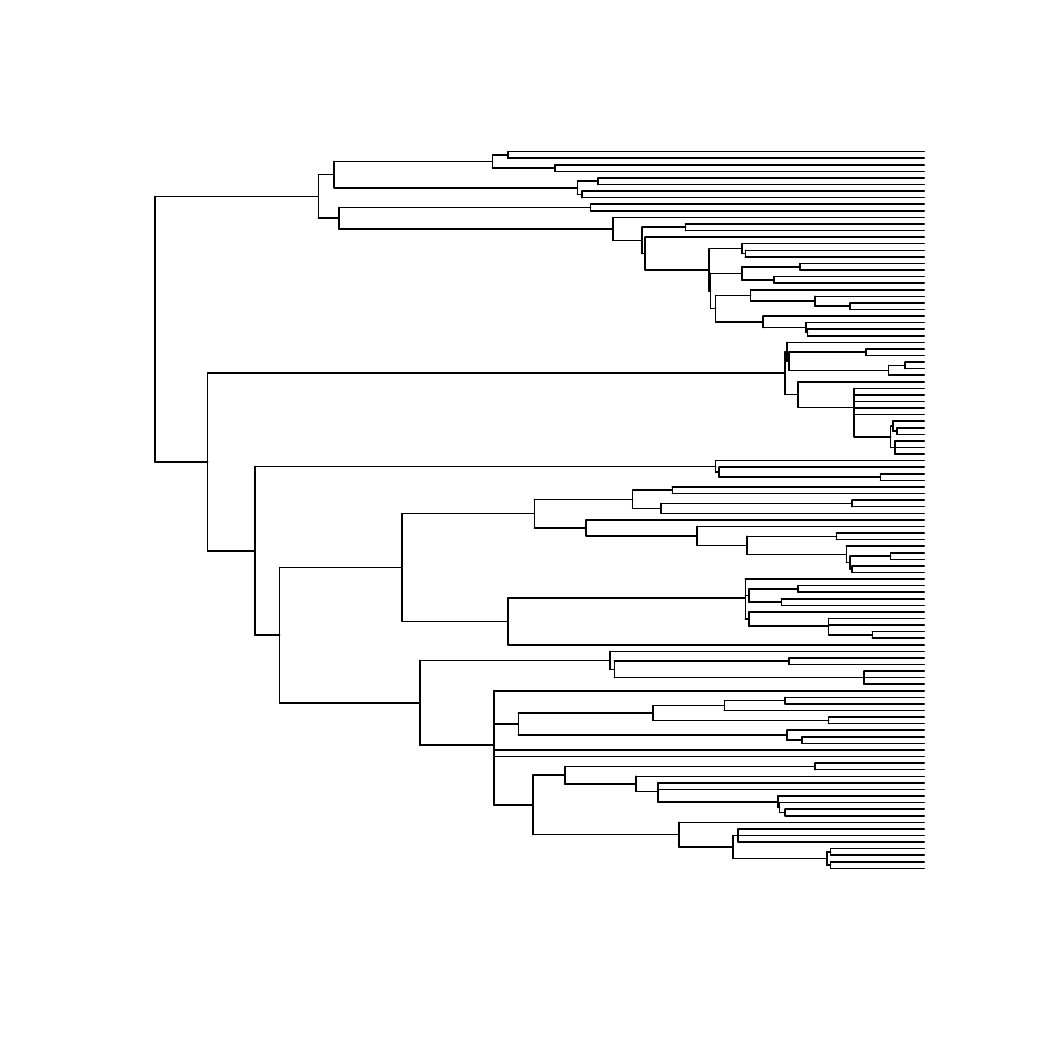}}
  \hspace{-1.2cm}
  \subfloat[][\(T\)]{\includegraphics[width=0.37\textwidth]{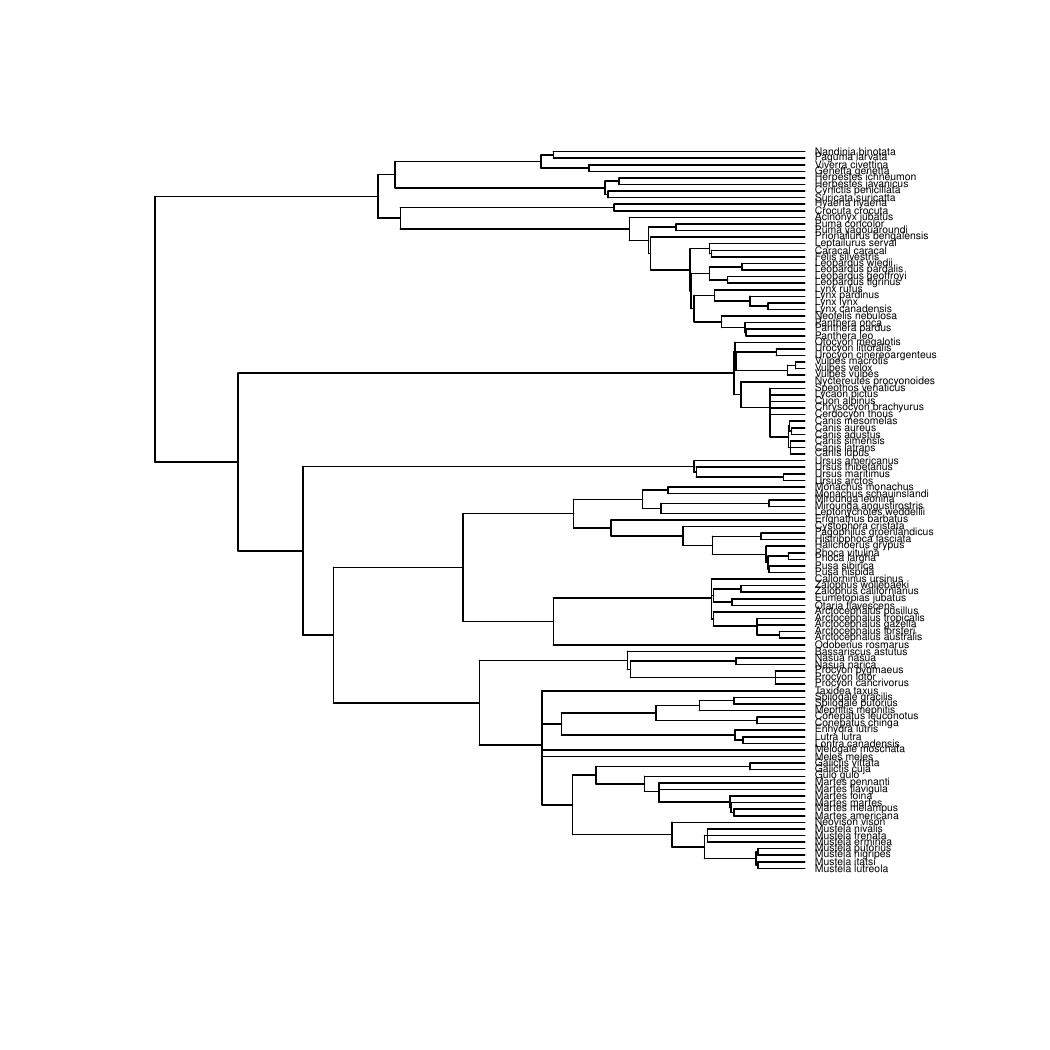}}
  \hspace{-1.2cm}
\subfloat[][\(EB(T,-0.02)\)]{\includegraphics[width=0.37\textwidth]{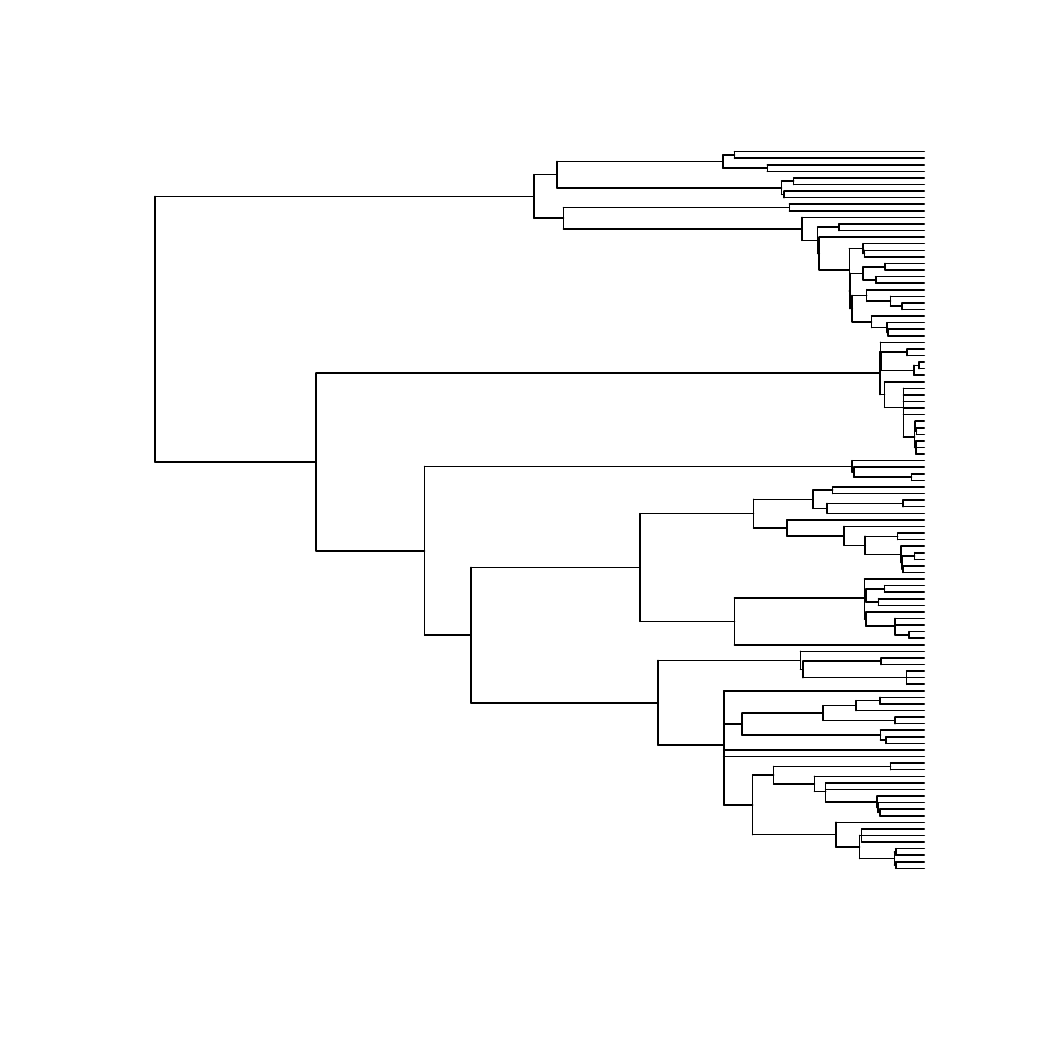}}  
  \caption{Examples of the early-burst transformation in the Carnivora subset of GMPD.}
  \label{fig:EB-Tree-example}
\end{figure}

  This supports the assumption that scaled phylogenies, based on explicit models of niche or trait evolution, can result in better predictions. The EB model allows evolutionary change to accelerate or decelerate through time, for example, evolutionary change may be fastest early in a clades history, but slows through time. The rate of change in the EB model is adjusted by a single parameter \(\eta \in \mathbb{R}\), with positive values of \(\eta\) indicating that evolution is faster earlier in history, while negative values suggests the opposite. Figure \ref{fig:EB-Tree-example} illustrates the EB model for different values of \(\eta\).

Under the EB model, the phylogenetic distance between a pair of hosts \((h,i)\) with a most recent common ancestor \(k\) is quantified as 
\begin{equation}\label{eq:early-burs}
  \phi(T_{hi}, \eta) = \phi(T_{hk}, \eta) + \phi(T_{ik}, \eta)= \frac{1}{\eta}(e^{\eta t_{h}} - e^{\eta t_{k}}) + \frac{1}{\eta}(e^{\eta t_{i}} - e^{\eta t_{k}}).
\end{equation}

Thus, for \(\eta=0\), EB reduces to the original tree distance as \(\phi(T_{hi}, 0) = T_{hi}\). While this represents one form of uncertainty in the phylogeny, future work may also incorporate uncertainty in tree topology as well as distance by using posterior distributions of trees resulting from Bayesian phylogenetic inference.  

\subsubsection{Full model}\label{sec:full-model}

Species interactions can be predicted using phylogenetic trees, though not completely, since interactions can also be driven by traits that are independent of phylogeny. In general, many network-based models assume that edge probabilities are driven by independent node affinity parameters, for example \textcite{chung2006complex, bickel2009nonparametric} and many others. Here we model the conditional probability of an interaction by combining both sources of information: phylogenetic distances and individual species affinities. Affinity parameters govern the general propensity for each organism to interact with members of the other class, larger affinities correlate with higher likelihood that an organism will interact. Let \yh\(>0\) be the affinity parameter of host \(h\), and \pj\(>0\) of parasite \(j\). The full conditional model is then
\begin{equation}\label{eq:full-model}
  \P(z_{hj}=1 \mid \Z_{-(hj)}) = 1- \exp\big (-\gamma_{h}\rho_{j}\delta_{hj}(\eta)\big ),
\end{equation}
with \(\delta_{hj}(\eta)\) as in \eqref{eq:phylogeny-only} under the EB transformation, and \(\Z_{-(hj)}\) is the interaction matrix \(\Z\) excluding \(z_{hj}\). The default value is \(\delta_{hj}(\eta)=1\) if no neighbouring interactions exist, reducing to the affinity-only model for this interaction. Alternative defaults are possible, such as the average pairwise distances in \(T\).

The affinity-only model results in a workable network prediction model, which has been shown in the literature on exchangeable random networks \parencite{hoff2002latent}. However, affinity-only models tend to generate adjacency matrices with many hyperactive columns and rows. This is because whenever a node has a sufficiently high affinity parameter it forms edges with almost all other nodes, which is likely to be unrealistic for most ecological networks. In Section \ref{sec:CrossValid} we show that both models, the affinity-only and phylogeny-only, independently result in useful predictive models that represent some variation in the data. However, each model captures different characteristics of the network and by layering them we obtain a non-trivial improvement.

Finally, we find it advantageous to use latent variables in modelling the binary variables \zhj. This facilitates the construction of the network joint distribution while accounting for the Markov network dependency imposed by \(\delta_{hj}(\eta)\). In addition, the latent variable construction becomes essential in addressing the ambiguity associated with the case when \(z_{hj} =0\), which entails two possibilities: a yet to be observed positive interaction, or a true absence of interaction due to incompatibility (implemented in Section \ref{sec:Uncertainty}). 

Thus, for each \(z_{hj}\) we define latent score \shj \(\in \mathbb R\) such that 
\begin{equation} \label{eq:Z-as-S}
z_{hj}=
\begin{cases}
1 & \quad \text{if } s_{hj}>0 \\
0 & \quad \text{otherwise, } \\
\end{cases} 
\end{equation}
where \shj \(\in \mathbb R\) is a continuous random variable acting as a latent score determining the probability of \(z_{hj}\) being an interaction. Although unobserved, \shj completely determines the binary variables \zhj. Therefore, the conditional model in \eqref{eq:full-model} can be completely specified in terms of the latent score as 
\begin{equation}\label{eq:latentScore}
\begin{aligned}
  \P(z_{hj}=1\mid \Z_{-(hj)}) =\E[\mathbb{I}_{\{s_{hj}>0\}}\mid \Z_{-(hj)}] =\P(s_{hj}>0 \mid \S_{-(hj)}) 
\end{aligned}
\end{equation}
where \(\S_{-(hj)}\) represents the interaction matrix \(\S\) excluding \(s_{hj}\), it replaces \(\Z\) as it carries the same probability events in its sign distribution.

The current formulation is flexible in the choice of distribution for \(s_{hj} \mid \S_{-(hj}\), the only imposed requirements is absolutely continuity with exponentially decaying tail probability as in \eqref{eq:full-model}. One possible choice is the Gumbel with mean parameter \(\log(\gamma_{h}\rho_{j}\delta_{hj}(\eta))\) and a scale of 1. Since we are only interested in positive reals, we use a zero-inflated Gumbel distribution for the latent score with the following density
\begin{equation}\label{eq:truncated-gumbel-density}
 \p(s_{hj}\mid \S_{-(hj)}) =  \tau_{hj}\exp(-s_{hj}-\tau_{hj}e^{-s_{hj}})\I_{\{s_{hj}>0\}} + \exp(-\tau_{hj}) \I_{\{s_{hj}= 0\}},
\end{equation}
where \(\tau_{hj} = \gamma_h \rho_j \delta_{hj}(\eta)\). Hence, the conditional joint distribution becomes 
\begin{equation}\begin{aligned}\label{eq:more-tractable-joint-dist}
  \P(z_{hj}=1,s_{hj}\mid \Z_{-(hj)}) &= \P(z_{hj}=1\mid s_{hj})\p(s_{hj}\mid \S_{-(hj)}) \\ &= \p(s_{hj}\mid \S_{-(hj)}) \mathbb{I}_{\{s_{hj}>0\}}.
\end{aligned}
\end{equation}

  The choice of a zero-truncated Gumbel was made to facilitate the construction of the joint distribution, in a manner similar to the Swendsen-Wang algorithm \parencite{Swendsen87} where a product of densities transforms to a sum in the exponential scale, improving the tractability of posteriors. Alternatively, the truncated exponential distribution can be used,  as \(s_{hj} \sim \min\{1, \text{Exp}(\tau_{hj})\}\), having the tail distribution in \eqref{eq:phylogeny-only-basic}, though it does not admit the direct interpretability as a latent score as with the Gumbel distribution.

The proposed latent score model, though intricate in formulation, is no more inferentially complex than the auto-logit model of \cite{besag1974spatial}. The reasons we use our model are: i) it exhibits a simple joint distribution for each row of \(\Z\) conditional on all others; ii) there are simple posterior distributions for the affinity parameters; and iii) we have the ability to correct for uncertainty using the latent score formulation (shown in Section \ref{sec:Uncertainty}). We could adopt other network-based conditional models, for example the family of auto-models by \cite{besag1986statistical}. One particular example is the multivariate Gaussian latent variable with a conditional mean structure of \textcite{jiang2011network}, which has a similar complexity to the phylogeny-only model in \eqref{eq:phylogeny-only}, modelling each column of \(\Z\) independently without affinity parameters. Other auto-models would also require the development of an efficient sampling scheme that makes sense for the conditional interaction probability, as done here.

\subsection{Prior and Posterior distribution of choice parameters}\label{sec:Prior}

By the Hammersley-Clifford theorem \citep{robert2013monte}, it is straightforward to verify that the joint distribution exists, as briefly shown in Online Supplement Section \ref{app:hammersley-clifford}. Even though the form is complicated, we do not need to access the joint density and instead may use a Gibbs sampler as in \citet{geman1984stochastic}. An iterative algorithm can then be used to sample from conditionally independent components of the joint distribution, with the posterior sample obtained by averaging. This approach is similar in spirit to the iterated conditional modes (ICM) algorithm of \cite{besag1986statistical}.

In the proposed model the joint distribution of rows are conditionally independent given the rest. Let \(\Z_{-(h.)}\) be \(\Z\) excluding the \(h\)-th row \(\z_{h.}\). With similar notations for \(\S\), the joint distribution of the \(h\)-th row is
\begin{equation} 
  \begin{aligned}
    \P(\z_{h.}, \s_{h.}\mid \Z_{-(h.)}) & = \gamma_{h}^{n_{h}}\bigg [\prod_{j=1}^{J}\big (\rho_{j}\delta_{hj}(\eta)\big)^{z_{hj}} \bigg ]\exp \bigg (-\sum_{j=1}^{J}s_{hj}+\gamma_{h}\rho_{j}\delta_{hj}(\eta) e^{-s_{hj}} \bigg )
\end{aligned}
\end{equation}
where \(n_{h} = \sum_{j=1}^{J}z_{hj}\) such that the row-wise joint posterior distribution is 
\begin{equation}\label{eq:postSimple}
\P(\s_{h.}, \gamma_{h}, \pv, \eta \mid \Z) \propto \P(\z_{h.}\mid \s_{h.})\P(\s_{h.} \mid\S_{-(h.)}, \gamma_{h}, \pv, \eta) \P(\gamma_{h}) \P(\pv) \P(\eta), 
\end{equation}
where \(\P(\z_{h.} \mid \s_{h.}) = \prod_{j=1}^{J}\P(z_{hj}\mid s_{hj})=1\), and \(\pv\) is the parasite affinity parameter set.

In a sweeping manner for \(h=1, \dots, H\) rows of \(\Z\), one samples \(\gamma_{h}\) from its full posterior, and \(\pv^{{(h)}} = (\rho^{(h)}_{1}\dots, \rho^{(h)}_{J})\) and \(\eta^{(h)}\) from their \(h\)-th row conditional posteriors. Obtaining an MCMC sample of \(\pv\) and \(\eta\) is done by averaging over the \(H\) samples from the row posteriors.

For prior specifications we choose a gamma distribution for both affinity parameters because of their conjugacy property. Thus, let \yh \(\stackrel{\text{iid}}{\sim}\) Gamma(\(\alpha_\gamma, \tau_\gamma\)) and \pj \(\stackrel{\text{iid}}{\sim}\) Gamma(\(\alpha_\rho, \tau_\rho\)). The full posterior distributions of \yh and the \(h\)-row partial posterior of \(\rho^{(h)}_j\), respectively, are 
\begin{equation}\label{eq:GammaRhoPost}
\begin{aligned}
\rho_j^{(h)} \mid \z_{h.},\s_{h.}  &\sim \text{ Gamma}\Bigg( \alpha_\rho +z_{hj}, \tau_\rho + \gamma_h\delta_{hj}(\eta)e^{-s_{hj}}\Bigg ),\\
\gamma_h\mid \z_{h.},\s_{h.} &\sim \text{ Gamma}\Bigg( \alpha_\gamma + n_h, \tau_\gamma +  \sum_{j=1}^J\rho_j\delta_{hj}(\eta)e^{-s_{hj}} \Bigg ). \\
\end{aligned}
\end{equation}

In the case of the scaling parameter \(\eta\) we assume a constant prior for simplicity and computational stability, although this could be readily modified to any subjective prior. 

The latent score is updated, given all other parameters as
\begin{equation}\label{eq:sampling-of-S}
  s_  {hj} \mid z_{hj}, \S_{-(hj)} \sim 
  \begin{cases}
    \chi_0 \quad &  \text{if } z_{hj} =0 \\
    \text{tGumbel}\bigg (\log\gamma_h\rho_j + \log\delta_{hj}(\eta),1, 0\bigg )\quad  &  \text{if } z_{h j} =1, \\
  \end{cases}
\end{equation}
where \(\chi_0\) is an atomic measure at zero and tGumbel\(\big (\tau, 1, 0\big )\) is the zero-truncated Gumbel with density 
\[\frac{\exp(-(s-\tau + e^{-(s-\tau)} )}{1-\exp(-e^\tau)}\chi_{(0,\infty)}(s).\]

The adaptive Metropolis-Hastings algorithm \parencite{haario2001adaptive} within Gibbs is used to update the model parameter. For additional details on the model and the MCMC method sampling algorithm refer to Online Supplement Section \ref{sec:model-gener-sett}.

\section{Uncertainty in unobserved interactions}\label{sec:Uncertainty}
In ecological networks it is unlikely that all potential links will be represented or observed. Some unobserved exist but are undocumented due to limited or biased sampling, while others may be true absences or ``forbidden" links \parencite{MoralesCastilla2015}. Evidence used to support an interaction will vary depending on the nature of the system, but it is often assumed that an interaction exists if at least one piece of evidence indicates so \parencite{Jordano2015}. 

This raises concern about the uncertainty of interactions in two ways. The first is due to uncertainty in documented interactions as false positive detection errors may occur, potentially as a result of species misidentification, sample contamination, or for parasites, unanticipated cross-reactions in serological tests \parencite{Aguirre2007}. We believe it would be useful for the scientific community to identify weakly supported interactions that may require additional supporting evidence, however our primary motivation is identification of ``novel" interactions, which is complicated by uncertainty in unobserved interactions.

The second concern arises when unobserved associations are by default assumed to be true absences. As discussed earlier, ecological networks are often under-sampled, and some fraction of unobserved interactions may occur but are currently undocumented, or represent potential interactions that are likely to occur given sufficient opportunity. Based on this assumption we build a measure of uncertainty in unobserved interactions by modifying our proposed model in \eqref{eq:full-model}. In \eqref{eq:Z-as-S}, we have assumed that \(\P(z_{hj}=1\mid s_{hj}>0)\) is degenerate at 1 given \(s_{hj}\). Thus we have only sampled positive scores for the case when \(z_{hj}=1\), as shown in \eqref{eq:sampling-of-S}. As a result, the posterior predictive distribution is only considered for the case when a pair has no documented associations (\(z_{hj}=0)\), underlining the assumption that the data is complete and trusted. In presence-only data, the objective is to model the non-trivial object \(\P(z_{hj}=1, \text{``a missing link"}\mid s_{hj}>0 )\). To account for such uncertainty, we attempt to approximate the proportion of interactions that are missing links in the latent space by measuring the percentage of positive scores where the input is 0 (\(z_{hj}=0\)) as
\begin{equation}\label{eq:uncertain-prop-function}
\p(z_{hj} =0 \mid s_{hj}, g) = 
\begin{cases}   1,   &\text{if } s_{hj} =0, \\
                g,   &\text{if }  s_{hj} > 0 ,\\
\end{cases}
\end{equation}
where \(g\) is the probability that an interaction is unobserved when the latent score indicates an interaction should exist. If \(g\) is large and close to 1, it is likely that many of the unobserved interactions could or should exist. Introducing \(g\) to the model affects all parameter estimates and the notion of \(\Z\). Therefore, the posterior predictive distribution is now considered for both cases. For the case of a documented association, the probability of an interaction is defined in \eqref{eq:full-model}, and for the case of no documentation the same probability is weighted by \(g\) as shown in detail in \eqref{eq:CondSampleS}.

Here we implicitly assume that \(g\) is common to all pairs of interactions. It is possible to assign a different parameter to groups of interactions. Nonetheless,
by the principle of parsimony, we favoured simplicity. This kind of construction has been used earlier by \textcite{WeirPett2000} when modelling spatial distributions to account for uncertainty in regions with unobserved statistics, and later by \textcite{jiang2011network} in modelling uncertainty in protein functions. 

\subsection{Markov Chain Monte Carlo algorithm}\label{sec:MCMC2}
Introducing  a measure of uncertainty in the model does not alter the MCMC sampling schemes introduced in Section \ref{sec:Prior}. The variables \(\yv, \pv\) and \(\eta\) are still only associated with \(\S\), nonetheless, by introducing the measure of uncertainty, the conditional sampling of each individual \(s_{hj}\)  is now  
{\small
\begin{equation}
  \label{eq:CondSampleS}
  \p(s_{hj} \mid \S_{-(hj)}, \Z, g) =
  \begin{cases}
    \frac{1}{\psi(\bar s_{hj})} \tau_{hj}\exp \bigg ({-(s_{hj} +\tau_{hj}e^{-s_{hj}})} \bigg ), & s_{hj}  > 0,\quad  z_{hj}=1, \\ 
    0,                                   & s_{hj}  = 0,\quad z_{hj}=1, \\ 
    \frac{g}{\theta(g,\bar s_{hj})}\tau_{hj}\exp \bigg ({-(s_{hj} + \tau_{hj}e^{-s_{hj}})} \bigg ), & s_{hj} >0,\quad z_{hj}=0,\\
    \frac{1}{\theta(g,\bar s_{hj})} 1-\psi(\bar s_{hj}),& s_{hj} =0,\quad z_{hj}=0, \\
  \end{cases}
\end{equation}}
where \(\tau_{hj} = \gamma_h\rho_j\delta_{hj}^\eta\), \(\psi(\bar s_{hj}) = \int_0^\infty \p(s \mid \S_{-(hj)}) \d s = 1 - \exp \big({-{\gamma_h\rho_j}{\delta_{hj}^\eta}} \big )\), and \(\theta(g, \bar s_{hj}) = g\psi(\bar s_{hj}) + 1-\psi(\bar s_{hj}) \).  

Sampling the uncertainty parameter is performed using the row-wise conditional distribution as 
\begin{equation}
  \label{eq:Sampleg}
  \P(g \mid \s_{h.}, \z_{h.}) \propto  \P(\z_{h.}\mid \s_{h.},g)\P(\s_{h.} \mid \S_{-(h.)})\;.\;\P(g) \propto g^{N_{-+}}(1-g)^{N_{++}}, 
\end{equation}
where \(N_{-+} = \#\{(h,j): \z_{hj}=0, s_{hj}>0 \},\; N_{++} = \#\{(h,j): z_{hj}=1, s_{hj}>0\}\), and \(P(g)\) is a uniform. Since the sampling is done by iteratively cycling through the rows of \(\Z\), in analogy to the ICM method, a sample of \(g\) is the average of the \(H\) row samples.

\section{Alternative models and comparison by cross-validation}\label{sec:CrossValid}
To validate the predictive performance of the proposed latent score full model, we compare it to the two submodels of Section \ref{sec:Network} (the affinity-only and the phylogeny-only models), and to the bilinear latent-distance model with two of its submodels (the bilinear and the latent-distance models) \citep{hoff2002latent,hoff2005bilinear}. The bilinear model excludes phylogenetic information, and assumes a logit formulation with an intercept and an affinity coefficient for each node, hence, it correlates with the affinity-only model in interpretation. The latent-distance mode assumes a one-dimensional latent variable for each node, and pairwise distances between nodes are the Euclidean distance between their respective latent variables, aligning it with the phylogeny-only model in interpretation. Therefore, the latent-distance model excludes explicit phylogenetic information, and distances are not informed by the association matrix, as in the case of our phylogeny-only model. Additional latent dimensions can be added to the latent-distance model, which might improve prediction, though at an extra cost of interpretation. The bilinear latent-distance model combines both former components, and all three variates of this model are implemented using {\sf latentnet} {\sf R}-package \citep{latentnetPaper, latentnetPackage}, as \verb ergmm(Z~rsociality) , \verb ergmm(Z~euclidean(d=1)) , and \verb ergmm(Z~rsociality+euclidean(d=1)) , respectively. The {\sf latentnet} package readily provide alternative forms of distances, though, for this dataset, we found that the Euclidean distance has a better performance.

Finally, we also compare our proposed model to a nearest-neighbour (NN) algorithm, in which we set the distances between hosts proportional to the number of parasite species they share, also known as the Jaccard distance. This form of distance does not require additional data other than \(\Z\). Hence, for this algorithm, we let the probability of a host-parasite interaction be equal to the average number of host-neighbours with documented association to the parasite, within the \(k\)-closest host-neighbours. A host can share different sets of parasites with different hosts, though at times the size of the different sets might be the same, yielding exact Jaccard pairwise distances to multiple hosts. Therefore, we let \(k\) be driven by the number of shared parasites, excluding the parasite of interest. For example, \(k=2\) would define a neighbourhood of all hosts that have at least the 2nd highest number of shared parasites for a host of interest. In brief, \(k\) is chosen by cross-validation; the details of the optimization criterion is discussed in later.

Link probabilities in many network models, as the one proposed here, are driven by the count of links of their respective nodes. Hence, in cross-validation, it is natural to hold a random portion of the observed links out from the training set and validate with them. In our settings, the predictive performance of each model is evaluated using the average of 5-fold cross-validations, where in each fold we set a different set of the observed interactions (\(z_{hj}=1\)) in \(\Z\) to unknowns (\(z_{hj}=0\)) while attempting to predict them using the remaining interactions. The same folds are used across all evaluated models. The predictive performance of each fold is assessed methodologically, using the proper scoring rules proposed by \citet{gneiting2007strictly} and \citet{ehm2016quantiles}, graphically, using the receiver operating characteristic (ROC) curves, and numerically, using the percentage of recovered interactions.

The recent work of \citet{ehm2016quantiles} has shown that, under unimportant regularity conditions, every score (loss) function consistent for the probability of binary events admits a representation as a mixture of the form
\[ L(p, y) = \int_{0}^{1}L_{\theta} (p,y) \mathrm{d} \mathbf{H}(\theta), \]
with \(\mathbf H\) being a non-negative measure, and
\begin{equation} \label{scoring-rule}L_{\theta}(p,y) =
  \begin{cases} 
    \theta, & y=0,\; p > \theta, \\
    1 - \theta, & y=1, \; p \leq \theta, \\
          0, & \text{otherwise,}
  \end{cases}
\end{equation}
for a predictive probability \(p\) of binary event \(y\), and \(\theta \in [0,1]\). The choice of the mixing measure \(\mathbf H\) determines the score function. For example,
when \(\mathbf H\) is twice the Lebesgue measure, \(L\) is the ubiquitous Brier score with \(L(p,0) = p^{2}\) and \(L(p,1) = (1-p)^{2}\). For alternative score functions of dichotomous events, please refer to Table 1 in \cite{gneiting2007strictly}.

In applied problems \(\theta\) in \eqref{scoring-rule} has an economic interpretation, for example in binary settings, \(\theta\) can represent the cost of a false positive prediction, \(1-\theta\) is the cost of a false negative, while true positive has no cost. Hence, for a fixed \(\theta\), an optimal strategy is to predict positive events with probability \(> \theta\) and negative events with probability \(< \theta\). This has a direct implication on model comparison;
if a model receives consistently a lower mean score for every \(\theta\) in comparison to alternative models, then the model dominates in predictive power. The choice of a proper scoring function becomes irrelevant in this case as the model would dominate for any other proper scoring rule \citep{ehm2016quantiles}.

In empirical settings one can compare competing models graphically by plotting the so-called Murphy's diagrams, which displays, for each model considered, the mean of the elementary score function \(L_{\theta}\) over different values of \(\theta \in [0,1]\). In our settings, for a fixed value of \(\theta\), we calculate the average of \(L_{\theta}\) over the test set of each cross-validation fold, with posterior predictive from its training set. The final score curve for each model is the average of scores over cross-validation folds.


The ROC curves is a popular graphical tool for the assessment of discrimination ability in binary prediction problems. For each model, an ROC curve is obtained by thresholding the predictive probabilities of the full unknowns in each cross-validation fold, calculating the true and false positive rates on each fold, and then averaging them over the 5-folds. With this process, the posterior predictive interaction matrix is obtained at the threshold value that maximizes the area under the ROC curve (AUC). Moreover, for each fold, the \(k\) parameter of the NN algorithm is chosen as the value that maximizes the AUC over the training set of that fold.

The phylogeny-only model in \eqref{eq:phylogeny-only} is ill-formulated for the case of single-host parasites, since \(\delta_{hj}(\eta) =0\). Therefore, for comparison across the models, each held-out portion is constructed to ensure that at least two interactions are kept in each column of \(\Z\). By this restriction, each held-out portion is approximately 11\% and 13\% of documented associations for the datasets with and without single-host parasites, respectively.

\section{Results} \label{sec:Results}

\subsection{Parameter estimation for the latent score full model}
For the GMPD we first fit the model proposed in Section \ref{sec:Network}. We run 20000 MCMC iterations and the same for burn-in for posterior estimates. In total we have \(J + H + 1\) parameters to estimate: an affinity parameter for each host and each parasite, and a tree scaling parameter for the host phylogeny.

Standard convergence diagnostics showed that all parameters had converged (For convergence and diagnostic plots refer to Online Supplement Section \ref{app:Diagnostic-Plots-Extra}). It is worth noting that the posterior distributions of the host parameters (\(\gamma\)) show large variation, which reflects that some hosts are more likely to interact with parasites, or have been more intensively studied. The magnitude of the unit-free scaling parameter \(\eta\) is found to concentrate around 1.702 with 95\% credible interval as (0.391, 5.805), indicating accelerating evolution compared to the original tree. 
\begin{figure}[ht!]
  \centering
\subfloat[][Murphy's diagrams]{\includegraphics[width=0.5\textwidth]{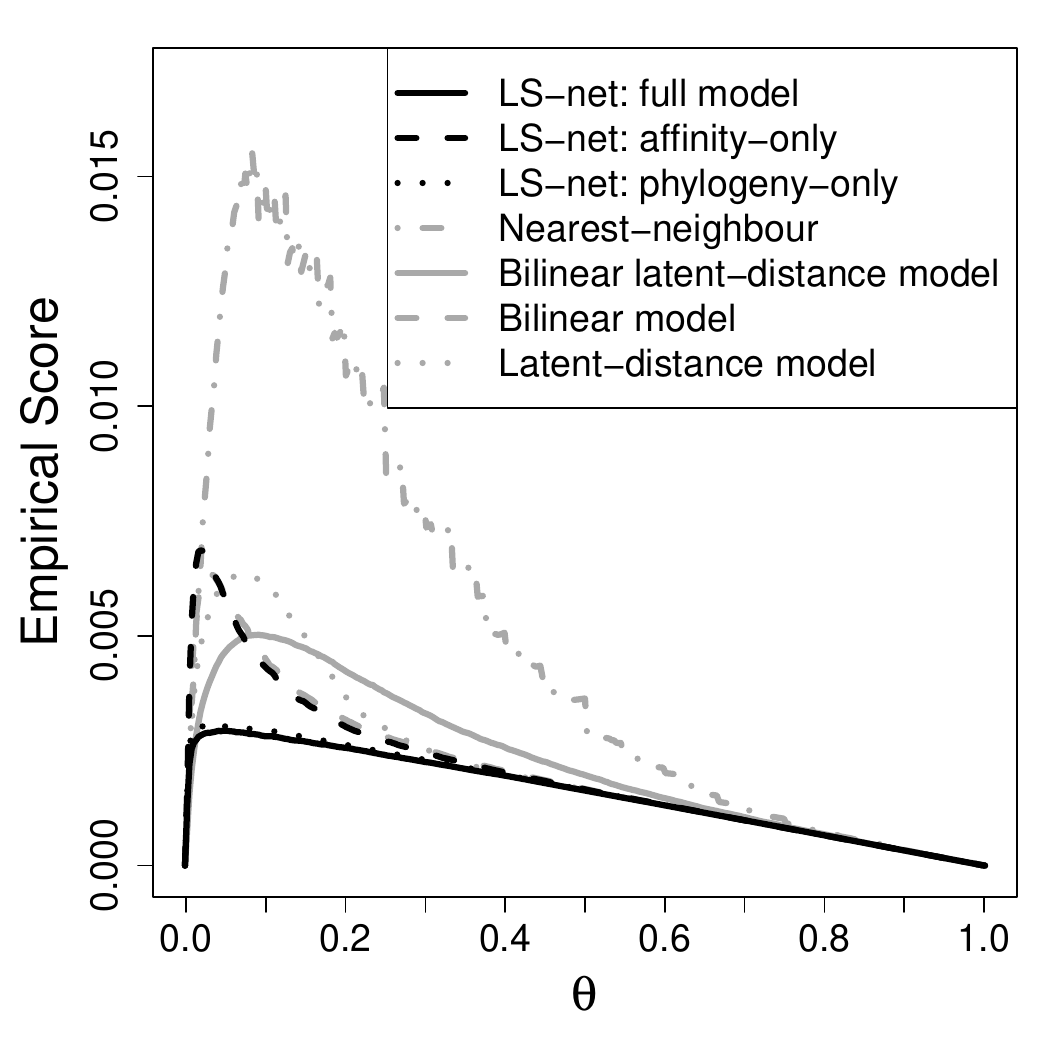}\label{fig:murphy-full}}
\subfloat[][ROC curves]{  \includegraphics[width=0.5\textwidth]{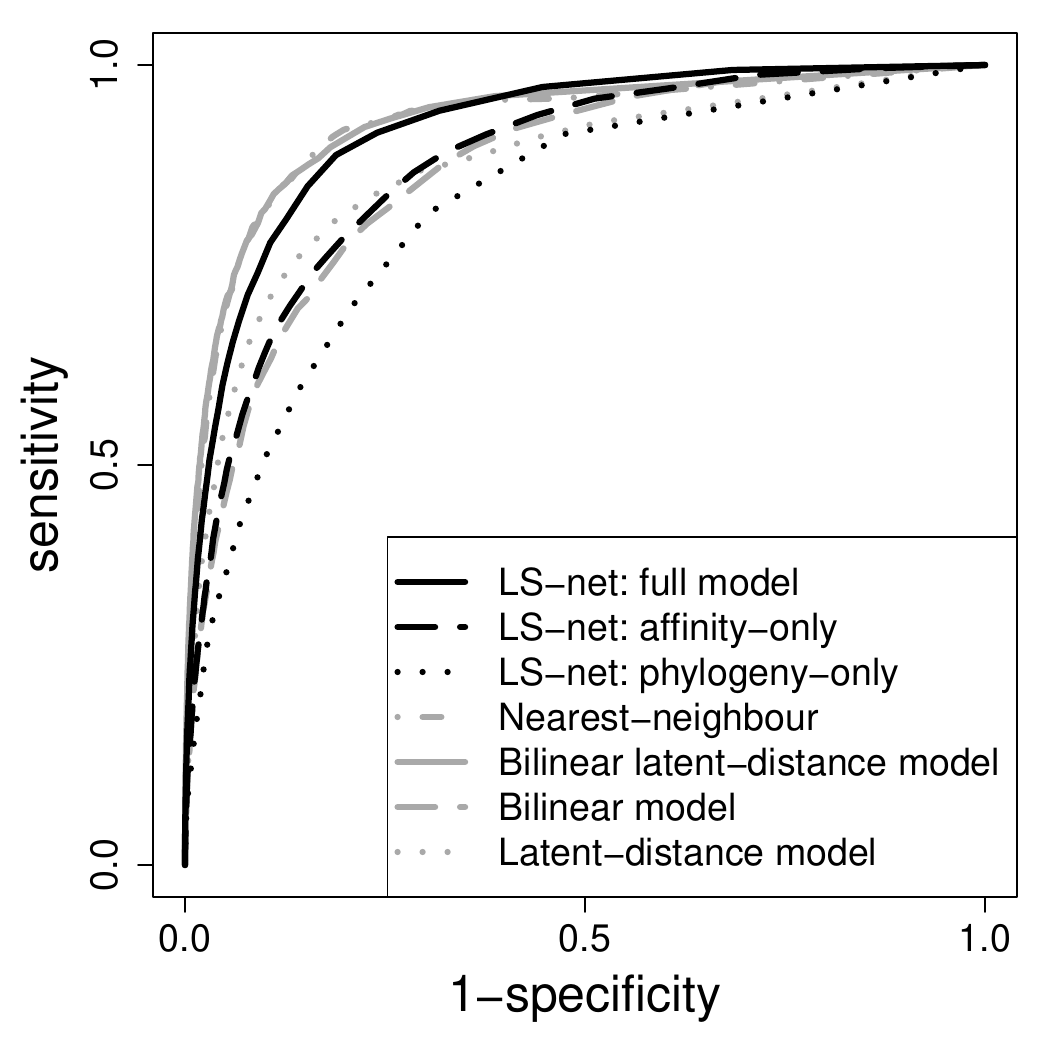}  \label{fig:10foldCV}}
\caption{
  Murphy's diagrams and ROC curves of the latent score network (LS-net) model and two of its submodels, in comparison to competing models, the NN algorithms,
the bilinear latent-distance models and two of its submodels (bilinear and latent-distance). Results are based on an average of 5-fold cross-validations on GMPD excluding single-host parasites.}  
\end{figure}

From Figure \ref{fig:murphy-full}, the predictive performance of the proposed LS-net full model dominates its competitors, with the NN algorithm performing the least well. All other models, except the phylogeny-only, have mixed performance making it harder to infer predictive dominance. Nonetheless, it is worth noting that phylogeny-only model performs equivalently to LS-net full model, contrary to other neighbourhood-based conditional models. The weaker performance of the Jaccard-based neighbourhood models (NN and latent-distance) in comparison to the phylogeny-only model suggest phylogeny may provide more power over Jaccard distances in predicting host-parasite interactions. Jaccard distances based on parasite sharing should in principle mimic evolutionary distances for hosts and parasites that are relatively well-studied, and show phylogenetic structure among hosts, as in the GMPD. As a result, phylogeny-based models may be more favourable than the NN algorithm for sparser datasets. Murphy's diagrams on GMPD including single-host-parasites follow similar pattern, and are depicted in Online Supplement Figure \ref{fig:murphy-supp}.

Evident from 5-fold average ROC curves in Figure \ref{fig:10foldCV}, the LS-net full model outperforms its two submodels and their counterparts, which confirms the notion that each of the simpler models captures different characteristics of the data, and layering them yields better results. The NN algorithm and the bilinear latent-distance model seem to have equivalent performance to the LS-net full model. Although the performance of the phylogeny-only model is subpar to its counterparts (NN and the latent-distance model), it outperforms significantly in Murphy's diagrams (fig. \ref{fig:murphy-full}), which is a stronger indicator of predictive performance than ROC curves.

\begin{figure}[ht!]
  \centering
\subfloat[][affinity-only]{\includegraphics[width=0.34\textwidth]{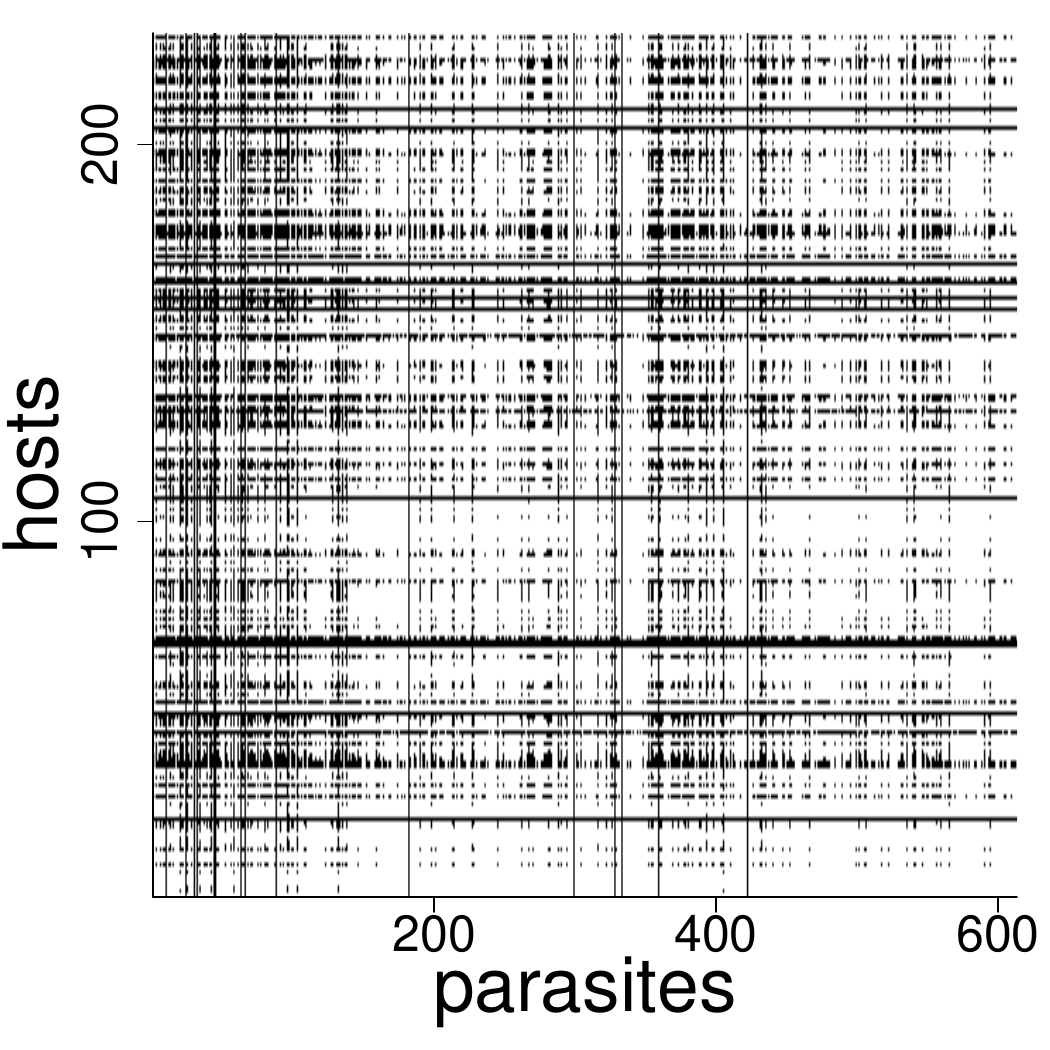}\label{fig:affinity-only-GMP}}
\subfloat[][phylogeny-only]{\includegraphics[width=0.34\textwidth]{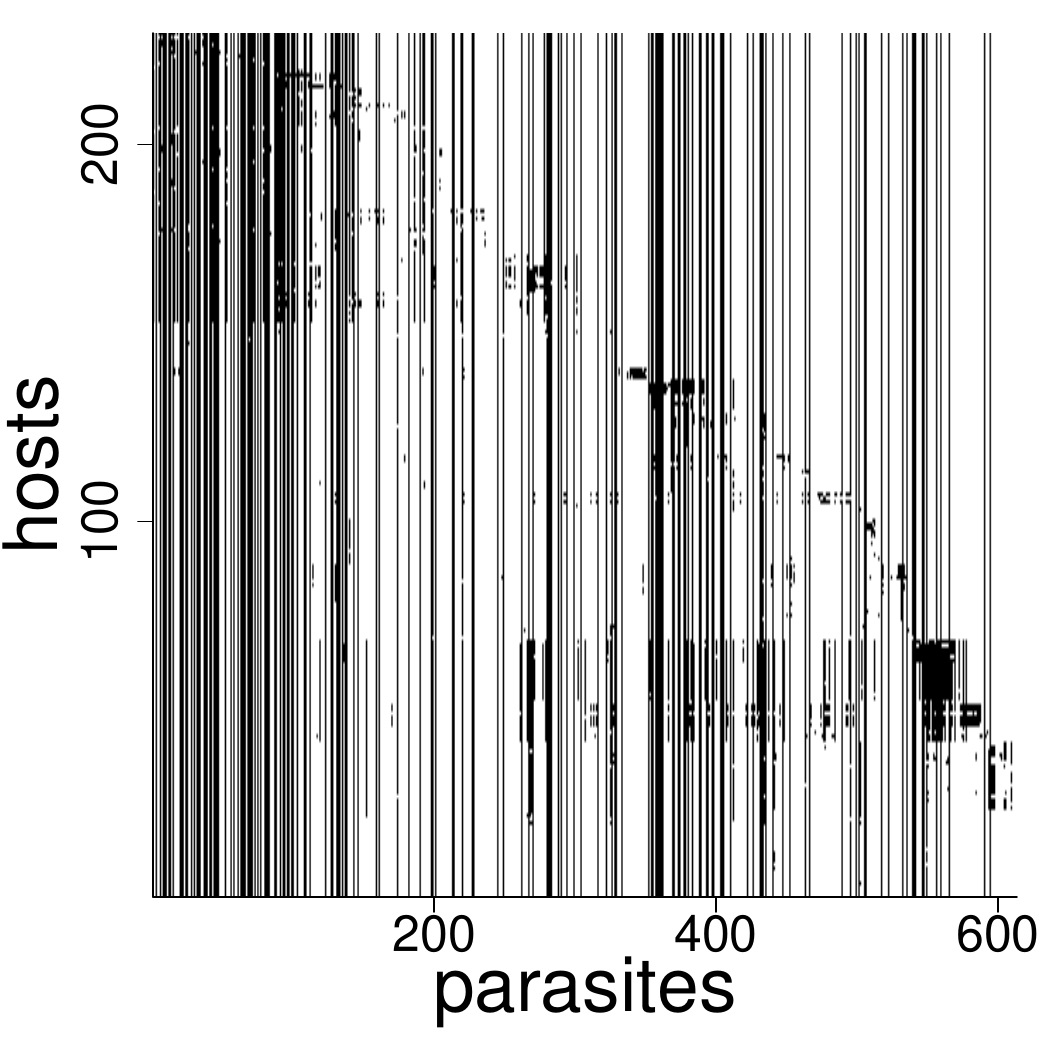}\label{fig:phylogeny-only-GMP}}
\subfloat[][full model]{\includegraphics[width=0.34\textwidth]{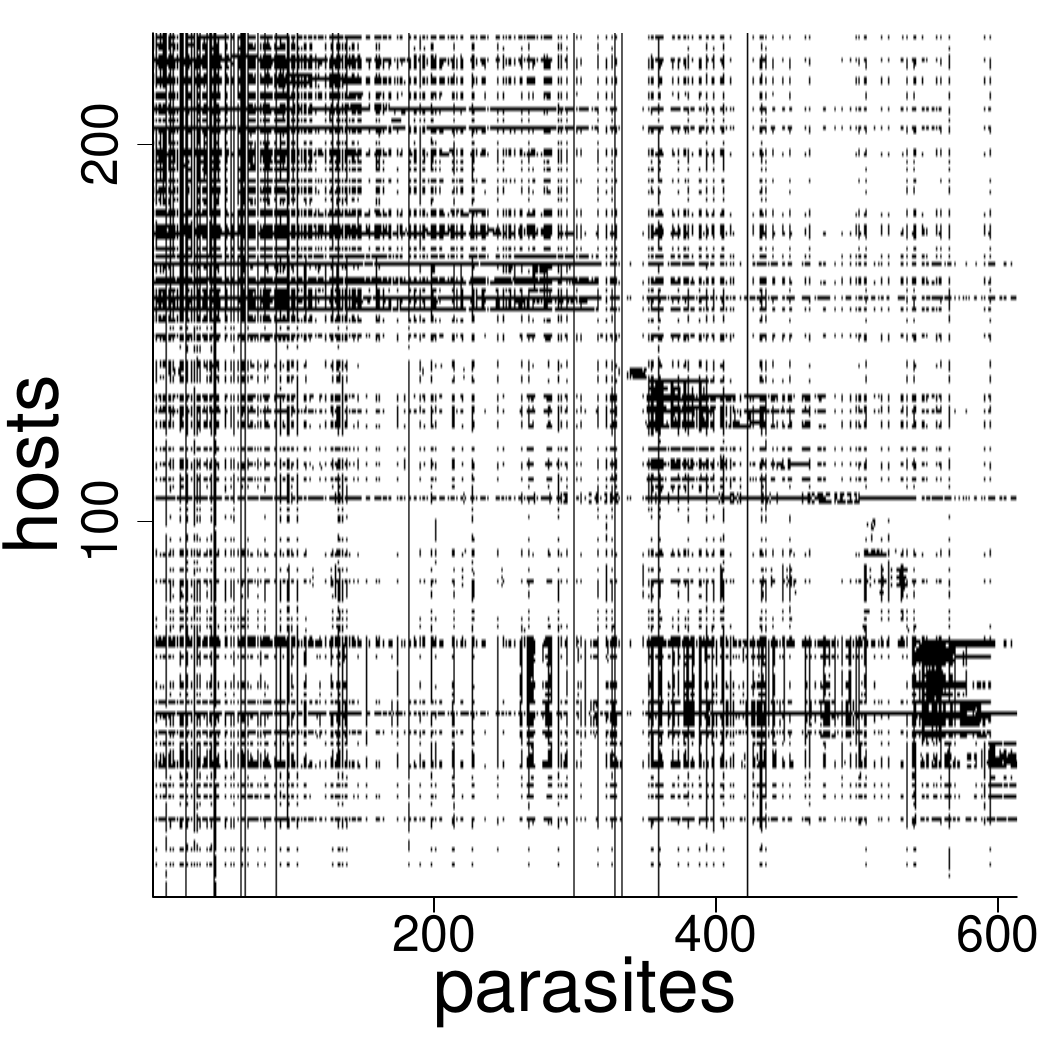}\label{fig:full-model-only-GMP}}
\caption{Posterior associations matrix comparison for the GMPD: between the affinity-only (left), phylogeny-only (middle) and full model (right), for GMPD excluding single-host parasites.}
\label{fig:Z-GMP-Progression}
\end{figure}

For a visual interpretation, Figure \ref{fig:Z-GMP-Progression} illustrates posterior predictive matrices for the affinity-only (\ref{fig:affinity-only-GMP}), phylogeny-only (\ref{fig:phylogeny-only-GMP}) and the full model (\ref{fig:full-model-only-GMP}). To show the full effect of different models, posterior predictive probabilities for all interactions in \(\Z\), observed and unobserved, are used to generate the matrices in Figure \ref{fig:Z-GMP-Progression}. From these figures, the affinity-only model does not appear to account for any neighbouring structure and results in hyperactive hosts, while the phylogeny-only model results in greater differences among parasites. The overall shape of the original \(\Z\) in Figure \ref{fig:GMP-Z} is best captured by the full model. In addition, the full model generates clear blocks of interacting hosts and parasites, reflecting interactions among particular host clades. For predictive matrices of competing models, please refer to Online Supplement Figure \ref{fig:Z-GMP-Progression-extra-models}.

For an analytical comparison, we followed the recommendation of \cite{Demsar} to use the single-sided paired Wilcoxon signed rank test on the 5-fold cross-validations rather than a fully Bayesian method (which could be implemented using a Wilcoxon-like statistic derived from posterior predictive samples). The paired test version is used since all models are tested using the same folds. For the GMPD excluding single-host parasites, we obtain a \(p\)-value less than 0.035 when comparing the full model to all other models except the bilinear latent-distance model and the NN algorithm. This indicates, for a 5\% level of significance, that the full model outperforms its two submodels, and the bilinear and latent-distance models. The test also suggests that our full model, bilinear latent-distance model, and the NN algorithm are of equivalent statistical performance.

\setlength\tabcolsep{3pt}
\renewcommand{\arraystretch}{0.85}
{\small
\begin{table}[ht!]
\centering
\caption{Area under the curve and prediction values for tested models}
\label{tb:AUC-GMP}
\begin{tabular}{l cc cc}
  & \multicolumn{2}{c}{no single-host parasites} & \multicolumn{2}{c}{with single-host parasites}\\

\input{tbAUC.tex} 
\end{tabular}
\end{table}
}

Single-host parasites comprise a non-negligible portion of the total interactions (\(\approx 17 \% \)), and including them in the calculation of host affinity parameters increases predictive performance, even though they are not included in the cross-validation set. To assess the effect of including single-host parasites on model performance, we repeated all analyses while keeping these in the original data. Table \ref{tb:AUC-GMP} shows the 5-fold average AUC and true positive prediction results when the single-host parasites are kept or removed from the GMPD. The predictive strength of the full model is now more evident. The increase in AUC for the full model is directly attributed to the inclusion of single-host parasites, since both the AUC and the percent of 1's recovered increased for the same held-out portion. This pattern is also more evident in the phylogeny-only model. For the other competing models, the AUC increase is coupled with a weaker improvement in the recovery of positive interactions, suggesting a stronger explanatory power of phylogenetic distances over Jaccard-based distances.

Since the single-host parasites are not part of the hold-out set, we infer that the improved AUC is due to the increased proportion of zeros in the larger database, as the held-out portion is kept constant. For the GMPD including single-host parasites, the single-sided paired Wilcoxon signed rank test results in a \(p\)-value less than 0.035 when comparing the full model to all other models. This indicates a stronger performance in comparison to the results of the GMPD excluding single-host parasites in terms of all measures -- the proper scoring rules, ROC curves, and percent of 1's recovered interactions.

Computationally, we found that our ICM method, implemented in {\sf R}, runs at least as fast with the {\sf latentnet} {\sf R}-package, and most of the time twice as fast. For more details refer to Online Supplement Table \ref{tb:sim-times}.

\subsection{Uncertainty in unobserved interactions}
\begin{figure}[!ht]
 \captionsetup[subfigure]{labelformat=empty}
  \centering
\subfloat[]{\includegraphics[width=0.5\textwidth]{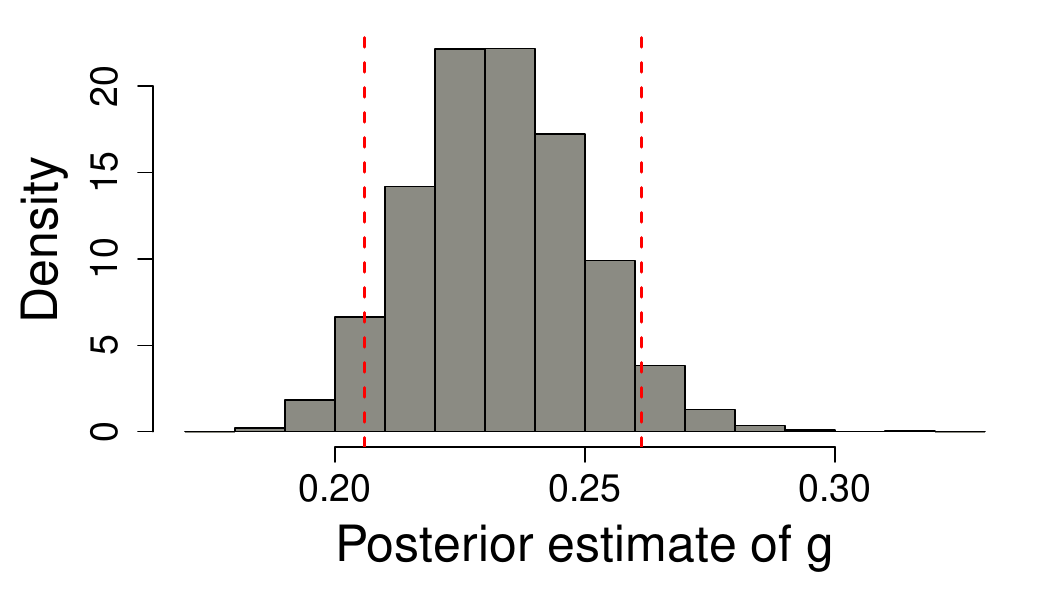}\label{fig:GMP-Post-g}}
\subfloat[][GMPD]{\includegraphics[width=0.5\textwidth]{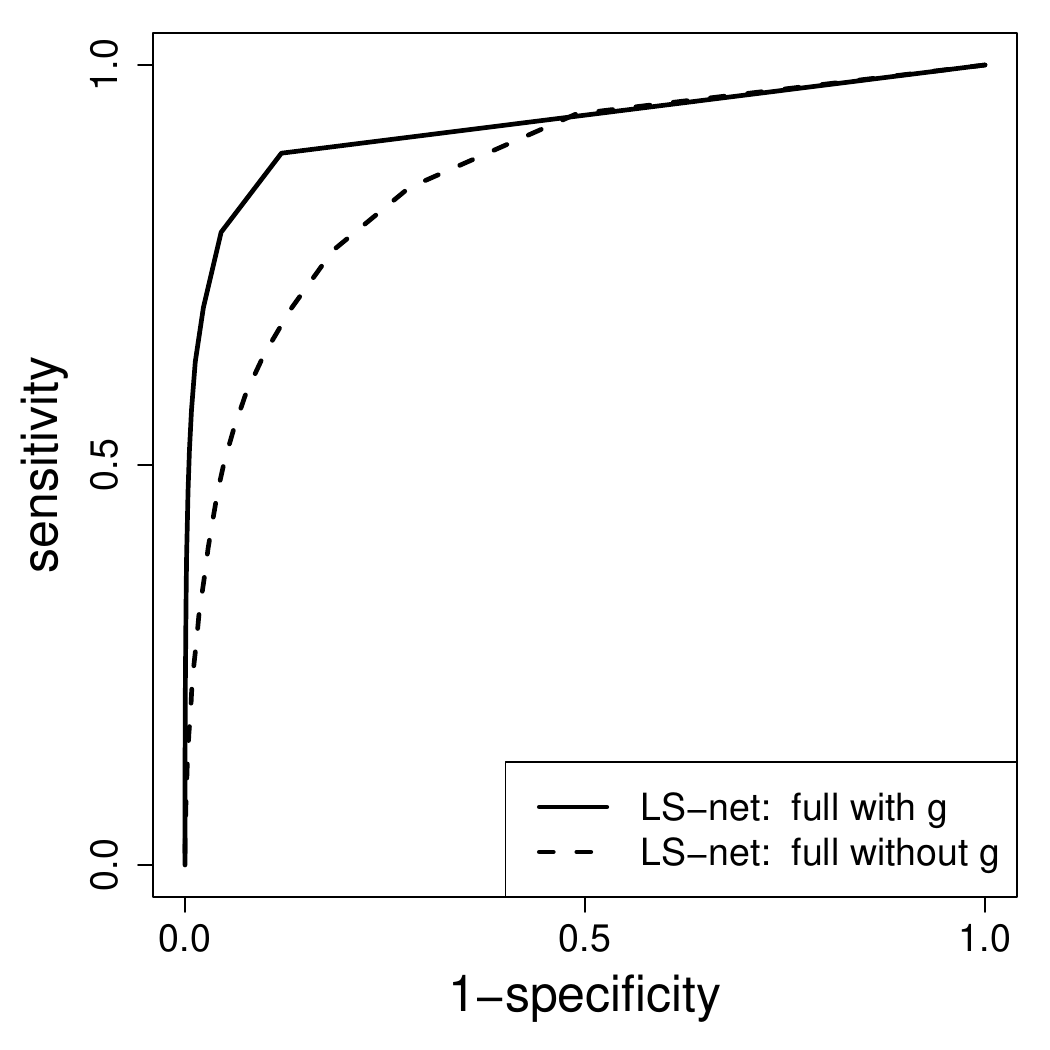}\label{fig:GMP-ROC}}
\label{fig:GMP-Post-and-ROC}
\caption{Posterior histogram for \(g\) (left) for GMPD, and comparison of ROC curves (right) for the full model with $g$ and without $g$ for GMPD excluding single-host parasites.}
\end{figure}

We improve our latent score model by accounting for uncertainty in unobserved interactions, as shown in Section \ref{sec:Uncertainty}. This addition increases the posterior predictive accuracy by estimating the proportion of missing interactions in the latent space, and reducing scores for unobserved interactions. Using the model in Section \ref{sec:Uncertainty}, we infer the uncertainty parameter \(g\), using 20000 MCMC iterations with half as burn-in. The posterior mean of \(g\) is found to be 0.232 (posterior histograms in Figure \ref{fig:GMP-Post-g}). Documented associations in the GMPD are identified through systematic searches of peer-reviewed articles that support an interaction, therefore, we expect those associations to be of high confidence, reflecting the relatively low value of \(g\). 

Introducing \(g\) to the model affects all interactions, including known ones. Therefore, to measure the predictive accuracy, we require a different cross-validation method to the one of Section \ref{sec:CrossValid}. We divide the GMPD into two sets, a training and a validation set. Since associations in the GMPD are sourced only from peer-reviewed articles, we were able to use information on article publication dates to create the training and test datasets. This mimics the discovery of interactions in the system rather than random hold-out of observations. Taking the earliest annotated year for each association we set the training set as all associations documented prior to and including 2006, and the validation set as all associations up to 2010. There are 3755 pairs of documented associations in the GMPD, including single-host parasites, up to and including 2006. By 2010, the associations increased to 4178, with 236 hosts and 1308 parasites, approximately a 10\% increase. For the training sets using the GMPD up to 2006, we used an average of 5-fold cross-validations, constructed as in Section \ref{sec:CrossValid}, to estimate the parameters of the model, where each fold ran for 20000 iterations with half as burn-in. Since the full model is used, cross-validation is no longer restricted to multi-host parasites as in Section \ref{sec:CrossValid}, nonetheless, to avoid empty columns at least one interaction is kept for each parasite.

Figure \ref{fig:GMP-ROC} illustrates the improvement in potential predictive accuracy between the models with or without \(g\). Essentially, incorporating uncertainty results in probability estimates for all interactions, undocumented and documented, where the former is penalized proportional to \(g\). This reduces the overlap in posterior probability densities between interacting and non-interacting pairs, refer to Online Supplement Figure \ref{fig:GMP-OBS-UNK1} for the posterior histogram of both categories. 

The model with \(g\) outperforms the full model on both AUC and proportion of positive interactions predicted, including and excluding the single-host parasites (Table \ref{tb:GMP-AUC}). These results represent the evaluation on the whole dataset, up to 2010, not only the held-out and undocumented portions as in Section \ref{sec:CrossValid}. The model with \(g\) is able to predict  90.90\% of the documented interactions in the 2010 GMPD, approximately 3798 out of 4178 interactions, where the model without \(g\) predicts approximately 194 fewer interactions. 
{\small
\begin{table}[!ht]
\centering
\caption{Area under the curve and prediction values for the model with(out) \(g\)}
\label{tb:GMP-AUC} 
\begin{tabular}{l cc cc}
    & \multicolumn{2}{c}{no single-host parasites} & \multicolumn{2}{c}{with single-host parasites}\\
    & AUC & \% 1's recovered & AUC & \% 1's recovered \\
  \hline                            
  with $g$    & 0.924 & 88.98 & 0.944 & 90.90 \\
  without $g$ & 0.865 & 76.80 & 0.918 & 86.26 \\
  \hline
\end{tabular}
\end{table}
}

\begin{figure}[ht!]
  \centering
\subfloat[][recovery power]{\includegraphics[width=0.5\textwidth]{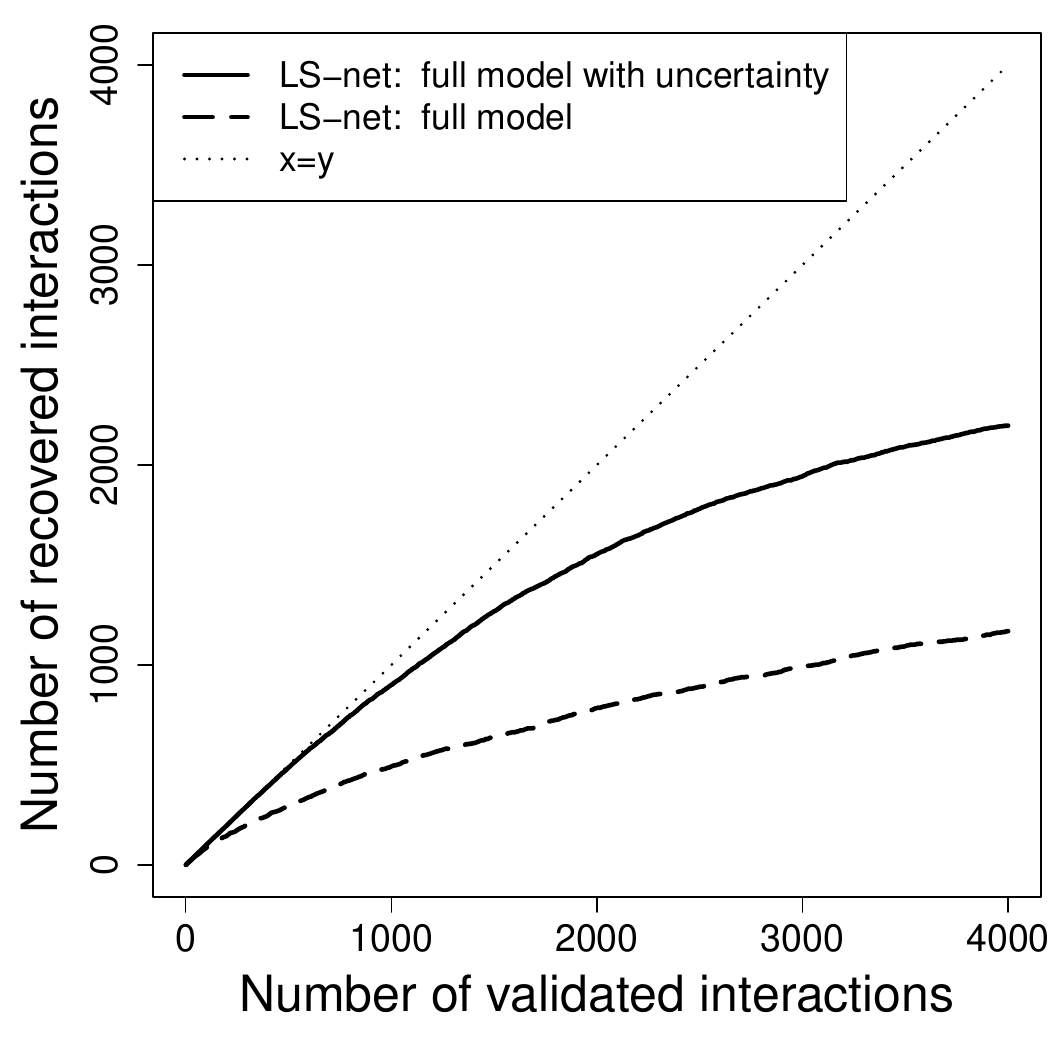}\label{fig:TopM-GMPD}}
\subfloat[][GMPD with \(g\) no single-host parasites]{\includegraphics[width=0.5\textwidth]{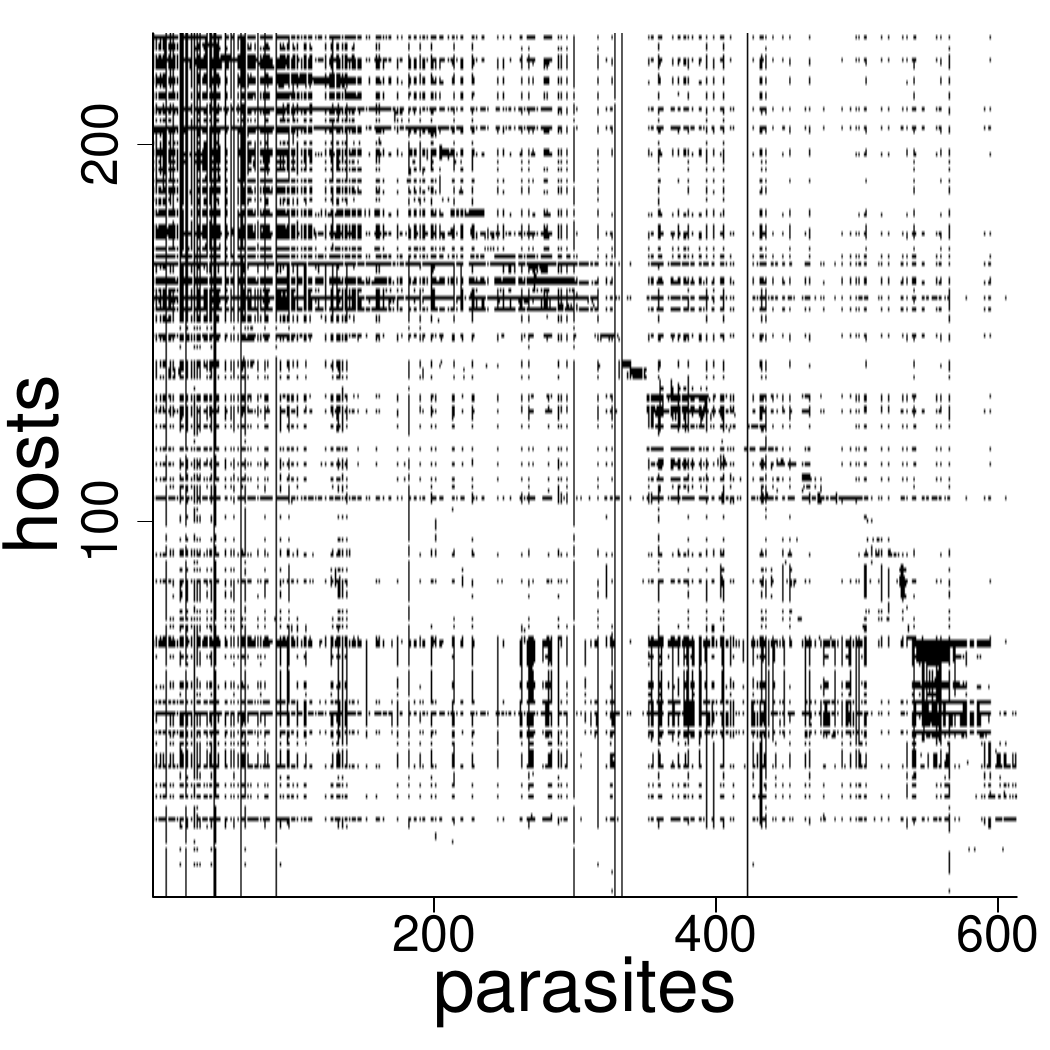}\label{fig:full-model-GMP}\label{GMPD-G}}
\caption{Number of pairwise recovered interactions from the original 2010 GMPD data (left), and the posterior interaction matrix for the 2010 GMPD, excluding single-host parasite, using the model that accounts for uncertainty with \(g\).}
\end{figure}

Another method of model comparison is through the proportion of recovered interactions from the full data. This can be quantified by sorting all pairwise interactions based on their posterior predictive probabilities, and examining the top \(x\) pairs with the highest predictive probabilities as they represent interactions with highest confidence. By counting the number of true interactions recovered in those \(x\) selected pairs, and by scaling \(x\) from 1 to 4000 we find the model with \(g\) again outperforms the full model by recovering more than double the number of interactions (Figure \ref{fig:TopM-GMPD}). 

Finally, for comparison with Figure \ref{fig:Z-GMP-Progression}, the posterior interaction matrix for the model with \(g\) excluding single-host parasites is shown in Figure \ref{GMPD-G}.

\begin{figure}[ht!]
  \centering
\subfloat[][with single-host]{\includegraphics[width=0.45\textwidth]{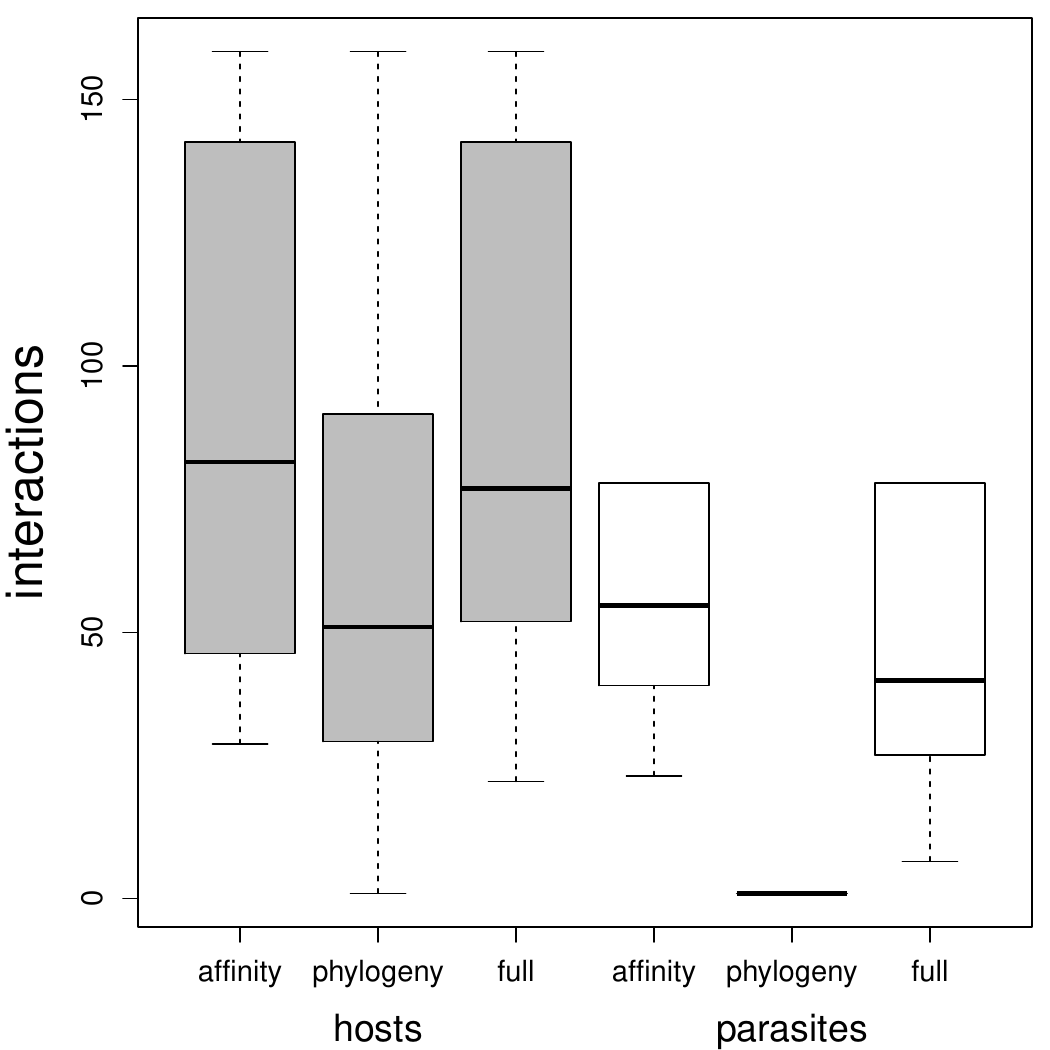}}
\subfloat[][without single-host]{\includegraphics[width=0.45\textwidth]{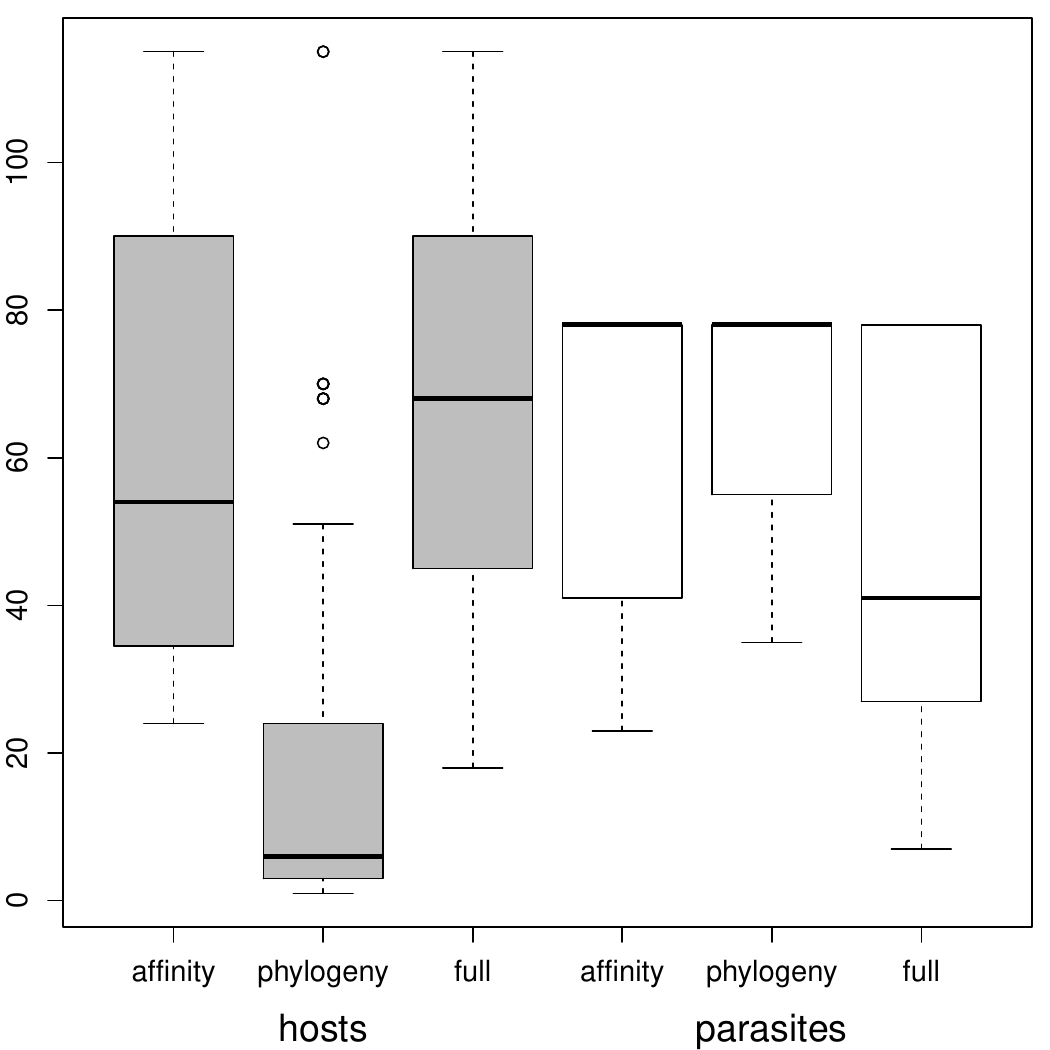}}
\caption{Comparison of the number of documented interactions (node degree) for both hosts (grey) and parasites (white) included in the 100 most probable, yet previously undocumented interactions across each of the three sub-models. Results are split into models a) with single-host parasites, and b) without single-host parasites.}
\label{fig:GMPD-boxplotdegree}
\end{figure}

Incorporating phylogenetic information identifies interactions that would not be considered likely under the affinity-only model. To illustrate this, we plot the number of documented interactions (node degree) for both hosts and parasites included in the 100 most probable yet previously undocumented interactions for each model (Fig. \ref{fig:GMPD-boxplotdegree}). When fitting the model including single-host parasites, we find that the top 100 predicted links for the phylogeny-only model tend to include hosts and parasites with fewer observed links in the original data. In fact, all top 100 novel predictions made by the phylogeny-only model include parasites that have 1 documented interaction, all of which would be given low probability by preferential attachment models (including our affinity only model). By contrast, in the top 100 predictions made by the affinity only model, the parasite with the fewest number of observed interactions has 26 known host species. This suggests that the inclusion of phylogenetic information allows the identification of highly probable interactions for rare or understudied species.

\section{Discussion}
We introduced a latent score model for link prediction in ecological networks and illustrate it using a recently published global database of host-parasite interactions. The proposed model is a combination of two separate models, an affinity based exchangeable random network model overlaid with a Markov network dependency informed by phylogeny \eqref{eq:phylogeny-only}. The affinity-only model is characterized by independent affinity parameters for each species, while the phylogeny-only model is characterized by a scaled species similarity matrix. Both parts perform reasonably well alone, and by overlaying them the posterior prediction is significantly improved.

Many advantages arise from integrating host phylogenies. By utilizing known evolutionary models, phylogenies provide remarkable predictive power comparable to state-of-the-art latent-distance models \citep{hoff2002latent,hoff2005bilinear}, but with added biological interpretation. Such tree-scaling models could also be integrated in existing link prediction frameworks, such as that outlined by \citet{Chiu2011} and elaborated by \citet{Ovaskainen2017}. However, computational issues might arise as the full joint distribution becomes intractable and is not guaranteed to exist. 

To our knowledge, our framework is the first to attempt to incorporate this type of evolutionary information in link prediction models. Computational issues arise from integrating this procedure, and we solve this by imposing minimal conditions on the latent variable, which produces promising results. We used the Early-Burst model to scale the phylogeny, but any other evolutionary model that scales the species covariance matrix could be fit.

While we incorporated phylogeny as the dependence structure, the model can easily accommodate different similarity matrices or types of dependence in an additive manner. For host-parasite networks, host traits or geographic overlap, or parasite similarity based on phylogeny, taxonomy, or traits may improve prediction \parencite{Pedersen2005,Davies2008,Luis2015}. Introducing different similarity measures affects the model characteristics in two ways: it changes the topology of the probability domain, and it increases the number of parameters to estimate due to introduced scaling parameters. The latter is easily integrated since the number of estimated parameters increases by one for each new scaling parameter. It is also possible to introduce different tree scaling parameters for different host-groups, allowing for a richer representation and added flexibility with minimum cost, which should improve performance. In addition, covariate data such as species traits can easily be integrated in the model in an additive manner. For example, set \(\tau_{hj} = \gamma_h \rho_j\delta(\eta) \exp(-\beta_i x_i - \beta_j x_j)\). Alternatively, they could be included in a hierarchical manner as a function of the affinity parameter. Each case represents a different interpretation, with covariates in the former driving the interaction probability directly, while the latter influencing the affinity parameters.

A particular dependence structure that does not require additional data is a similarity based on the number of shared interactions, as used in the NN algorithm (Section \ref{sec:CrossValid}). In host-parasite networks, parasite community similarity is often well predicted by evolutionary distance among hosts \parencite{Gilbert2007,Davies2008}. In this case, the NN similarity is likely capturing some of the phylogenetic structure in the network and could be a reasonable approach if a reliable phylogeny is unavailable. However, as phylogeny is estimated independently from the interaction data, it will likely be more robust to incomplete sampling of the original network than NN type dependence structures. 

Many ecological and other real world networks display power-law degree distributions \parencite{Albert2002}. This is also the case with the host-parasite database used in this paper, where both hosts and parasites exhibit power-law degree distributions. The affinity-only version of the proposed model in \eqref{eq:full-model} has been shown to generate a power-law behaviour when a Generalized Gamma process is used \citep{brix1999generalized,lijoi2007controlling,caron2014sparse}. In fact, when \(\gamma_h =\gamma\) for all \(h\), the affinity-only model behaves much like the Stable Indian Buffet process of \citet{teh2009indian} that has a power-law behaviour. Nonetheless, we find the full model to show a significant improvement in predictive accuracy over the affinity-only model, though it does not yield a degree distribution with a power-law. However, when accounting for uncertainty in the full model, the posterior predictions we regain a power-law degree distribution for hosts and parasites (Online Supplement Figure \ref{fig:GMP-Degree-G}). It would be interesting in future work to explore which other network properties are maintained using this model.

While the intent of this research is to identify undocumented interactions, this model can also account for uncertainty in observed interactions. In this case, our model may be used to identify weakly supported interactions that are false positives or sampling artefacts in the literature that may benefit from additional investigation. In the case of host-parasite interactions, our approach could form an integral component of proactive surveillance systems for emerging diseases \parencite{Farrell2013}. However, the framework illustrated here is not limited to host-parasite networks, but is well suited to multiple ecological networks such as plant-herbivore, flower-pollinator, or predator-prey interactions.

\section{Acknowledgements}
 We like to thank the McGill Statistics-Biology Exchange Group (S-BEX) and organizers Russell Steele, Zofia Taranu, and Amanda Winegardner for fostering an environment that led to this collaboration. We also thank Jonathan Davies and his lab for critical feedback throughout model development and writing, and the Macroecology of Infectious Disease Research Coordination Network (funded by NSF DEB 1316223) for providing early versions of the GMPD. MJF was funded by an NSERC Vanier CGS, and ME by FQRNT.

\bibliography{references}
\bibliographystyle{imsart-nameyear}

\newpage
\bigskip
\begin{center}
{\large\bf SUPPLEMENTARY MATERIAL}
\end{center}
\setlength{\parskip}{0mm}
\input{Supp}

\end{document}

%% file: tbAUC.tex
Model               & AUC   & \% 1's recovered & AUC   & \% 1's recovered \\
\hline
  LS-net: full model             & 0.921 & 87.46 &  0.959 & 92.56 \\ 
  LS-net: affinity-only          & 0.876 & 80.99 &  0.933 & 88.79 \\ 
  LS-net: phylogeny-only         & 0.823 & 79.75 &  0.917 & 90.34 \\ 
  Nearest-neighbour              & 0.926 & 88.61 &  0.948 & 91.10 \\ 
  Bilinear latent-distance model & 0.929 & 86.35 &  0.936 & 86.40 \\ 
  Bilinear model                 & 0.868 & 78.55 &  0.914 & 84.05 \\ 
  Latent-distance model          & 0.872 & 77.32 &  0.890 & 78.69 \\
\hline

%% file: Supp.tex
\begin{appendices} 
  \section{Model general settings}\label{sec:model-gener-sett}
 Following the settings and notations of Section \ref{sec:Network} and the conditional full model in \eqref{eq:full-model}, let \(\tau_{hj} =\gamma_{h}\rho_{j}\delta_{hj}(\eta)\), and \(\delta_{hj}(\eta)\) as in \eqref{eq:phylogeny-only}. Define a latent score \(s_{hj}\) as in \eqref{eq:Z-as-S}. Such a characterization prompts a conditional joint distribution of the form in \eqref{eq:more-tractable-joint-dist}. Moreover, it can be verified that 
 \[
   \p(s_{hj} \mid z_{hj}, \Z_{-(hj)}) =
\begin{cases}
  \frac{1}{1-\exp(-\tau_{hj})}\p(s_{hj}\mid \S_{-(hj)})\mathbb{I}_{\{s_{hj}>0\}} & z_{hj}=1 \\
  \frac{1}{\exp(-\tau_{hj})}\p(s_{hj}\mid \S_{-(hj)})\mathbb{I}_{\{s_{hj}=0\}} & z_{hj}=0.
\end{cases}
\]

It remains to define the distribution of \(s_{hj}\mid \Z_{-(hj)}\) to satisfy the property that
\[ \P(z_{hj}=1\mid \Z_{-(hj)}) = 1-\exp(-\tau_{hj}) = \int_{\mathbb{R}}\p(s\mid \S_{-(hj)}) \mathbb{I}_{\{s>0\}}\d s.\]

One possible choice is the zero-inflated Gumbel density as in \eqref{eq:truncated-gumbel-density}. The latent score is used only as a modelling for extra tractability, as in \eqref{eq:more-tractable-joint-dist}.

\section{Latent score sampling with uncertainty}
By modelling the uncertainty parameter \(g\) as  in \eqref{eq:uncertain-prop-function}, one arrives at the conditional joint distribution
\[\begin{aligned}
    \P(z_{hj} = 1 ,s_{hj}\mid g, \Z_{-(hj)}) &= \p(s_{hj}\mid \Z_{-(hj)}) \I_{\{s_{hj}>0\}},\\
    \P(z_{hj} = 0 ,s_{hj}\mid g, \Z_{-(hj)}) &= \p(s_{hj}\mid \Z_{-(hj)}) \Big [g\I_{\{s_{hj}>0\}}+\I_{\{s_{hj}=0\}} \Big ].
\end{aligned}
\]
The conditional sampling of the latent truncated score variable \(s_{hj}\) becomes
\[\begin{aligned} 
\p(s_{hj}\mid z_{hj}, \Z_{-(hj)}, g)  &= \frac{\P(z_{hj} \mid s_{hj}, g )\; .\;\p(s_{hj}\mid \Z_{-(hj)}) }{\int\P(z_{hj} \mid s, g )\;. \;\p(s\mid \Z_{-(hj)})\d s} = C \;.\; \p(s_{hj}\mid \Z_{-(hj)}),
\end{aligned}\]
such that
{\small
\[\begin{aligned}
C &=  \frac{\P(z_{hj} \mid s_{hj}, g )}{\int\P(z_{hj} \mid s, g )\;. \;\p(s\mid \Z_{-(hj)})\d s}\\
  &=  \frac{\P(z_{hj} \mid s_{hj}, g )}{\int_{s>0}\P(z_{hj} \mid s, g )\;. \;\p(s\mid \Z_{-(hj)})\d s+\int_{s\leq 0}\P(z_{hj} \mid s, g )\;. \;\p(s\mid \Z_{-(hj)})\d s} \\
  & = \begin{cases}
    \frac{\P(z_{hj} \mid s_{hj}, g )}{\int_{s>0}1\; .\;\p(s\mid \Z_{-(hj)})\d s+\int_{s\leq 0}0\; .\; \p(s\mid \Z_{-(hj)})\d s }, & \text{ when } z_{hj} = 1, \\
    \frac{\P(z
_{hj} \mid s_{hj}, g )}{\int_{s>0}g\; .\;\p(s\mid \Z_{-(hj)})\d s+\int_{s\leq 0}1\; .\; \p(s\mid \Z_{-(hj)} )\d s },  & \text{ when } z_{hj} =0, \\
    \end{cases} \\
  & =  \begin{cases}
    \frac{1}{\psi(\bar s_{hj})} , & s_{hj}  > 0,\quad  z_{hj}=1, \\ 
    0,                             & s_{hj}  = 0,\quad z_{hj}=1, \\ 
    \frac{g}{g\psi(\bar s_{hj}) + 1-\psi(\bar s_{hj}) }, & s_{hj} >0,\quad z_{hj}=0,\\
    \frac{1}{  g\psi(\bar s_{hj}) + 1-\psi(\bar s_{hj}) },& s_{hj} =0,\quad z_{hj}=0, \\
  \end{cases}
\end{aligned}\]
}
\section{Existence of the joint distribution}\label{app:hammersley-clifford}
\begin{mytheorem}Hammersley-Clifford,\citep{robert2013monte}. \\
Under marginal positively conditions, the joint distribution of random variables \(X = (x_1, x_2,  \dots, x_n)\) is  proportional to 
\[\frac{\P(X)}{\P(X^*)}= \prod_{i=1}^n\frac{\P(x_i \mid x_1, \dots, x_{i-1}, x^*_{i+1}, \dots, x^*_n)}{\P(x^*_i \mid x_1, \dots, x_{i-1}, x^*_{i+1}, \dots, x^*_n)}
\]
where \(x^*_i\) are fixed observations, for example \(x^{*}_{i}=1\).
\end{mytheorem}

In regards to conditional probability in \eqref{eq:full-model}, assume the phylogeny-only model where \(\tau_{hj} = \delta_{hj}(\eta)\), and \(\delta_{hj}(\eta)\) as in \eqref{eq:phylogeny-only}. Since each column of \(\Z\) is independent, it suffices to show that the joint distribution exists for each column. Applying the Hammersley-Clifford theorem by setting \(z^{*}_{hj}=1\), we have
\[ \frac{\P(z_{hj} \mid z_{1j}, \dots, z_{(h-1)j}, z^*_{(h+1)j}, z^*_{Hj})}{\P(z^*_{hj} \mid z_{1j}, \dots, z_{(h-1)j}, z^{*}_{(h+1)j}, z^{*}_{Hj})} = \Big [ \frac{\exp(-\bar{\tau}_{hj})}{1-\exp(-\bar{\tau}_{hj})}\Big ]^{1-z_{hj}}, \]
\[ \bar{\tau}_{hj} = \sum_{i=1}^{h-1}\frac{z_{ij}}{\phi(T_{hi}, \eta)}  + \sum_{i=h+1}^H \frac{1}{\phi(T_{hi}, \eta)},\]
\[\bar{\tau}_{1j} = \sum_{i=2}^{H}\frac{1}{\phi(T_{1i}, \eta)}, \quad
\bar{\tau}_{Hj} = \sum_{i=1}^{H-1}\frac{z_{ij}}{{\phi(T_{Hi}, \eta)}}.\]

Essentially, by removing the event of no interactions, as \(\z_{h.} = (0,0, \dots, 0)\), and setting \(\bar{\tau}_{hj}=1\) whenever it is 0, the joint distribution exists.
\newpage
\section{Additional results for GMPD} \label{app:Diagnostic-Plots-Extra}
A section so additional GMPD results and analysis.
\begin{figure}[t!]
  \centering
  \includegraphics[width=0.9\textwidth]{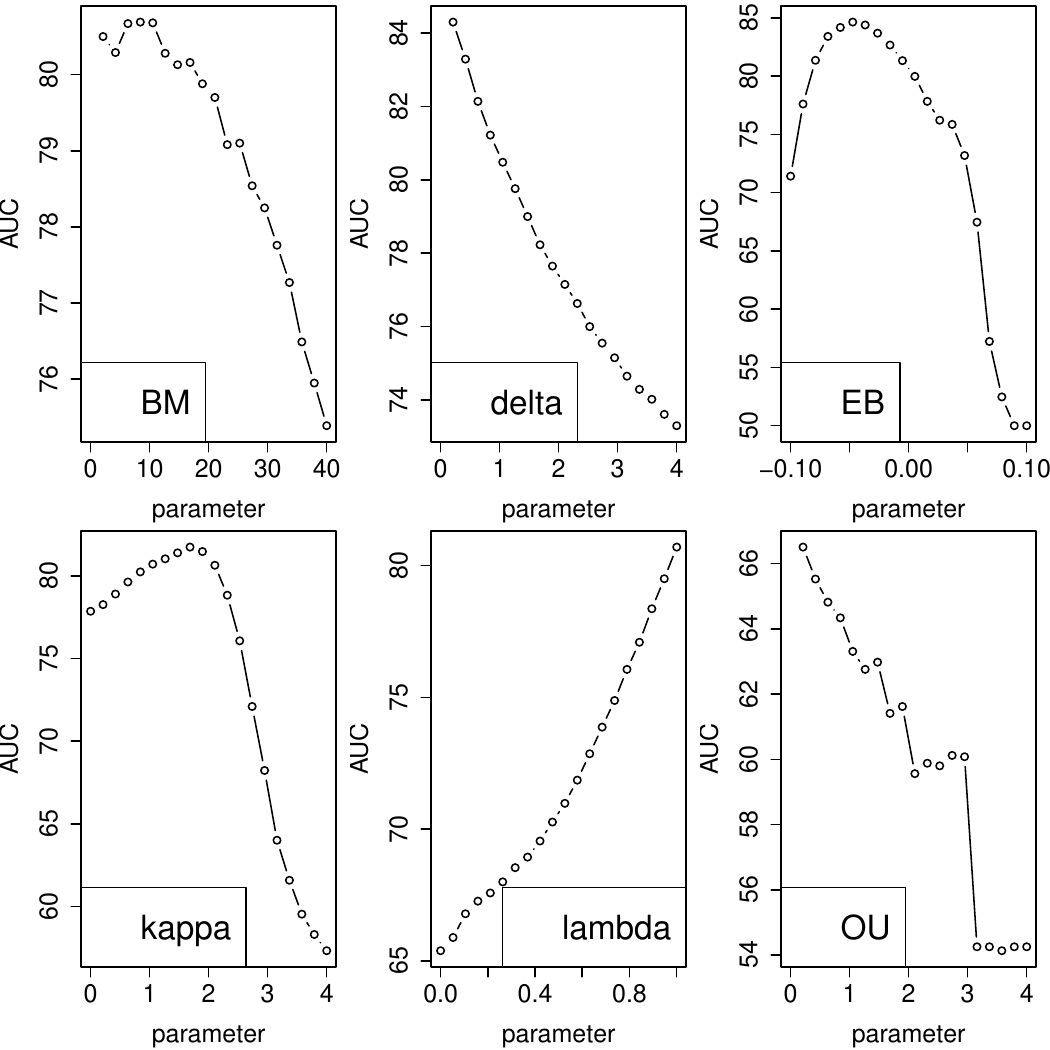}
  \caption{Grid search over the tree transformation parameter for basic AUC results under the phylogeny-only model (paper Eq. \eqref{eq:phylogeny-only}) with GMPD (excluding single-host parasites)for different phylogeny transformational models: delta, early-bust (EB), kappa, lambda, and the Ornstein-Uhlenbeck (OU). The kappa model was discarded as it was designed to represent a speciational model of evolution with change occurring at speciation events, which makes the transformation highly sensitive to missing species in the phylogeny.}    \label{fig:AUC-transformational-models}
\end{figure}

\begin{figure}[t!]
  \centering
  \subfloat[][trace and ACF plots]{\includegraphics[width=1\textwidth]{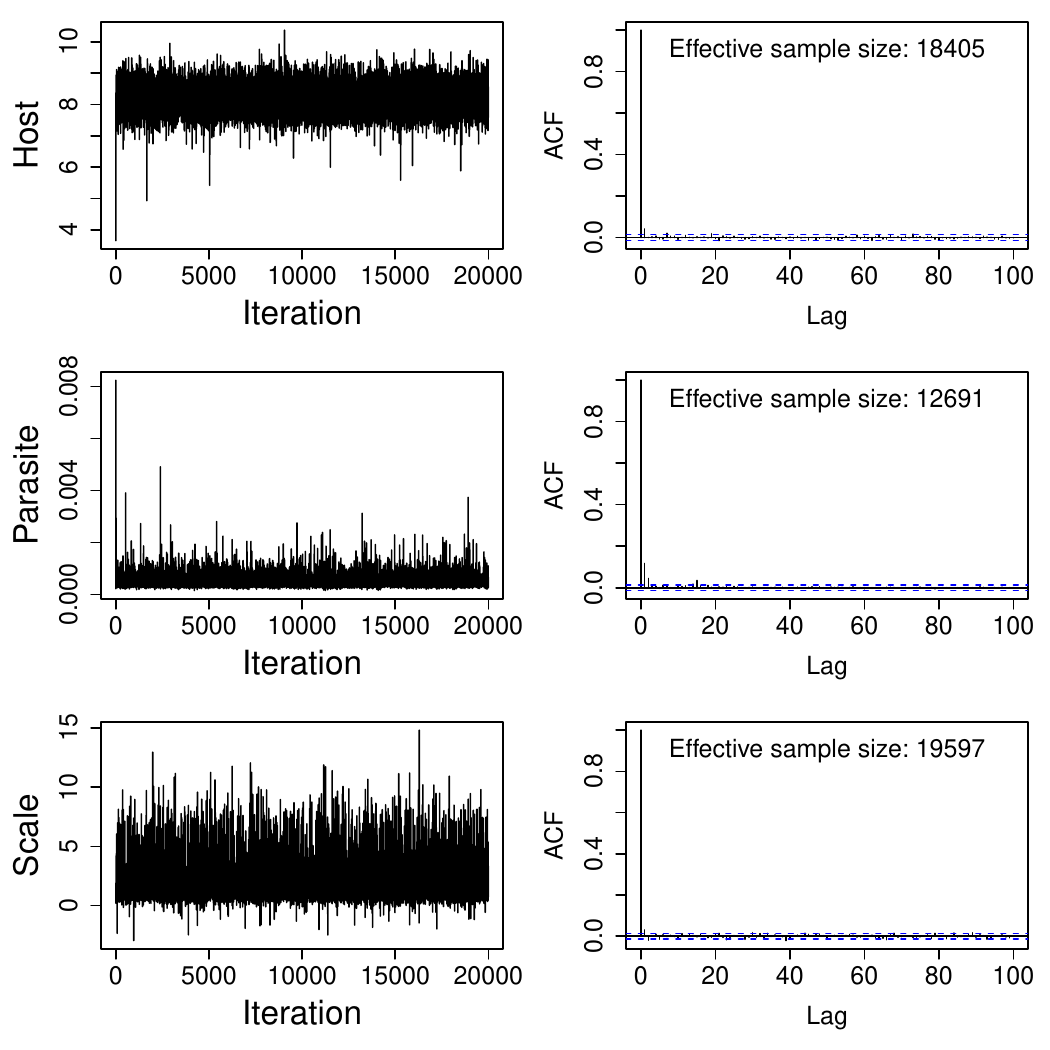}}
  \caption{Trace and auto-correlation plots with effective sample sizes: host (top) and parasite (middle) of highest median posterior, and phylogeny EB model parameter (bottom), for GMPD including single-host parasites.}
  \label{fig:tracePlots}
\end{figure}

\begin{figure}[t!]
  \centering
\subfloat[][nearest-neighbour]{\includegraphics[width=0.34\textwidth]{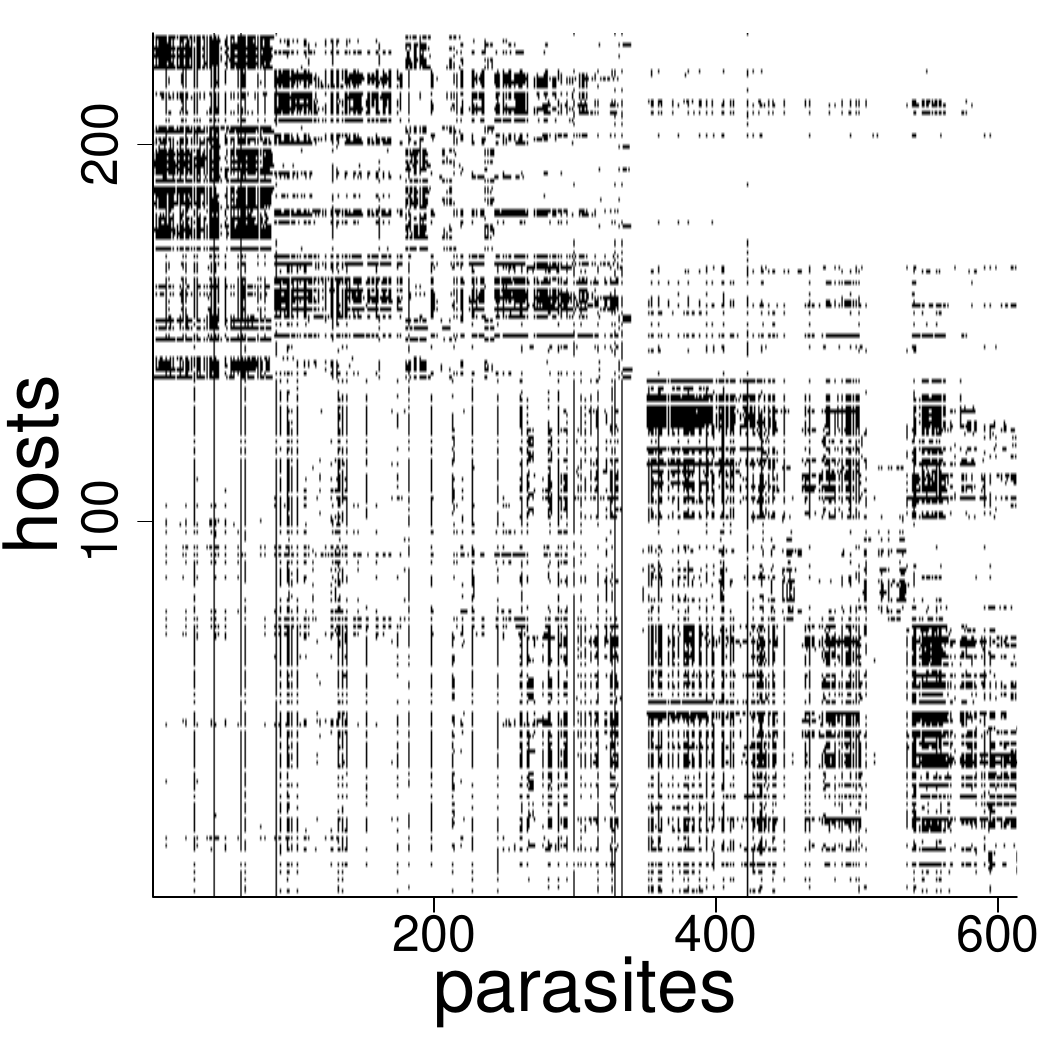}\label{fig:NN-GMP}}
\subfloat[][bilinear latent-distance]{\includegraphics[width=0.34\textwidth]{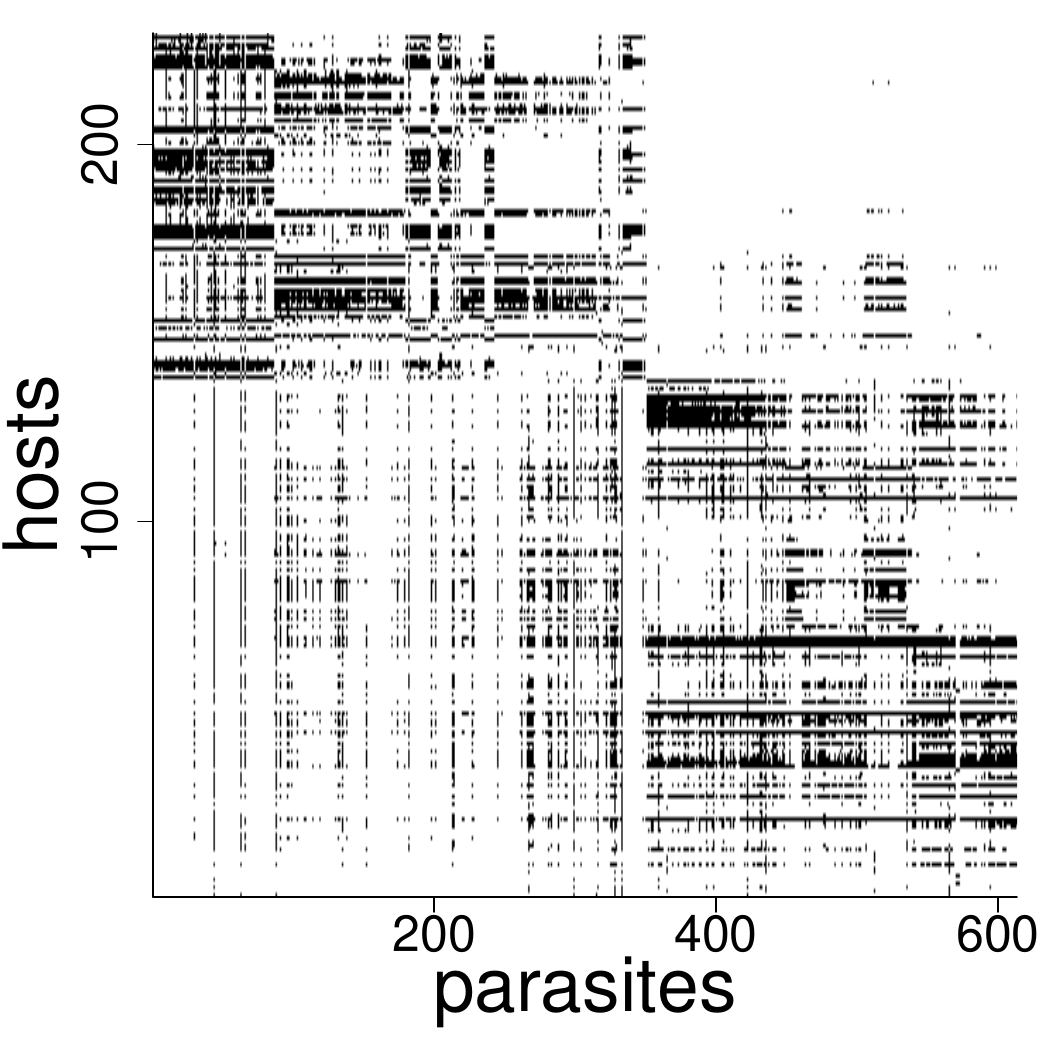}\label{fig:latent-GMP}}
\subfloat[][latent-distance]{\includegraphics[width=0.34\textwidth]{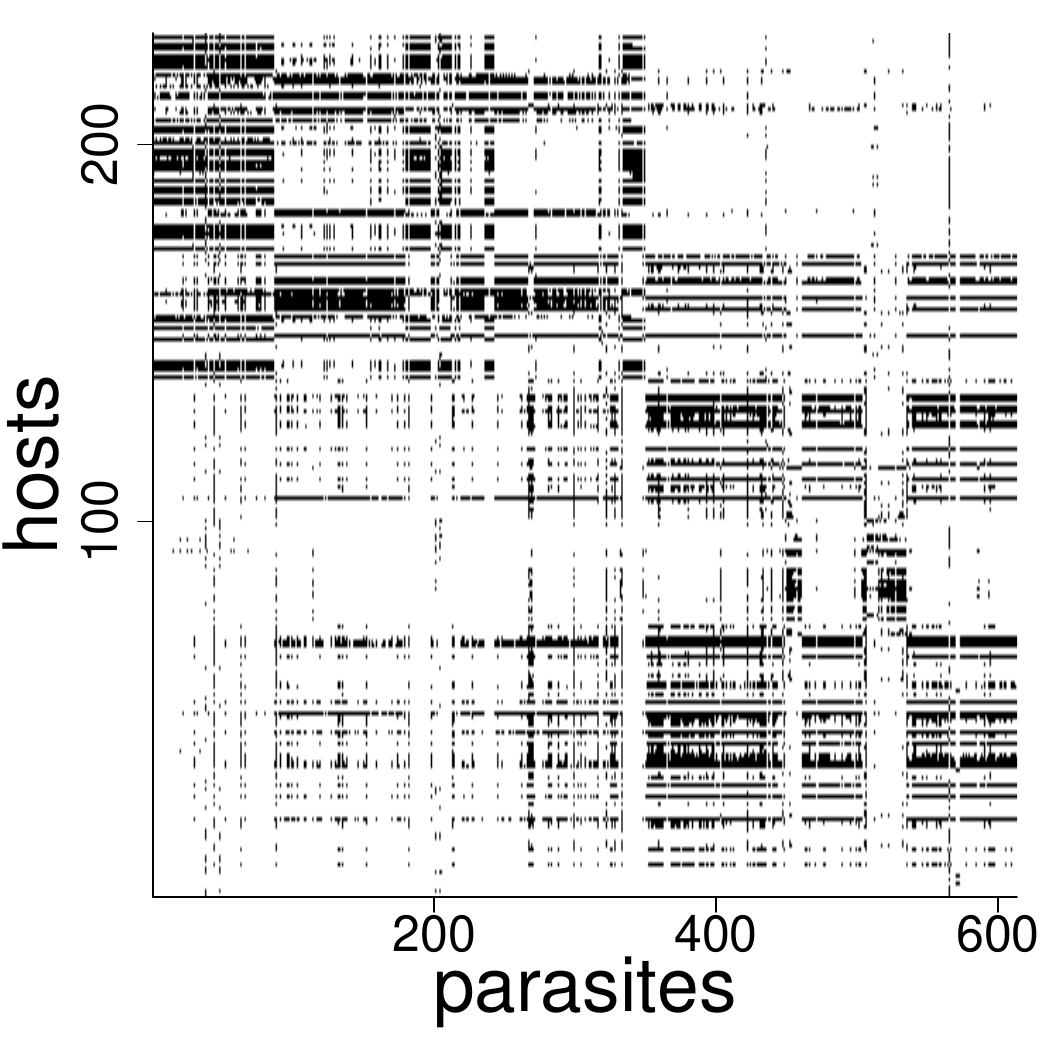}\label{fig:latent-with-euc-GMP}}
\caption{Posterior associations matrix comparison for the GMPD: between the nearest−neighbour(left), bilinear latent-distance (middle) and latent-distance model (right), for GMPD excluding single-host parasites. For more details please refer to Section \ref{sec:CrossValid}.}
\label{fig:Z-GMP-Progression-extra-models}
\vspace{-5em}
\end{figure}

\begin{figure}[t!]
  \centering
\subfloat[][without \(g\)]{\includegraphics[width=0.5\textwidth]{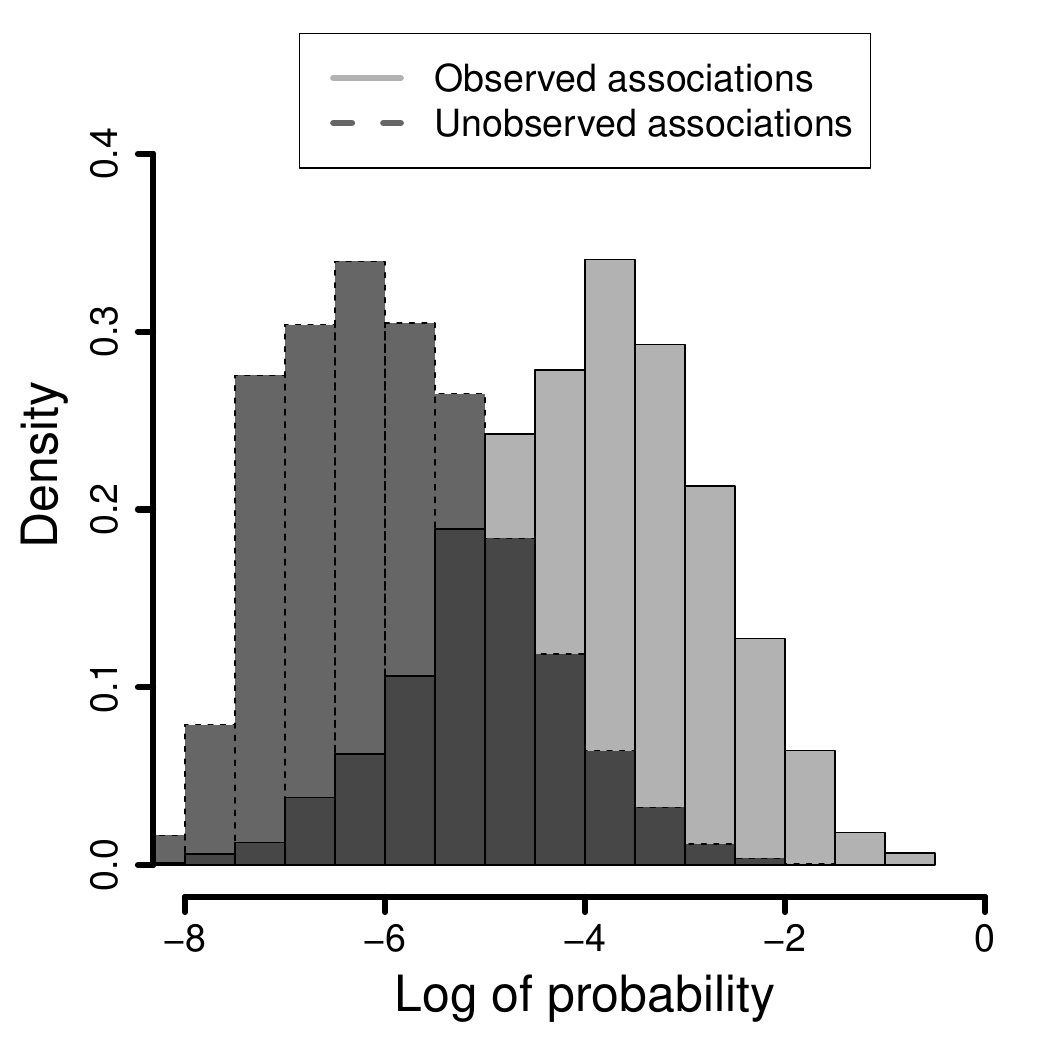}}   
\subfloat[][with \(g\)]{\includegraphics[width=0.5\textwidth]{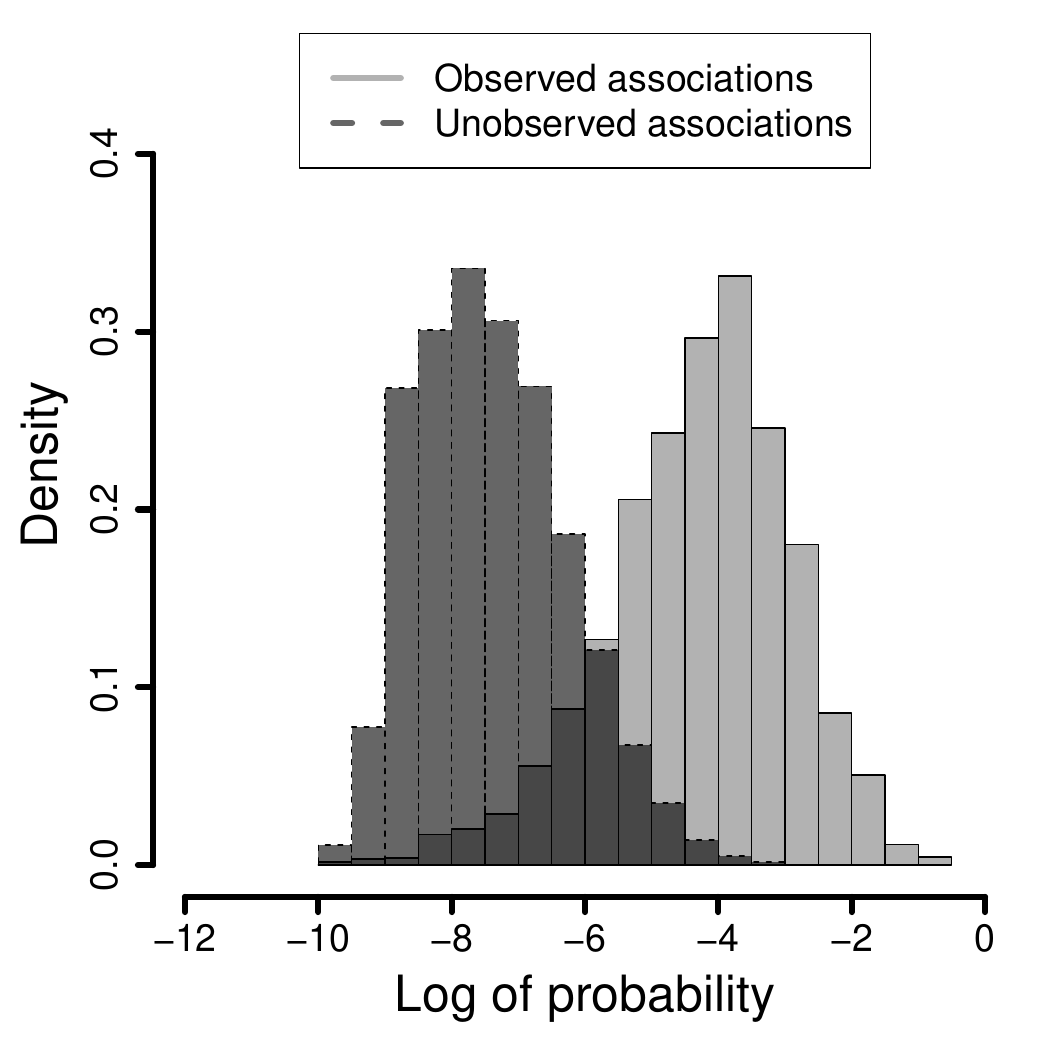}}
\caption{Comparison in posterior log-probability between observed and unobserved interactions; model without $g$ (left) and with $g$ (right), for GMPD with single-host parasites.}\label{fig:GMP-OBS-UNK1}
\vspace{-5em}
\end{figure}


\begin{figure}[t!]
  \centering
  \includegraphics[width=1\textwidth]{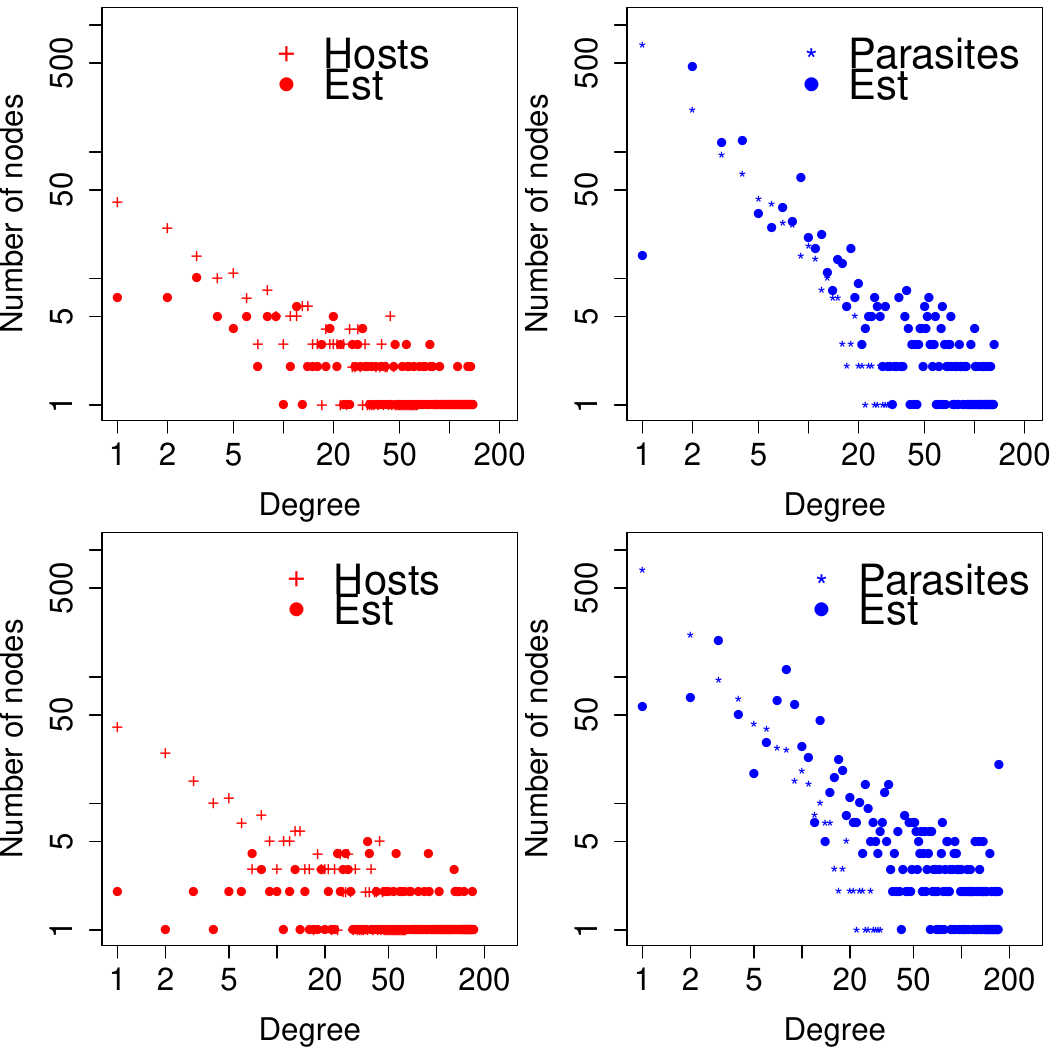}
  \caption{Comparison of degree distribution on log-scale, for the full model (without accounting for uncertainty)  and the model with $g$, 2010 GMPD with single-host parasites.}
\label{fig:GMP-Degree-G}
\end{figure}
\clearpage
\begin{figure}[ht!]
  \centering
\subfloat[][full unknowns]{\includegraphics[width=0.5\textwidth]{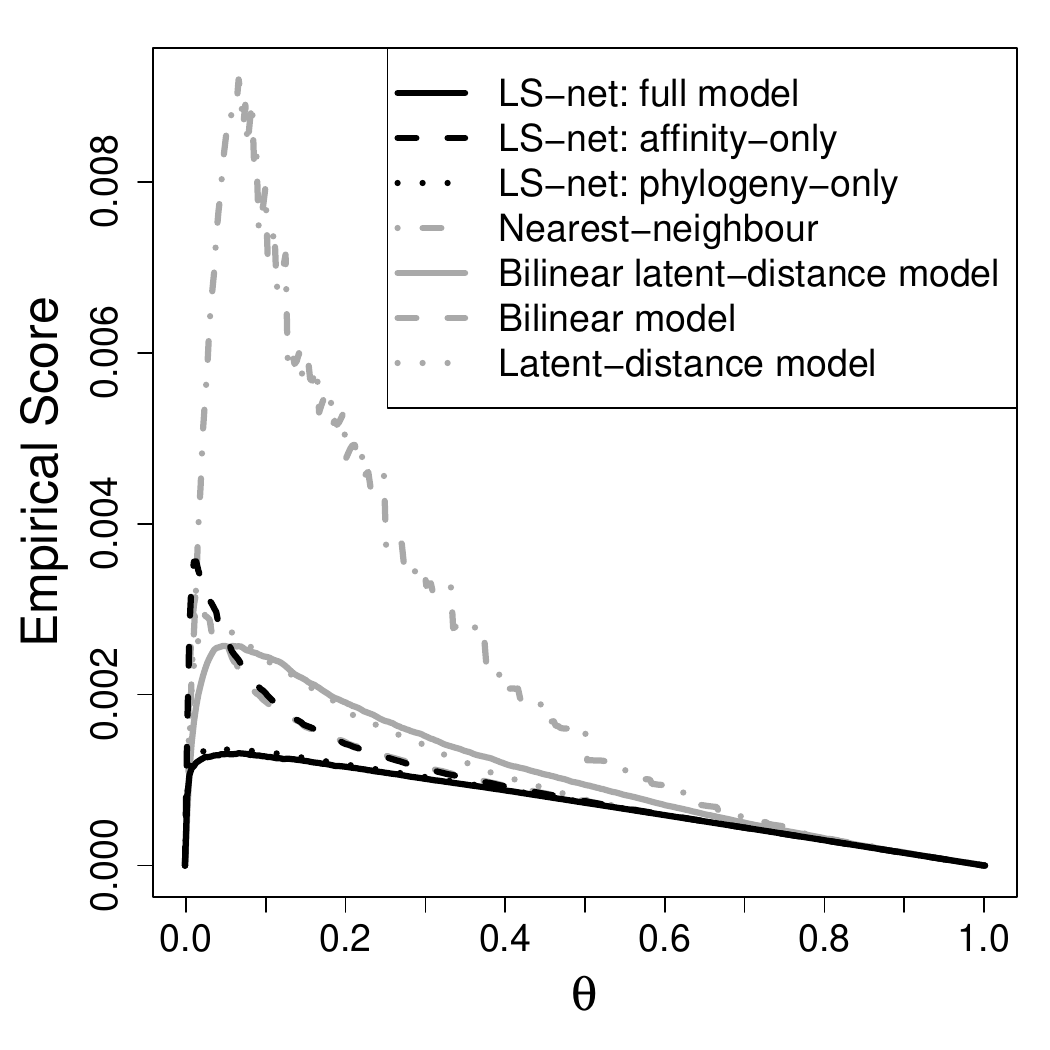}\label{fig:murphy-full-supp}}
\subfloat[][held-out 1's (presence-only)]{\includegraphics[width=0.5\textwidth]{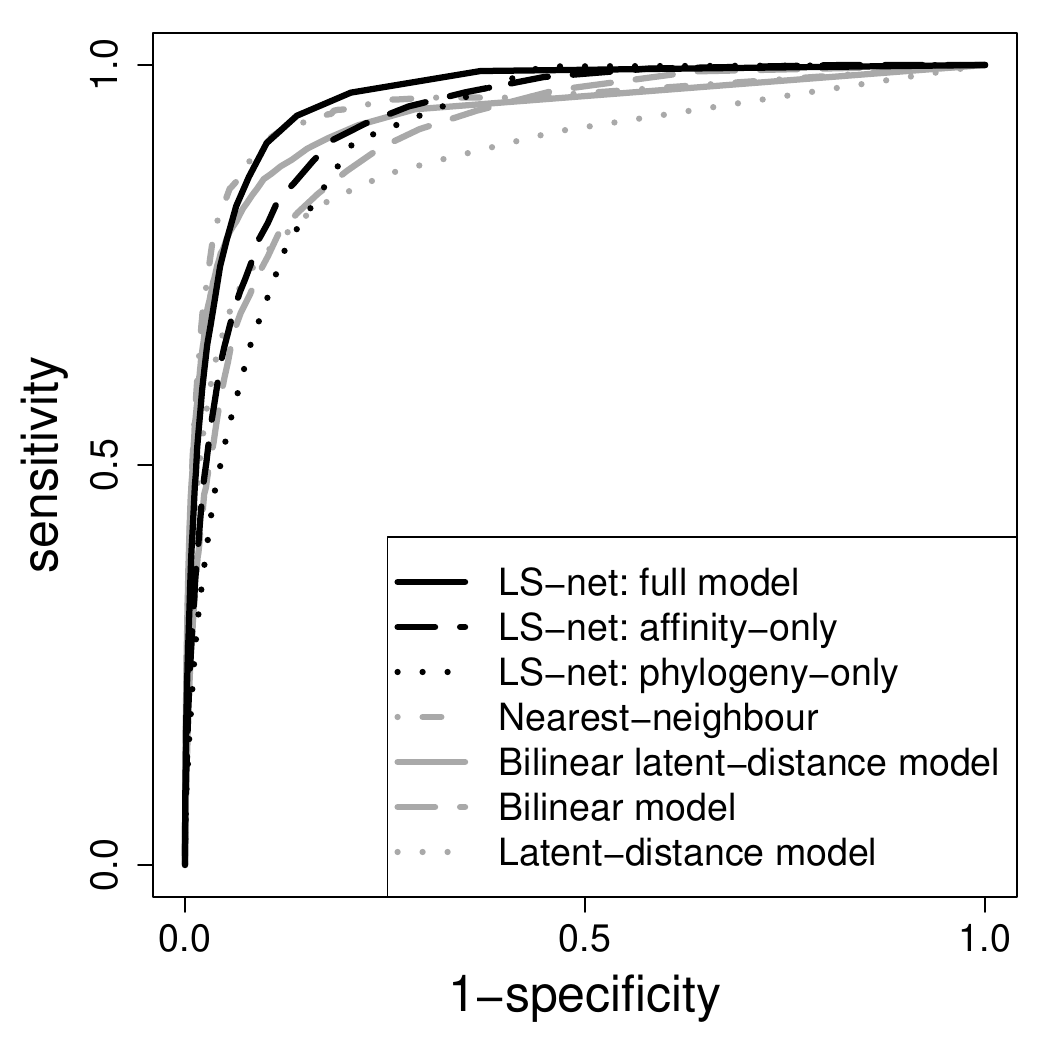}\label{fig:murphy-1000-supp}}
\caption{
  Murphy's diagrams and ROC curves of the latent score network (LS-net) model and two of its submodels, in comparison to competing models, the NN algorithms,
  the bilinear latent-distance models and two of its submodels (bilinear and latent-distance). Results are based on an average of 5-fold cross-validations on GMPD including single-host parasites.}
\label{fig:murphy-supp}
\end{figure}

{\small
\begin{table}[b!]
\centering
\caption{Simulation times for different models. LS-net models are ICM-based, implement in {\sf R} and run for 20,000 MCMC iterations. \citet{hoff2005bilinear}'s models are estimated with the official {\sf latentnet} {\sf R}-package \citep{latentnetPackage,latentnetPaper}, with option {\sf ergmm.control(mle.maxit=10)} . All simulations are run on a single core AMD Opteron 6380.}
\label{tb:sim-times}
\begin{tabular}{lllrl}
Size of GMPD                                       & Model                                             & time (hours) \\
\hline
\(229\times 613\) (with single-host parasites)     & LS-net full                                       & 3.85         \\
                                                   & LS-net phylogeny-only                             & 1.68         \\
                                                   & Bilinear latent-distance \citep{hoff2005bilinear} & 5.71         \\
                                                   & latent-distance \citep{hoff2005bilinear}          & 5.49         \\
\(236\times 1308\) (without single-host parasites) & LS-net full                                       & 2.10         \\
                                                   & LS-net phylogeny-only                             & 0.94        \\ 
                                                   & Bilinear latent-distance \citep{hoff2005bilinear} & 2.09         \\
                                                   & latent-distance          \citep{hoff2005bilinear} & 2.08         \\
\end{tabular}
\end{table}
}

\end{appendices}


%% file: HP-arxiv.bbl
\begin{thebibliography}{61}

\bibitem[\protect\citeauthoryear{Aguirre et~al.}{2007}]{Aguirre2007}
\begin{barticle}[author]
\bauthor{\bsnm{Aguirre},~\bfnm{A.~Alonso}\binits{A.~A.}},
  \bauthor{\bsnm{Keefe},~\bfnm{Thomas~J.}\binits{T.~J.}},
  \bauthor{\bsnm{Reif},~\bfnm{John~S.}\binits{J.~S.}},
  \bauthor{\bsnm{Kashinsky},~\bfnm{Lizabeth}\binits{L.}},
  \bauthor{\bsnm{Yochem},~\bfnm{Pamela~K.}\binits{P.~K.}},
  \bauthor{\bsnm{Saliki},~\bfnm{Jeremiah~T.}\binits{J.~T.}},
  \bauthor{\bsnm{Stott},~\bfnm{Jeffrey~L.}\binits{J.~L.}},
  \bauthor{\bsnm{Goldstein},~\bfnm{Tracey}\binits{T.}},
  \bauthor{\bsnm{Dubey},~\bfnm{J.~P.}\binits{J.~P.}},
  \bauthor{\bsnm{Braun},~\bfnm{Robert}\binits{R.}} \AND
  \bauthor{\bsnm{Antonelis},~\bfnm{George}\binits{G.}}
(\byear{2007}).
\btitle{INFECTIOUS DISEASE MONITORING OF THE ENDANGERED HAWAIIAN MONK SEAL}.
\bjournal{Journal of Wildlife Diseases}
\bvolume{43}
\bpages{229-241}.
\bdoi{10.7589/0090-3558-43.2.229}
\end{barticle}
\endbibitem

\bibitem[\protect\citeauthoryear{Albert and Barabasi}{2002}]{Albert2002}
\begin{barticle}[author]
\bauthor{\bsnm{Albert},~\bfnm{R{\'{e}}ka}\binits{R.}} \AND
  \bauthor{\bsnm{Barabasi},~\bfnm{Albert~Laszlo}\binits{A.~L.}}
(\byear{2002}).
\btitle{{Statistical mechanics of complex networks}}.
\bjournal{Reviews of Modern Physics}
\bvolume{74}
\bpages{47--97}.
\end{barticle}
\endbibitem

\bibitem[\protect\citeauthoryear{Bartomeus}{2013}]{Bartomeus2013}
\begin{barticle}[author]
\bauthor{\bsnm{Bartomeus},~\bfnm{Ignasi}\binits{I.}}
(\byear{2013}).
\btitle{{Understanding Linkage Rules in Plant-Pollinator Networks by Using
  Hierarchical Models That Incorporate Pollinator Detectability and Plant
  Traits}}.
\bjournal{PLoS ONE}
\bvolume{8}
\bpages{e69200}.
\bdoi{10.1371/journal.pone.0069200}
\end{barticle}
\endbibitem

\bibitem[\protect\citeauthoryear{Bastazini et~al.}{2017}]{Bastazini2017}
\begin{barticle}[author]
\bauthor{\bsnm{Bastazini},~\bfnm{Vinicius A.~G.}\binits{V.~A.~G.}},
  \bauthor{\bsnm{Ferreira},~\bfnm{Pedro M.~A.}\binits{P.~M.~A.}},
  \bauthor{\bsnm{Azambuja},~\bfnm{Beth{\^{a}}nia~O.}\binits{B.~O.}},
  \bauthor{\bsnm{Casas},~\bfnm{Grasiela}\binits{G.}},
  \bauthor{\bsnm{Debastiani},~\bfnm{Vanderlei~J.}\binits{V.~J.}},
  \bauthor{\bsnm{Guimar{\~{a}}es},~\bfnm{Paulo~R.}\binits{P.~R.}} \AND
  \bauthor{\bsnm{Pillar},~\bfnm{Val{\'{e}}rio~D.}\binits{V.~D.}}
(\byear{2017}).
\btitle{{Untangling the Tangled Bank: A Novel Method for Partitioning the
  Effects of Phylogenies and Traits on Ecological Networks}}.
\bjournal{Evolutionary Biology}
\bvolume{44}
\bpages{312--324}.
\bdoi{10.1007/s11692-017-9409-8}
\end{barticle}
\endbibitem

\bibitem[\protect\citeauthoryear{Besag}{1974}]{besag1974spatial}
\begin{barticle}[author]
\bauthor{\bsnm{Besag},~\bfnm{Julian}\binits{J.}}
(\byear{1974}).
\btitle{Spatial interaction and the statistical analysis of lattice systems}.
\bjournal{Journal of the Royal Statistical Society. Series B (Methodological)}
\bvolume{36}
\bpages{192--236}.
\end{barticle}
\endbibitem

\bibitem[\protect\citeauthoryear{Besag}{1986}]{besag1986statistical}
\begin{barticle}[author]
\bauthor{\bsnm{Besag},~\bfnm{Julian}\binits{J.}}
(\byear{1986}).
\btitle{On the statistical analysis of dirty pictures}.
\bjournal{Journal of the Royal Statistical Society. Series B (Methodological)}
\bvolume{48}
\bpages{259--302}.
\end{barticle}
\endbibitem

\bibitem[\protect\citeauthoryear{Bickel and
  Chen}{2009}]{bickel2009nonparametric}
\begin{barticle}[author]
\bauthor{\bsnm{Bickel},~\bfnm{Peter~J.}\binits{P.~J.}} \AND
  \bauthor{\bsnm{Chen},~\bfnm{Aiyou}\binits{A.}}
(\byear{2009}).
\btitle{A nonparametric view of network models and {Newman--Girvan} and other
  modularities}.
\bjournal{Proceedings of the National Academy of Sciences}
\bvolume{106}
\bpages{21068--21073}.
\end{barticle}
\endbibitem

\bibitem[\protect\citeauthoryear{Braga, Razzolini and Boeger}{2015}]{Braga2014}
\begin{barticle}[author]
\bauthor{\bsnm{Braga},~\bfnm{Mariana~P.}\binits{M.~P.}},
  \bauthor{\bsnm{Razzolini},~\bfnm{Emanuel}\binits{E.}} \AND
  \bauthor{\bsnm{Boeger},~\bfnm{Walter~A.}\binits{W.~A.}}
(\byear{2015}).
\btitle{{Drivers of parasite sharing among Neotropical freshwater fishes}}.
\bjournal{Journal of Animal Ecology}
\bvolume{84}
\bpages{487--497}.
\bdoi{10.1111/1365-2656.12298}
\end{barticle}
\endbibitem

\bibitem[\protect\citeauthoryear{Brix}{1999}]{brix1999generalized}
\begin{barticle}[author]
\bauthor{\bsnm{Brix},~\bfnm{Anders}\binits{A.}}
(\byear{1999}).
\btitle{Generalized gamma measures and shot-noise {Cox} processes}.
\bjournal{Advances in Applied Probability}
\bvolume{31}
\bpages{929--953}.
\end{barticle}
\endbibitem

\bibitem[\protect\citeauthoryear{Caron and Fox}{2017}]{caron2014sparse}
\begin{barticle}[author]
\bauthor{\bsnm{Caron},~\bfnm{François}\binits{F.}} \AND
  \bauthor{\bsnm{Fox},~\bfnm{Emily~B}\binits{E.~B.}}
(\byear{2017}).
\btitle{Sparse graphs using exchangeable random measures}.
\bjournal{Journal of the Royal Statistical Society: Series B (Statistical
  Methodology)}
\bvolume{79}
\bpages{1295-1366}.
\end{barticle}
\endbibitem

\bibitem[\protect\citeauthoryear{Chiu and Westveld}{2011}]{Chiu2011}
\begin{barticle}[author]
\bauthor{\bsnm{Chiu},~\bfnm{Grace~S.}\binits{G.~S.}} \AND
  \bauthor{\bsnm{Westveld},~\bfnm{Anton~H.}\binits{A.~H.}}
(\byear{2011}).
\btitle{A unifying approach for food webs, phylogeny, social networks, and
  statistics}.
\bjournal{Proceedings of the National Academy of Sciences}
\bvolume{108}
\bpages{15881--15886}.
\bdoi{10.1073/pnas.1015359108}
\end{barticle}
\endbibitem

\bibitem[\protect\citeauthoryear{Chung and Lu}{2006}]{chung2006complex}
\begin{bbook}[author]
\bauthor{\bsnm{Chung},~\bfnm{Fan}\binits{F.}} \AND
  \bauthor{\bsnm{Lu},~\bfnm{Linyuan}\binits{L.}}
(\byear{2006}).
\btitle{Complex graphs and networks}.
\bseries{107}.
\bpublisher{American Mathematical Society}.
\end{bbook}
\endbibitem

\bibitem[\protect\citeauthoryear{Cleaveland, Laurenson and
  Taylor}{2001}]{Cleaveland2001}
\begin{barticle}[author]
\bauthor{\bsnm{Cleaveland},~\bfnm{S}\binits{S.}},
  \bauthor{\bsnm{Laurenson},~\bfnm{M~K}\binits{M.~K.}} \AND
  \bauthor{\bsnm{Taylor},~\bfnm{L~H}\binits{L.~H.}}
(\byear{2001}).
\btitle{{Diseases of humans and their domestic mammals: pathogen
  characteristics, host range and the risk of emergence.}}
\bjournal{Philosophical transactions of the Royal Society of London. Series B,
  Biological sciences}
\bvolume{356}
\bpages{991--999}.
\bdoi{10.1098/rstb.2001.0889}
\end{barticle}
\endbibitem

\bibitem[\protect\citeauthoryear{Dallas, Park and
  Drake}{2017}]{dallas2017predicting}
\begin{barticle}[author]
\bauthor{\bsnm{Dallas},~\bfnm{Tad}\binits{T.}},
  \bauthor{\bsnm{Park},~\bfnm{Andrew~W}\binits{A.~W.}} \AND
  \bauthor{\bsnm{Drake},~\bfnm{John~M}\binits{J.~M.}}
(\byear{2017}).
\btitle{Predicting cryptic links in host-parasite networks}.
\bjournal{PLOS Computational Biology}
\bvolume{13}
\bpages{1-15}.
\end{barticle}
\endbibitem

\bibitem[\protect\citeauthoryear{Davies and Pedersen}{2008}]{Davies2008}
\begin{barticle}[author]
\bauthor{\bsnm{Davies},~\bfnm{T~Jonathan}\binits{T.~J.}} \AND
  \bauthor{\bsnm{Pedersen},~\bfnm{Amy~B}\binits{A.~B.}}
(\byear{2008}).
\btitle{{Phylogeny and geography predict pathogen community similarity in wild
  primates and humans.}}
\bjournal{Proceedings of the Royal Society - Biological sciences}
\bvolume{275}
\bpages{1695--701}.
\bdoi{10.1098/rspb.2008.0284}
\end{barticle}
\endbibitem

\bibitem[\protect\citeauthoryear{Dem{\v{s}}ar}{2006}]{Demsar}
\begin{barticle}[author]
\bauthor{\bsnm{Dem{\v{s}}ar},~\bfnm{Janez}\binits{J.}}
(\byear{2006}).
\btitle{Statistical Comparisons of Classifiers over Multiple Data Sets}.
\bjournal{Journal of Machine Learning Research}
\bvolume{7}
\bpages{1--30}.
\end{barticle}
\endbibitem

\bibitem[\protect\citeauthoryear{Ehm et~al.}{2016}]{ehm2016quantiles}
\begin{barticle}[author]
\bauthor{\bsnm{Ehm},~\bfnm{Werner}\binits{W.}},
  \bauthor{\bsnm{Gneiting},~\bfnm{Tilmann}\binits{T.}},
  \bauthor{\bsnm{Jordan},~\bfnm{Alexander}\binits{A.}} \AND
  \bauthor{\bsnm{Kr{\"u}ger},~\bfnm{Fabian}\binits{F.}}
(\byear{2016}).
\btitle{Of quantiles and expectiles: consistent scoring functions, Choquet
  representations and forecast rankings}.
\bjournal{Journal of the Royal Statistical Society: Series B (Statistical
  Methodology)}
\bvolume{78}
\bpages{505--562}.
\end{barticle}
\endbibitem

\bibitem[\protect\citeauthoryear{Farrell, Berrang-Ford and
  Davies}{2013}]{Farrell2013}
\begin{barticle}[author]
\bauthor{\bsnm{Farrell},~\bfnm{Maxwell~J}\binits{M.~J.}},
  \bauthor{\bsnm{Berrang-Ford},~\bfnm{Lea}\binits{L.}} \AND
  \bauthor{\bsnm{Davies},~\bfnm{T~Jonathan}\binits{T.~J.}}
(\byear{2013}).
\btitle{{The study of parasite sharing for surveillance of zoonotic diseases}}.
\bjournal{Environmental Research Letters}
\bvolume{8}
\bpages{015036}.
\end{barticle}
\endbibitem

\bibitem[\protect\citeauthoryear{Farrell et~al.}{2015}]{JANE:JANE12342}
\begin{barticle}[author]
\bauthor{\bsnm{Farrell},~\bfnm{Maxwell~J.}\binits{M.~J.}},
  \bauthor{\bsnm{Stephens},~\bfnm{Patrick~R.}\binits{P.~R.}},
  \bauthor{\bsnm{Berrang-Ford},~\bfnm{Lea}\binits{L.}},
  \bauthor{\bsnm{Gittleman},~\bfnm{John~L.}\binits{J.~L.}} \AND
  \bauthor{\bsnm{Davies},~\bfnm{T.~Jonathan}\binits{T.~J.}}
(\byear{2015}).
\btitle{The path to host extinction can lead to loss of generalist parasites}.
\bjournal{Journal of Animal Ecology}
\bvolume{84}
\bpages{978--984}.
\bdoi{10.1111/1365-2656.12342}
\end{barticle}
\endbibitem

\bibitem[\protect\citeauthoryear{Fritz, Bininda-Emonds and
  Purvis}{2009}]{Fritz2009}
\begin{barticle}[author]
\bauthor{\bsnm{Fritz},~\bfnm{Susanne~A}\binits{S.~A.}},
  \bauthor{\bsnm{Bininda-Emonds},~\bfnm{Olaf R~P}\binits{O.~R.~P.}} \AND
  \bauthor{\bsnm{Purvis},~\bfnm{Andy}\binits{A.}}
(\byear{2009}).
\btitle{{Geographical variation in predictors of mammalian extinction risk: big
  is bad, but only in the tropics.}}
\bjournal{Ecology Letters}
\bvolume{12}
\bpages{538--549}.
\bdoi{10.1111/j.1461-0248.2009.01307.x}
\end{barticle}
\endbibitem

\bibitem[\protect\citeauthoryear{Geman and Geman}{1984}]{geman1984stochastic}
\begin{barticle}[author]
\bauthor{\bsnm{Geman},~\bfnm{Stuart}\binits{S.}} \AND
  \bauthor{\bsnm{Geman},~\bfnm{Donald}\binits{D.}}
(\byear{1984}).
\btitle{Stochastic relaxation, {Gibbs} distributions, and the {Bayesian}
  restoration of images}.
\bjournal{IEEE Transactions on Pattern Analysis and Machine Intelligence}
\bvolume{6}
\bpages{721--741}.
\end{barticle}
\endbibitem

\bibitem[\protect\citeauthoryear{Gilbert and Webb}{2007}]{Gilbert2007}
\begin{barticle}[author]
\bauthor{\bsnm{Gilbert},~\bfnm{Gregory~S}\binits{G.~S.}} \AND
  \bauthor{\bsnm{Webb},~\bfnm{Campbell~O}\binits{C.~O.}}
(\byear{2007}).
\btitle{{Phylogenetic signal in plant pathogen-host range.}}
\bjournal{Proceedings of the National Academy of Sciences of the United States
  of America}
\bvolume{104}
\bpages{4979--4983}.
\bdoi{10.1073/pnas.0607968104}
\end{barticle}
\endbibitem

\bibitem[\protect\citeauthoryear{Gneiting and
  Raftery}{2007}]{gneiting2007strictly}
\begin{barticle}[author]
\bauthor{\bsnm{Gneiting},~\bfnm{Tilmann}\binits{T.}} \AND
  \bauthor{\bsnm{Raftery},~\bfnm{Adrian~E}\binits{A.~E.}}
(\byear{2007}).
\btitle{Strictly proper scoring rules, prediction, and estimation}.
\bjournal{Journal of the American Statistical Association}
\bvolume{102}
\bpages{359--378}.
\end{barticle}
\endbibitem

\bibitem[\protect\citeauthoryear{G{\'{o}}mez, Verd{\'{u}} and
  Perfectti}{2010}]{Gomez2010}
\begin{barticle}[author]
\bauthor{\bsnm{G{\'{o}}mez},~\bfnm{Jos{\'{e}}~M}\binits{J.~M.}},
  \bauthor{\bsnm{Verd{\'{u}}},~\bfnm{Miguel}\binits{M.}} \AND
  \bauthor{\bsnm{Perfectti},~\bfnm{Francisco}\binits{F.}}
(\byear{2010}).
\btitle{{Ecological interactions are evolutionarily conserved across the entire
  tree of life.}}
\bjournal{Nature}
\bvolume{465}
\bpages{918--21}.
\bdoi{10.1038/nature09113}
\end{barticle}
\endbibitem

\bibitem[\protect\citeauthoryear{Gravel et~al.}{2013}]{Gravel2013}
\begin{barticle}[author]
\bauthor{\bsnm{Gravel},~\bfnm{Dominique}\binits{D.}},
  \bauthor{\bsnm{Poisot},~\bfnm{Timoth{\'{e}}e}\binits{T.}},
  \bauthor{\bsnm{Albouy},~\bfnm{Camille}\binits{C.}},
  \bauthor{\bsnm{Velez},~\bfnm{Laure}\binits{L.}} \AND
  \bauthor{\bsnm{Mouillot},~\bfnm{David}\binits{D.}}
(\byear{2013}).
\btitle{{Inferring food web structure from predator-prey body size
  relationships}}.
\bjournal{Methods in Ecology and Evolution}
\bvolume{4}
\bpages{1083--1090}.
\bdoi{10.1111/2041-210X.12103}
\end{barticle}
\endbibitem

\bibitem[\protect\citeauthoryear{Haario, Saksman and
  Tamminen}{2001}]{haario2001adaptive}
\begin{barticle}[author]
\bauthor{\bsnm{Haario},~\bfnm{Heikki}\binits{H.}},
  \bauthor{\bsnm{Saksman},~\bfnm{Eero}\binits{E.}} \AND
  \bauthor{\bsnm{Tamminen},~\bfnm{Johanna}\binits{J.}}
(\byear{2001}).
\btitle{An adaptive Metropolis algorithm}.
\bjournal{Bernoulli}
\bvolume{7}
\bpages{223--242}.
\end{barticle}
\endbibitem

\bibitem[\protect\citeauthoryear{Harmon et~al.}{2010}]{harmon2010early}
\begin{barticle}[author]
\bauthor{\bsnm{Harmon},~\bfnm{Luke~J}\binits{L.~J.}},
  \bauthor{\bsnm{Losos},~\bfnm{Jonathan~B}\binits{J.~B.}},
  \bauthor{\bsnm{Jonathan~Davies},~\bfnm{T}\binits{T.}},
  \bauthor{\bsnm{Gillespie},~\bfnm{Rosemary~G}\binits{R.~G.}},
  \bauthor{\bsnm{Gittleman},~\bfnm{John~L}\binits{J.~L.}},
  \bauthor{\bsnm{Bryan~Jennings},~\bfnm{W}\binits{W.}},
  \bauthor{\bsnm{Kozak},~\bfnm{Kenneth~H}\binits{K.~H.}},
  \bauthor{\bsnm{McPeek},~\bfnm{Mark~A}\binits{M.~A.}},
  \bauthor{\bsnm{Moreno-Roark},~\bfnm{Franck}\binits{F.}},
  \bauthor{\bsnm{Near},~\bfnm{Thomas~J}\binits{T.~J.}} \betal{et~al.}
(\byear{2010}).
\btitle{Early bursts of body size and shape evolution are rare in comparative
  data}.
\bjournal{Evolution}
\bvolume{64}
\bpages{2385--2396}.
\end{barticle}
\endbibitem

\bibitem[\protect\citeauthoryear{Hastie and
  Fithian}{2013}]{hastie2013inference}
\begin{barticle}[author]
\bauthor{\bsnm{Hastie},~\bfnm{Trevor}\binits{T.}} \AND
  \bauthor{\bsnm{Fithian},~\bfnm{Will}\binits{W.}}
(\byear{2013}).
\btitle{Inference from presence-only data; the ongoing controversy}.
\bjournal{Ecography}
\bvolume{36}
\bpages{864--867}.
\end{barticle}
\endbibitem

\bibitem[\protect\citeauthoryear{Hoff}{2005}]{hoff2005bilinear}
\begin{barticle}[author]
\bauthor{\bsnm{Hoff},~\bfnm{Peter~D}\binits{P.~D.}}
(\byear{2005}).
\btitle{Bilinear mixed-effects models for dyadic data}.
\bjournal{Journal of the American Statistical Association}
\bvolume{100}
\bpages{286--295}.
\end{barticle}
\endbibitem

\bibitem[\protect\citeauthoryear{Hoff, Raftery and
  Handcock}{2002}]{hoff2002latent}
\begin{barticle}[author]
\bauthor{\bsnm{Hoff},~\bfnm{Peter~D}\binits{P.~D.}},
  \bauthor{\bsnm{Raftery},~\bfnm{Adrian~E}\binits{A.~E.}} \AND
  \bauthor{\bsnm{Handcock},~\bfnm{Mark~S}\binits{M.~S.}}
(\byear{2002}).
\btitle{Latent space approaches to social network analysis}.
\bjournal{Journal of the American Statistical Association}
\bvolume{97}
\bpages{1090--1098}.
\end{barticle}
\endbibitem

\bibitem[\protect\citeauthoryear{Huang et~al.}{2015}]{Huang2015}
\begin{barticle}[author]
\bauthor{\bsnm{Huang},~\bfnm{Shan}\binits{S.}},
  \bauthor{\bsnm{Drake},~\bfnm{John~M.}\binits{J.~M.}},
  \bauthor{\bsnm{Gittleman},~\bfnm{John~L.}\binits{J.~L.}} \AND
  \bauthor{\bsnm{Altizer},~\bfnm{Sonia}\binits{S.}}
(\byear{2015}).
\btitle{{Parasite diversity declines with host evolutionary distinctiveness: A
  global analysis of carnivores}}.
\bjournal{Evolution}
\bvolume{69}
\bpages{621--630}.
\bdoi{10.1111/evo.12611}
\end{barticle}
\endbibitem

\bibitem[\protect\citeauthoryear{Jiang, Gold and
  Kolaczyk}{2011}]{jiang2011network}
\begin{barticle}[author]
\bauthor{\bsnm{Jiang},~\bfnm{Xiaoyu}\binits{X.}},
  \bauthor{\bsnm{Gold},~\bfnm{David}\binits{D.}} \AND
  \bauthor{\bsnm{Kolaczyk},~\bfnm{Eric~D}\binits{E.~D.}}
(\byear{2011}).
\btitle{Network-based Auto-probit Modeling for Protein Function Prediction}.
\bjournal{Biometrics}
\bvolume{67}
\bpages{958--966}.
\end{barticle}
\endbibitem

\bibitem[\protect\citeauthoryear{Jordano}{2016}]{Jordano2015}
\begin{barticle}[author]
\bauthor{\bsnm{Jordano},~\bfnm{Pedro}\binits{P.}}
(\byear{2016}).
\btitle{{Sampling networks of ecological interactions}}.
\bjournal{Functional Ecology}
\bvolume{30}
\bpages{1883-1893}.
\end{barticle}
\endbibitem

\bibitem[\protect\citeauthoryear{Krivitsky and Handcock}{2008}]{latentnetPaper}
\begin{barticle}[author]
\bauthor{\bsnm{Krivitsky},~\bfnm{Pavel~N.}\binits{P.~N.}} \AND
  \bauthor{\bsnm{Handcock},~\bfnm{Mark~S.}\binits{M.~S.}}
(\byear{2008}).
\btitle{Fitting position latent cluster models for social networks with
  latentnet}.
\bjournal{Journal of Statistical Software}
\bvolume{24}.
\end{barticle}
\endbibitem

\bibitem[\protect\citeauthoryear{Krivitsky and
  Handcock}{2017}]{latentnetPackage}
\begin{bmanual}[author]
\bauthor{\bsnm{Krivitsky},~\bfnm{Pavel~N.}\binits{P.~N.}} \AND
  \bauthor{\bsnm{Handcock},~\bfnm{Mark~S.}\binits{M.~S.}}
(\byear{2017}).
\btitle{latentnet: Latent Position and Cluster Models for Statistical Networks}
\bpublisher{The Statnet Project (\url{http://www.statnet.org})}
\bnote{R package version 2.8.0}.
\end{bmanual}
\endbibitem

\bibitem[\protect\citeauthoryear{{La Salle}, Williams and
  Moritz}{2016}]{LaSalle2016}
\begin{barticle}[author]
\bauthor{\bsnm{{La Salle}},~\bfnm{John}\binits{J.}},
  \bauthor{\bsnm{Williams},~\bfnm{Kristen~J.}\binits{K.~J.}} \AND
  \bauthor{\bsnm{Moritz},~\bfnm{Craig}\binits{C.}}
(\byear{2016}).
\btitle{{Biodiversity analysis in the digital era}}.
\bjournal{Philosophical Transactions of the Royal Society B: Biological
  Sciences}
\bvolume{371}
\bpages{20150337}.
\end{barticle}
\endbibitem

\bibitem[\protect\citeauthoryear{Lijoi, Mena and
  Pr{\"u}nster}{2007}]{lijoi2007controlling}
\begin{barticle}[author]
\bauthor{\bsnm{Lijoi},~\bfnm{Antonio}\binits{A.}},
  \bauthor{\bsnm{Mena},~\bfnm{Rams{\'e}s~H}\binits{R.~H.}} \AND
  \bauthor{\bsnm{Pr{\"u}nster},~\bfnm{Igor}\binits{I.}}
(\byear{2007}).
\btitle{Controlling the reinforcement in {Bayesian} non-parametric mixture
  models}.
\bjournal{Journal of the Royal Statistical Society: Series B (Statistical
  Methodology)}
\bvolume{69}
\bpages{715--740}.
\end{barticle}
\endbibitem

\bibitem[\protect\citeauthoryear{Luis et~al.}{2015}]{Luis2015}
\begin{barticle}[author]
\bauthor{\bsnm{Luis},~\bfnm{Angela~D.}\binits{A.~D.}},
  \bauthor{\bsnm{O'Shea},~\bfnm{Thomas~J.}\binits{T.~J.}},
  \bauthor{\bsnm{Hayman},~\bfnm{David T~S}\binits{D.~T.~S.}},
  \bauthor{\bsnm{Wood},~\bfnm{James L~N}\binits{J.~L.~N.}},
  \bauthor{\bsnm{Cunningham},~\bfnm{Andrew~A.}\binits{A.~A.}},
  \bauthor{\bsnm{Gilbert},~\bfnm{Amy~T.}\binits{A.~T.}},
  \bauthor{\bsnm{Mills},~\bfnm{James~N.}\binits{J.~N.}} \AND
  \bauthor{\bsnm{Webb},~\bfnm{Colleen~T.}\binits{C.~T.}}
(\byear{2015}).
\btitle{{Network analysis of host-virus communities in bats and rodents reveals
  determinants of cross-species transmission}}.
\bjournal{Ecology Letters}
\bvolume{18}
\bpages{1153--1162}.
\bdoi{10.1111/ele.12491}
\end{barticle}
\endbibitem

\bibitem[\protect\citeauthoryear{Morales-Castilla
  et~al.}{2015}]{MoralesCastilla2015}
\begin{barticle}[author]
\bauthor{\bsnm{Morales-Castilla},~\bfnm{Ignacio}\binits{I.}},
  \bauthor{\bsnm{Matias},~\bfnm{Miguel~G}\binits{M.~G.}},
  \bauthor{\bsnm{Gravel},~\bfnm{Dominique}\binits{D.}} \AND
  \bauthor{\bsnm{Ara{\'u}jo},~\bfnm{Miguel~B}\binits{M.~B.}}
(\byear{2015}).
\btitle{Inferring biotic interactions from proxies}.
\bjournal{Trends in Ecology \& Evolution}
\bvolume{30}
\bpages{347--356}.
\end{barticle}
\endbibitem

\bibitem[\protect\citeauthoryear{Olival et~al.}{2017}]{Olival2017}
\begin{barticle}[author]
\bauthor{\bsnm{Olival},~\bfnm{Kevin~J}\binits{K.~J.}},
  \bauthor{\bsnm{Hosseini},~\bfnm{Parviez~R}\binits{P.~R.}},
  \bauthor{\bsnm{Zambrana-Torrelio},~\bfnm{Carlos}\binits{C.}},
  \bauthor{\bsnm{Ross},~\bfnm{Noam}\binits{N.}},
  \bauthor{\bsnm{Bogich},~\bfnm{Tiffany~L}\binits{T.~L.}} \AND
  \bauthor{\bsnm{Daszak},~\bfnm{Peter}\binits{P.}}
(\byear{2017}).
\btitle{Host and viral traits predict zoonotic spillover from mammals}.
\bjournal{Nature}
\bvolume{546}
\bpages{646--650}.
\end{barticle}
\endbibitem

\bibitem[\protect\citeauthoryear{Ovaskainen et~al.}{2016}]{Ovaskainen2016}
\begin{barticle}[author]
\bauthor{\bsnm{Ovaskainen},~\bfnm{Otso}\binits{O.}},
  \bauthor{\bsnm{Abrego},~\bfnm{Nerea}\binits{N.}},
  \bauthor{\bsnm{Halme},~\bfnm{Panu}\binits{P.}} \AND
  \bauthor{\bsnm{Dunson},~\bfnm{David}\binits{D.}}
(\byear{2016}).
\btitle{{Using latent variable models to identify large networks of
  species-to-species associations at different spatial scales}}.
\bjournal{Methods in Ecology and Evolution}
\bvolume{7}
\bpages{549--555}.
\bdoi{10.1111/2041-210X.12501}
\end{barticle}
\endbibitem

\bibitem[\protect\citeauthoryear{Ovaskainen et~al.}{2017}]{Ovaskainen2017}
\begin{barticle}[author]
\bauthor{\bsnm{Ovaskainen},~\bfnm{Otso}\binits{O.}},
  \bauthor{\bsnm{Tikhonov},~\bfnm{Gleb}\binits{G.}},
  \bauthor{\bsnm{Norberg},~\bfnm{Anna}\binits{A.}}, \bauthor{\bsnm{{Guillaume
  Blanchet}},~\bfnm{F.}\binits{F.}},
  \bauthor{\bsnm{Duan},~\bfnm{Leo}\binits{L.}},
  \bauthor{\bsnm{Dunson},~\bfnm{David}\binits{D.}},
  \bauthor{\bsnm{Roslin},~\bfnm{Tomas}\binits{T.}} \AND
  \bauthor{\bsnm{Abrego},~\bfnm{Nerea}\binits{N.}}
(\byear{2017}).
\btitle{{How to make more out of community data? A conceptual framework and its
  implementation as models and software}}.
\bjournal{Ecology Letters}
\bvolume{20}
\bpages{561--576}.
\bdoi{10.1111/ele.12757}
\end{barticle}
\endbibitem

\bibitem[\protect\citeauthoryear{Pagel}{1999}]{pagel1999inferring}
\begin{barticle}[author]
\bauthor{\bsnm{Pagel},~\bfnm{Mark}\binits{M.}}
(\byear{1999}).
\btitle{Inferring the historical patterns of biological evolution}.
\bjournal{Nature}
\bvolume{401}
\bpages{877--884}.
\end{barticle}
\endbibitem

\bibitem[\protect\citeauthoryear{Park et~al.}{2018}]{park2018characterizing}
\begin{barticle}[author]
\bauthor{\bsnm{Park},~\bfnm{AW}\binits{A.}},
  \bauthor{\bsnm{Farrell},~\bfnm{MJ}\binits{M.}},
  \bauthor{\bsnm{Schmidt},~\bfnm{JP}\binits{J.}},
  \bauthor{\bsnm{Huang},~\bfnm{S}\binits{S.}},
  \bauthor{\bsnm{Dallas},~\bfnm{TA}\binits{T.}},
  \bauthor{\bsnm{Pappalardo},~\bfnm{P}\binits{P.}},
  \bauthor{\bsnm{Drake},~\bfnm{JM}\binits{J.}},
  \bauthor{\bsnm{Stephens},~\bfnm{PR}\binits{P.}},
  \bauthor{\bsnm{Poulin},~\bfnm{R}\binits{R.}},
  \bauthor{\bsnm{Nunn},~\bfnm{CL}\binits{C.}} \betal{et~al.}
(\byear{2018}).
\btitle{Characterizing the phylogenetic specialism--generalism spectrum of
  mammal parasites}.
\bjournal{Proceedings of the Royal Society B: Biological Sciences}
\bvolume{285}
\bpages{20172613}.
\end{barticle}
\endbibitem

\bibitem[\protect\citeauthoryear{Parrish et~al.}{2008}]{Parrish2008}
\begin{barticle}[author]
\bauthor{\bsnm{Parrish},~\bfnm{Colin~R}\binits{C.~R.}},
  \bauthor{\bsnm{Holmes},~\bfnm{Edward~C}\binits{E.~C.}},
  \bauthor{\bsnm{Morens},~\bfnm{David~M}\binits{D.~M.}},
  \bauthor{\bsnm{Park},~\bfnm{Eun-Chung}\binits{E.-C.}},
  \bauthor{\bsnm{Burke},~\bfnm{Donald~S}\binits{D.~S.}},
  \bauthor{\bsnm{Calisher},~\bfnm{Charles~H}\binits{C.~H.}},
  \bauthor{\bsnm{Laughlin},~\bfnm{Catherine~a}\binits{C.~a.}},
  \bauthor{\bsnm{Saif},~\bfnm{Linda~J}\binits{L.~J.}} \AND
  \bauthor{\bsnm{Daszak},~\bfnm{Peter}\binits{P.}}
(\byear{2008}).
\btitle{{Cross-species virus transmission and the emergence of new epidemic
  diseases.}}
\bjournal{Microbiology and Molecular Biology Reviews}
\bvolume{72}
\bpages{457--70}.
\bdoi{10.1128/MMBR.00004-08}
\end{barticle}
\endbibitem

\bibitem[\protect\citeauthoryear{Pearse and Altermatt}{2013}]{Pearse2013}
\begin{barticle}[author]
\bauthor{\bsnm{Pearse},~\bfnm{Ian~S.}\binits{I.~S.}} \AND
  \bauthor{\bsnm{Altermatt},~\bfnm{Florian}\binits{F.}}
(\byear{2013}).
\btitle{{Predicting novel trophic interactions in a non-native world}}.
\bjournal{Ecology Letters}
\bvolume{16}
\bpages{1088--1094}.
\bdoi{10.1111/ele.12143}
\end{barticle}
\endbibitem

\bibitem[\protect\citeauthoryear{Pedersen et~al.}{2005}]{Pedersen2005}
\begin{barticle}[author]
\bauthor{\bsnm{Pedersen},~\bfnm{Amy~B}\binits{A.~B.}},
  \bauthor{\bsnm{Altizer},~\bfnm{Sonia}\binits{S.}},
  \bauthor{\bsnm{Poss},~\bfnm{Mary}\binits{M.}},
  \bauthor{\bsnm{Cunningham},~\bfnm{Andrew~A}\binits{A.~A.}} \AND
  \bauthor{\bsnm{Nunn},~\bfnm{Charles~L}\binits{C.~L.}}
(\byear{2005}).
\btitle{{Patterns of host specificity and transmission among parasites of wild
  primates.}}
\bjournal{International Journal for Parasitology}
\bvolume{35}
\bpages{647--57}.
\bdoi{10.1016/j.ijpara.2005.01.005}
\end{barticle}
\endbibitem

\bibitem[\protect\citeauthoryear{Pedersen et~al.}{2007}]{Pedersen2007}
\begin{barticle}[author]
\bauthor{\bsnm{Pedersen},~\bfnm{Amy~B}\binits{A.~B.}},
  \bauthor{\bsnm{Jones},~\bfnm{Kate~E}\binits{K.~E.}},
  \bauthor{\bsnm{Nunn},~\bfnm{Charles~L}\binits{C.~L.}} \AND
  \bauthor{\bsnm{Altizer},~\bfnm{Sonia}\binits{S.}}
(\byear{2007}).
\btitle{{Infectious diseases and extinction risk in wild mammals.}}
\bjournal{Conservation Biology}
\bvolume{21}
\bpages{1269--79}.
\bdoi{10.1111/j.1523-1739.2007.00776.x}
\end{barticle}
\endbibitem

\bibitem[\protect\citeauthoryear{Petchey et~al.}{2008}]{Petchey2008}
\begin{barticle}[author]
\bauthor{\bsnm{Petchey},~\bfnm{O.~L.}\binits{O.~L.}},
  \bauthor{\bsnm{Beckerman},~\bfnm{A.~P.}\binits{A.~P.}},
  \bauthor{\bsnm{Riede},~\bfnm{J.~O.}\binits{J.~O.}} \AND
  \bauthor{\bsnm{Warren},~\bfnm{P.~H.}\binits{P.~H.}}
(\byear{2008}).
\btitle{{Size, foraging, and food web structure}}.
\bjournal{Proceedings of the National Academy of Sciences}
\bvolume{105}
\bpages{4191--4196}.
\bdoi{10.1073/pnas.0710672105}
\end{barticle}
\endbibitem

\bibitem[\protect\citeauthoryear{Poelen, Simons and
  Mungall}{2014}]{POELEN2014148}
\begin{barticle}[author]
\bauthor{\bsnm{Poelen},~\bfnm{Jorrit~H.}\binits{J.~H.}},
  \bauthor{\bsnm{Simons},~\bfnm{James~D.}\binits{J.~D.}} \AND
  \bauthor{\bsnm{Mungall},~\bfnm{Chris~J.}\binits{C.~J.}}
(\byear{2014}).
\btitle{Global biotic interactions: An open infrastructure to share and analyze
  species-interaction datasets}.
\bjournal{Ecological Informatics}
\bvolume{24}
\bpages{148--159}.
\end{barticle}
\endbibitem

\bibitem[\protect\citeauthoryear{Robert and Casella}{2013}]{robert2013monte}
\begin{bbook}[author]
\bauthor{\bsnm{Robert},~\bfnm{Christian}\binits{C.}} \AND
  \bauthor{\bsnm{Casella},~\bfnm{George}\binits{G.}}
(\byear{2013}).
\btitle{{Monte Carlo} statistical methods}.
\bpublisher{Springer Science \& Business Media}.
\end{bbook}
\endbibitem

\bibitem[\protect\citeauthoryear{Stephens et~al.}{2017}]{Stephens2017}
\begin{barticle}[author]
\bauthor{\bsnm{Stephens},~\bfnm{Patrick~R}\binits{P.~R.}},
  \bauthor{\bsnm{Pappalardo},~\bfnm{Paula}\binits{P.}},
  \bauthor{\bsnm{Huang},~\bfnm{Shan}\binits{S.}},
  \bauthor{\bsnm{Byers},~\bfnm{James~E.}\binits{J.~E.}},
  \bauthor{\bsnm{Farrell},~\bfnm{Maxwell~J.}\binits{M.~J.}},
  \bauthor{\bsnm{Gehman},~\bfnm{Alyssa}\binits{A.}},
  \bauthor{\bsnm{Ghai},~\bfnm{Ria~R.}\binits{R.~R.}},
  \bauthor{\bsnm{Haas},~\bfnm{Sarah~E.}\binits{S.~E.}},
  \bauthor{\bsnm{Han},~\bfnm{Barbara}\binits{B.}},
  \bauthor{\bsnm{Park},~\bfnm{Andrew~W.}\binits{A.~W.}},
  \bauthor{\bsnm{Schmidt},~\bfnm{John~P.}\binits{J.~P.}},
  \bauthor{\bsnm{Altizer},~\bfnm{Sonia}\binits{S.}},
  \bauthor{\bsnm{Ezenwa},~\bfnm{Vanessa~O.}\binits{V.~O.}} \AND
  \bauthor{\bsnm{Nunn},~\bfnm{Charles~L.}\binits{C.~L.}}
(\byear{2017}).
\btitle{{Global Mammal Parasite Database version 2.0}}.
\bjournal{Ecology}
\bvolume{98}
\bpages{1476-1476}.
\bdoi{10.1002/ecy.1799}
\end{barticle}
\endbibitem

\bibitem[\protect\citeauthoryear{Stock et~al.}{2017}]{Stock2017}
\begin{barticle}[author]
\bauthor{\bsnm{Stock},~\bfnm{Michiel}\binits{M.}},
  \bauthor{\bsnm{Poisot},~\bfnm{Timoth{\'{e}}e}\binits{T.}},
  \bauthor{\bsnm{Waegeman},~\bfnm{Willem}\binits{W.}} \AND \bauthor{\bsnm{{De
  Baets}},~\bfnm{Bernard}\binits{B.}}
(\byear{2017}).
\btitle{{Linear filtering reveals false negatives in species interaction
  data}}.
\bjournal{Scientific Reports}
\bvolume{7}
\bpages{1--8}.
\bdoi{10.1038/srep45908}
\end{barticle}
\endbibitem

\bibitem[\protect\citeauthoryear{Streicker et~al.}{2010}]{Streicker2010}
\begin{barticle}[author]
\bauthor{\bsnm{Streicker},~\bfnm{Daniel~G}\binits{D.~G.}},
  \bauthor{\bsnm{Turmelle},~\bfnm{Amy~S}\binits{A.~S.}},
  \bauthor{\bsnm{Vonhof},~\bfnm{Maarten~J}\binits{M.~J.}},
  \bauthor{\bsnm{Kuzmin},~\bfnm{Ivan~V}\binits{I.~V.}},
  \bauthor{\bsnm{McCracken},~\bfnm{Gary~F}\binits{G.~F.}} \AND
  \bauthor{\bsnm{Rupprecht},~\bfnm{Charles~E}\binits{C.~E.}}
(\byear{2010}).
\btitle{{Host phylogeny constrains cross-species emergence and establishment of
  rabies virus in bats.}}
\bjournal{Science}
\bvolume{329}
\bpages{676--9}.
\bdoi{10.1126/science.1188836}
\end{barticle}
\endbibitem

\bibitem[\protect\citeauthoryear{Swendsen and Wang}{1987}]{Swendsen87}
\begin{barticle}[author]
\bauthor{\bsnm{Swendsen},~\bfnm{Robert~H.}\binits{R.~H.}} \AND
  \bauthor{\bsnm{Wang},~\bfnm{Jian-Sheng}\binits{J.-S.}}
(\byear{1987}).
\btitle{Nonuniversal critical dynamics in {Monte Carlo} simulations}.
\bjournal{Physical Review Letters}
\bvolume{58}
\bpages{86--88}.
\end{barticle}
\endbibitem

\bibitem[\protect\citeauthoryear{Teh and Gorur}{2009}]{teh2009indian}
\begin{binproceedings}[author]
\bauthor{\bsnm{Teh},~\bfnm{Yee~W}\binits{Y.~W.}} \AND
  \bauthor{\bsnm{Gorur},~\bfnm{Dilan}\binits{D.}}
(\byear{2009}).
\btitle{Indian buffet processes with power-law behavior}.
In \bbooktitle{Advances in Neural Information Processing Systems}
\bvolume{22}
\bpages{1838--1846}.
\end{binproceedings}
\endbibitem

\bibitem[\protect\citeauthoryear{Wardeh et~al.}{2015}]{wardeh2015database}
\begin{barticle}[author]
\bauthor{\bsnm{Wardeh},~\bfnm{Maya}\binits{M.}},
  \bauthor{\bsnm{Risley},~\bfnm{Claire}\binits{C.}},
  \bauthor{\bsnm{McIntyre},~\bfnm{Marie~Kirsty}\binits{M.~K.}},
  \bauthor{\bsnm{Setzkorn},~\bfnm{Christian}\binits{C.}} \AND
  \bauthor{\bsnm{Baylis},~\bfnm{Matthew}\binits{M.}}
(\byear{2015}).
\btitle{Database of host-pathogen and related species interactions, and their
  global distribution}.
\bjournal{Scientific Data}
\bvolume{2}.
\end{barticle}
\endbibitem

\bibitem[\protect\citeauthoryear{Webb et~al.}{2002}]{Webb2002}
\begin{barticle}[author]
\bauthor{\bsnm{Webb},~\bfnm{Campbell~O.}\binits{C.~O.}},
  \bauthor{\bsnm{Ackerly},~\bfnm{David~D.}\binits{D.~D.}},
  \bauthor{\bsnm{McPeek},~\bfnm{Mark~A.}\binits{M.~A.}} \AND
  \bauthor{\bsnm{Donoghue},~\bfnm{Michael~J.}\binits{M.~J.}}
(\byear{2002}).
\btitle{{Phylogenies and Community Ecology}}.
\bjournal{Annual Review of Ecology and Systematics}
\bvolume{33}
\bpages{475--505}.
\bdoi{10.1146/annurev.ecolsys.33.010802.150448}
\end{barticle}
\endbibitem

\bibitem[\protect\citeauthoryear{Weir and Pettitt}{2000}]{WeirPett2000}
\begin{barticle}[author]
\bauthor{\bsnm{Weir},~\bfnm{I.~S.}\binits{I.~S.}} \AND
  \bauthor{\bsnm{Pettitt},~\bfnm{A.~N.}\binits{A.~N.}}
(\byear{2000}).
\btitle{Binary probability maps using a hidden conditional autoregressive
  {Gaussian} process with an application to {Finnish} common toad data}.
\bjournal{Journal of the Royal Statistical Society: Series C (Applied
  Statistics)}
\bvolume{49}
\bpages{473--484}.
\end{barticle}
\endbibitem

\bibitem[\protect\citeauthoryear{Wiens et~al.}{2010}]{Wiens2010}
\begin{barticle}[author]
\bauthor{\bsnm{Wiens},~\bfnm{John~J}\binits{J.~J.}},
  \bauthor{\bsnm{Ackerly},~\bfnm{David~D}\binits{D.~D.}},
  \bauthor{\bsnm{Allen},~\bfnm{Andrew~P}\binits{A.~P.}},
  \bauthor{\bsnm{Anacker},~\bfnm{Brian~L}\binits{B.~L.}},
  \bauthor{\bsnm{Buckley},~\bfnm{Lauren~B}\binits{L.~B.}},
  \bauthor{\bsnm{Cornell},~\bfnm{Howard~V}\binits{H.~V.}},
  \bauthor{\bsnm{Damschen},~\bfnm{Ellen~I}\binits{E.~I.}},
  \bauthor{\bsnm{{Jonathan Davies}},~\bfnm{T}\binits{T.}},
  \bauthor{\bsnm{Grytnes},~\bfnm{John-Arvid}\binits{J.-A.}},
  \bauthor{\bsnm{Harrison},~\bfnm{Susan~P}\binits{S.~P.}},
  \bauthor{\bsnm{Hawkins},~\bfnm{Bradford~a}\binits{B.~a.}},
  \bauthor{\bsnm{Holt},~\bfnm{Robert~D}\binits{R.~D.}},
  \bauthor{\bsnm{McCain},~\bfnm{Christy~M}\binits{C.~M.}} \AND
  \bauthor{\bsnm{Stephens},~\bfnm{Patrick~R}\binits{P.~R.}}
(\byear{2010}).
\btitle{{Niche conservatism as an emerging principle in ecology and
  conservation biology.}}
\bjournal{Ecology Letters}
\bvolume{13}
\bpages{1310--24}.
\bdoi{10.1111/j.1461-0248.2010.01515.x}
\end{barticle}
\endbibitem

\bibitem[\protect\citeauthoryear{Williams and Martinez}{2000}]{Williams2000}
\begin{barticle}[author]
\bauthor{\bsnm{Williams},~\bfnm{Richard~J.}\binits{R.~J.}} \AND
  \bauthor{\bsnm{Martinez},~\bfnm{Neo~D.}\binits{N.~D.}}
(\byear{2000}).
\btitle{{Simple rules yield complex food webs}}.
\bjournal{Nature}
\bvolume{404}
\bpages{180--183}.
\bdoi{10.1038/35004572}
\end{barticle}
\endbibitem

\end{thebibliography}
